\documentclass[12pt,fleqn]{article}
\usepackage[a4paper, top=2.4cm, bottom=2.4cm, left=2.4cm, right=2.4cm]{geometry}
\usepackage{natbib}
\setcitestyle{authoryear,round}
\usepackage{lipsum}
\makeatletter
\def\ps@pprintTitle{%
 \let\@oddhead\@empty
 \let\@evenhead\@empty
 \def\@oddfoot{}%
 \let\@evenfoot\@oddfoot}
\makeatother
\usepackage{amsmath}
\usepackage{graphicx}
\usepackage{ragged2e}
\usepackage{float}
\usepackage{booktabs}
\usepackage{enumerate}
\usepackage[normalem]{ulem}

\DeclareMathOperator*{\argmin}{arg\,min}
\usepackage{lscape}

\usepackage{arydshln}
\usepackage{mathtools}
\usepackage{afterpage}
\usepackage{amsfonts}
\usepackage{algorithm,algpseudocode}
\usepackage{arydshln}
\usepackage[scr]{rsfso}
\usepackage{multirow}
\usepackage[dvipsnames]{xcolor}
\usepackage[flushleft]{threeparttable}
\usepackage{siunitx}
\usepackage{caption}
\usepackage{rotating}
\usepackage{amsfonts}
\usepackage{booktabs}
\usepackage{siunitx}
\usepackage{tabularx}
\usepackage{siunitx}
\usepackage{amsthm}
\usepackage{amssymb,bm}
\usepackage{amsfonts}
\definecolor{applegreen}{rgb}{0.55, 0.71, 0.0}
\definecolor{ao(english)}{rgb}{0.0, 0.5, 0.0}
\usepackage{booktabs,caption}
\usepackage[pdftex]{hyperref}
\usepackage{setspace}
\usepackage{bbold}
\usepackage{appendix}
\DeclareMathOperator{\vecop}{vec}

\usepackage{dsfont}
\usepackage{comment}
\usepackage{color}
\usepackage[dvipsnames]{xcolor}
\usepackage{subfig}
\usepackage[normalem]{ulem}

\newcommand{\add}[1]{{\textcolor{purple}{#1}}}

\newcommand{\smout}[1]{\hbox{\sout{$#1$}}\;}
\newcommand{\abs}[1]{{\left\lvert#1\right\rvert}}
\newcommand{\norm}[1]{{\left\lVert#1\right\rVert}}
\DeclareMathOperator*{\plim}{plim}
\newcommand{\E}{\mathbb{E}}
\newcommand{\M}{\mathcal{M}}
\renewcommand{\P}{\mathcal{P}}
\theoremstyle{definition}

\newtheorem{assumption}{Assumption}

\newtheorem{remark}{Remark}
\theoremstyle{plain}
\newtheorem{theorem}{Theorem}
\newtheorem{lemma}{Lemma}

\newcommand{\GCing}{J}
\newcommand{\GCed}{I}
\newcommand{\nGCing}{N_J}
\newcommand{\nGCed}{N_I}
\newcommand{\nGCvars}{N_{GC}}
\newcommand{\nXGC}{N_X}
\newcommand{\GCnot}{-(\GCed \cup \GCing)}
\allowdisplaybreaks

\newcommand{\bA}{\bm A}
\newcommand{\ba}{\bm a}
\newcommand{\bB}{\bm B}
\newcommand{\bb}{\bm b}

\newcommand{\bc}{\bm c}

\newcommand{\bE}{\bm E}
\newcommand{\be}{\bm e}

\newcommand{\bG}{\bm G}

\newcommand{\bI}{\bm I}

\newcommand{\bM}{\bm M}

\newcommand{\bR}{\bm R}
\newcommand{\br}{\bm r}

\newcommand{\bU}{\bm U}
\newcommand{\bu}{\bm u}

\newcommand{\bX}{\bm X}
\newcommand{\bx}{\bm x}
\newcommand{\bY}{\bm Y}
\newcommand{\by}{\bm y}
\newcommand{\bZ}{\bm Z}

\newcommand{\bzero}{\bm 0}

\newcommand{\bbeta}{\bm \beta}
\newcommand{\bgamma}{\bm \gamma}
\newcommand{\bdelta}{\bm \delta}
\newcommand{\bepsilon}{\bm \epsilon}

\newcommand{\bnu}{\bm \nu}
\newcommand{\bxi}{\bm \xi}
\newcommand{\bpi}{\bm \pi}

\newcommand{\bsigma}{\bm \sigma}

\newcommand{\bGamma}{\bm \varGamma}

\newcommand{\bXi}{\bm \varXi}
\newcommand{\bPi}{\bm \varPi}
\newcommand{\bSigma}{\bm \varSigma}

\newcommand{\bOmega}{\bm \varOmega}

\title{Granger Causality Testing in High-Dimensional VARs:\\ a Post-Double-Selection Procedure}
\author{Alain Hecq \and Luca Margaritella\thanks{The second and third author thank NWO for financial support. Previous versions of this paper have been presented at NESG 2018, CFE-CM Statistics 2018, $EC^2$ 2018, 4th Vienna Workshop on High Dimensional Time Series in Macroeconomics and Finance 2019, QFFE 2019, IAAE 2019 and $22^{nd}$ Dynamic Econometrics Conference 2019. We gratefully acknowledge the comments by participants at these conferences. In addition, we thank Marcelo Medeiros, Anders B. Kock, Peter Pedroni, Etienne Wijler and Benoit Duvocelle for their valuable feedback. All remaining errors are our own.
Address correspondence to: Luca Margaritella, Department of Quantitative Economics, Maastricht University, P.O. Box 616, 6200 MD Maastricht, The Netherlands. E-mail: \href{mailto:l.margaritella@maastrichtuniversity.nl}{\textcolor{blue}{l.margaritella@maastrichtuniversity.nl}}} \and Stephan Smeekes}
\date{Department of Quantitative Economics\\ School of Business and Economics\\Maastricht University\\ \bigskip December 2020}

\begin{document}
\maketitle
\begin{abstract}
We develop an LM test for Granger causality in high-dimensional VAR models based on penalized least squares estimations. To obtain a test retaining the appropriate size after the variable selection done by the lasso, we propose a post-double-selection procedure to partial out effects of nuisance variables and establish its uniform asymptotic validity. We conduct an extensive set of Monte-Carlo simulations that show our tests perform well under different data generating processes, even without sparsity. We apply our testing procedure to find networks of volatility spillovers and we find evidence that causal relationships become clearer in high-dimensional compared to standard low-dimensional VARs.

\end{abstract}

\doublespacing

\section{Introduction}
Economics, statistics and finance have seen a rapid increase of applications involving time series in high-dimensional systems. Central to many of these applications is the vector autoregressive (VAR) model that allows for a flexible modelling of dynamic interactions between multiple time series. In this paper we develop a simple method to test for Granger causality in high-dimensional VARs (HD-VARs) with potentially many variables.

Many financial applications consider Granger causality analysis, especially for constructing high-dimensional networks. Networks of financial firms'  intedependencies are investigated in \citet{basu2015network}, \citet{gao2017efficient}, \citet{demirer2018estimating} and \citet{barigozzi2019nets}. Similarly, spillovers and contagion among stock returns are investigated in networks using Granger causality analysis in \citet{lin2017regularized}, \citet{vyrost2015granger} and \citet{corsi2018measuring}.

Most of the econometric literature has traditionally been focused on allowing for high dimensionality in VARs through the use of factor models \citep[see e.g.][]{bernanke2005measuring,chudik2016theory} or Bayesian methods \citep{banbura2010large}. For instance \citet{billio2012econometric} develops measures of connectedness to assess systemic risk propagation among institutions in the financial system using principal component analysis and Granger causality networks. Recent years have seen an increase in \textit{regularized}, or penalized, estimation of \textit{sparse} VARs based on popular methods from statistics such as the lasso \citep{tibshirani1996regression} and elastic net \citep{zou2005regularization}, which impose sparsity by setting a (data-driven) selection of the coefficients to zero.

Compared to factor models, such sparsity-seeking methods have often an advantage of interpretability, as in many economic applications, it appears natural to believe that the most important dynamic interactions among a large set of variables can be adequately captured by a relatively small -- but unknown -- number of `key' variables. As such, the use of these methods for estimating HD-VAR models has also increased significantly in recent years, see e.g. \citet{nicholson2017varx}, \citet{basu2019low}, \citet{billio2019bayesian}, \citet{wilms2018algorithm}, \citet{korobilis2019adaptive}).

Regularized estimation theory for high-dimensional time series and VAR models is now well established, see among others \citet{song2011large}, \citet{basu2015regularized}, \citet{kock2015oracle}, \citet{davis2016sparse}, \citet{medeiros2016l1}, \citet{audrino2018oracle} and \citet{masini2019regularized} and \citet{wong2020lasso};  \citet{kock2020penalized} provide a recent review. However, performing inference on HD-VARs, such as testing for Granger causality, still remains a non-trivial matter. As is well known, performing inference after model selection (post-selection inference) is complicated as the selection step invalidates `standard' inference where the uncertainty regarding the selection is ignored \citep[see][]{leeb2005model}. 
Complexities introduced by the temporal and cross-sectional dependencies in the VAR mean that most recently developed post-selection inference methods are not automatically applicable. 

Most existing literature on Granger causality testing in HD-VARs therefore has so far not considered post-selection inferential procedures. 
\citet{wilms2016predictive} propose a bootstrap Granger causality test in HD-VARs, but do not account for post-selection issues. Similarly, \citet{skripnikov2018joint} investigate the problem of jointly estimating multiple network Granger causal models in VARs with sparse transition matrices using lasso-type methods, but focus mostly on estimation rather than testing. \citet{song2019better} focus on statistical procedures for testing indirect/spurious causality in high-dimensional scenarios, but consider factor models rather than regularized regression techniques. \citet{lin2017regularized} consider high-dimensional multi-block VARs derived from a two-blocks recursive linear dynamical system and use a maximum likelihood (ML) estimator for Gaussian data. In order to obtain the ML estimates for the system transition matrices and the precision matrix, respectively the lasso and graphical lasso on the residuals are iterated until convergence. \citet{krampe2018bootstrap} develops bootstrap techniques for sparse VAR models combining a model-based bootstrap procedure and the de-sparsified lasso (see \citet{van2014asymptotically}) to perform inference on the autoregressive parameters. 
\citet{chaudhry2017uncertainty} look at de-biased estimators as in \citet{javanmard2014confidence}, for Gaussian and sub-Gaussian VAR processes with a focus on Granger-causality and control of the false discovery rate.

In this paper we build on the post-double-selection approach proposed by \citet{belloni2014inference}, to develop a valid post-selection test of Granger causality in HD-VARs. The finite-sample performance depends heavily on the exact implementation of the method. In particular, the tuning parameter selection in the penalized estimation is crucial. We therefore perform an extensive simulation study to investigate the finite-sample performance of the different ways to set up the test in order to be able to give some practical recommendations. 
In addition, we investigate the construction of networks of realized volatilities using a sample of 30 financial stocks modeled as a vector heterogeneous VAR \citep{corsi2009simple}. We are able to demonstrate how our approach allows for obtaining much sharper conclusions than standard low-dimensional VAR techniques.

The remainder of the paper is as follows:  Section \ref{sec:model} introduces the high-dimensional VAR model and Granger causality tests. In Section \ref{sec:est} we propose our estimation and inferential framework. Section \ref{sec:asy} establishes the asymptotic properties of our method and discusses the assumptions required for the theory to hold.
Section \ref{sec:sim} reports the results of the Monte Carlo simulations. We apply our method in Section \ref{sec:appl} to construct volatility spillover networks.
Section \ref{sec:conc} concludes. Proofs and supplemental results can be found in the appendix.

A few words on notation. For any $n$-dimensional vector $\bx$, we let $\norm{\bx}_p = \left(\sum_{i=1}^n \abs{x_i}^p \right)^{1/p}$ denote the $\ell_p$-norm. For any index set $S \subseteq \{1, \ldots, n\}$, let $\bx_{S}$ denote the sub-vector of $\bx_t$ containing only those elements $x_i$ such that $i \in S$. $\abs{S}$ denotes the cardinality of the set $S$. We use $\xrightarrow{p}$ and $\xrightarrow{d}$ to denote convergence in probability and distribution, respectively.

\section{High-dimensional Granger causality tests} \label{sec:model}
Loosely speaking, the notion of Granger causality captures predictability given a particular information set \citep{granger1969investigating,granger1980testing}. If the addition of variable $X$ to the given information set $\Omega$ alters the conditional distribution of another variable $Y$, and both $X$ and $\Omega$ are observed prior to $Y$, then $X$ improves predictability of $Y$, and is said to \emph{Granger cause} $Y$ with respect to $\Omega$. \citet{granger1969investigating} originally envisioned the information set $\Omega$ ``be all the information in the universe'' (p.~428), which is of course not a workable concept. Yet clearly the choice of information set has a major effect on the interpretation of the finding of (non-)Granger causality, as discussed in \citet{granger1980testing}. In particular, spurious Granger causality from $X$ to $Y$ may be found when both $X$ and $Y$ are Granger caused by $Z$, but $Z$ is omitted from $\Omega$. As such, one might want to include as many potentially relevant variables in the information set as possible in order to avoid finding spurious causality due to omitted variables, thereby moving as much as possible towards the universal information set envisioned by Granger. However, conditioning on so many variables leads to obvious problems of high-dimensionality rendering many standard statistical techniques invalid.

In this paper we focus on testing Granger causality in mean using linear models, in which setup the VAR model is the natural tool to investigate this problem.
However, to enlarge the information set means estimating a VAR with an increasing number of variables. The number of parameters in a VAR increases quadratically with the number of time series included; an unrestricted VAR($p$) has $K^2p$ coefficients to be estimated, where $K$ is the number of series and $p$ is the lag-length. As the time series dimension $T$ is typically fairly small for many economic applications, the data do not contain sufficient information to estimate the parameters and consequently standard least squares and maximum likelihood methods suffer from the curse of dimensionality, resulting in estimators with high variance that overfit the data.
\subsection{Granger causality testing in VAR models}
Let $\by_1,\ldots,\by_T$ be a $K$-dimensional multiple time series process, where $\by_t=(y_{1,t},\ldots,y_{K,t})^{'}$ is generated by a VAR($p$) process 
\begin{equation} \label{eq:dgp}
\by_t=\bA_1 \by_{t-1}+\cdots+\bA_{p}y_{t-{p}}+\bu_t, \qquad t=p+1,\ldots,T\;,
\end{equation}
where for notational simplicity we assume the variables have zero mean; if not they can be demeaned prior to the analysis, or equivalently a vector of intercepts is added. $\bA_1,\ldots,\bA_{p}$ are $K\times K$ parameter matrices and $\bu_t$ is a martingale difference sequence (mds) of error terms. We consider weakly stationary VAR models, as formalized in Assumption \ref{as:VAR} below. 
\begin{assumption}\label{as:VAR}
The VAR model in \eqref{eq:dgp} satisfies:
\begin{enumerate}[(a)]
\item $\{\bu_t\}_{t=1}^T$ is a weakly stationary mds with respect to $\mathcal{F}_t = \sigma(\by_t, \by_{t-1}, \by_{t-2}, \ldots)$ $\bu_t$ such that $\E (\bu_t| \mathcal{F}_{t-1}) = \bzero$ for all $t$ and $\bSigma_{u} = \E (\bu_t \bu_t^\prime)$ is positive definite. \item All roots of $\det(\bI_{K}-\sum_{j=1}^{p} \bA_j z^j)$ lie outside the unit disc, such that the lag polynomial is invertible.
\end{enumerate}
\end{assumption}
 
In the VAR model \eqref{eq:dgp} we are interested in testing whether variables in the set $\GCing$ Granger cause variables in the set $\GCed$ in mean, conditional on all the other variables, where $\GCing, \GCed \subset \{1,\ldots, K\}$ and $\GCing \cap \GCed = \emptyset$. Let $\nGCed = \abs{\GCed}$ and $\nGCing = \abs{\nGCing}$ denote the number of variables in $I$ and $J$ respectively. We describe our proecdure here in general form for testing blocks of variables. For any sets $S_1, S_2 \subseteq \{1, \ldots, K\}$ of variables define the best linear predictor in $L_2$-norm of $\by_{S_1,t}$ given $\bx_{S_2,t-1}^{(p)} = (\by_{S_2,t-1}^\prime, \ldots, \by_{S_2, t-p}^\prime)^\prime$ as $\mathcal{P} (\by_{S_1,t}|\bx_{S_2,t-1}^{(p)}) = \bGamma^{*} \bx_{S_2,t-1}^{(p)}$, where $\bGamma^* = \min_{\bGamma} \E \left[\left.\norm{\by_{S_1,t} - \bGamma \bx_{S_2, t-1}}_2^2 \right. \right]$. Then we say that $\by_{\GCing,t}$ \emph{does not Granger cause} $\by_{\GCed,t}$ conditionally on $\bx_{\GCing^c,t}$ if
\begin{equation} \label{eq:GCdef}
\mathcal{P} (\by_{\GCed,t} | \bx_{\GCing^c,t}^{(p)}) = \mathcal{P} (\by_{\GCed,t} | \bx_{t-1}^{(p)})
\end{equation}
for any value of $\bx_{\GCing^c,t}$. In other words, conditional on $\bx_{\GCing^c,t}$, addition of the lags of $\by_{\GCing,t}$ to the information set does not improve predictability of $\by_{\GCed,t}$. Note that Granger (non-)causality as defined in \eqref{eq:GCdef} is a property of the population. In the VAR \eqref{eq:dgp} this means that testing for Granger causality can be done via testing the joint significance of the blocks of coefficients in the matrices $\bA_1, \ldots, \bA_p$ corresponding to the impact of variables $\GCing$ on $\GCed$.

To illustrate, consider \eqref{eq:dgp} with $p=1$ lag, and assume without loss of generality that the variables in $\by_t$ are ordered such that $\by_t = \left(\by_{\GCed,t}^\prime, \by_{\GCing, t}^\prime, \by_{\GCnot, t}^\prime\right)^\prime$, where $\GCnot)$ refers to all variables not in $\GCing$ or $\GCed$. Then we can write
\begin{equation} \label{eq:dgp_p1}
\begin{bmatrix}
\by_{\GCed,t} \\ \by_{\GCing,t} \\ \by_{\GCnot,t}
\end{bmatrix}
=
\begin{bmatrix}
\bA_{\GCed,\GCed} & \bA_{\GCed, \GCing} & \bA_{\GCed, \GCnot} \\
\bA_{\GCing,\GCed} & \bA_{\GCing, \GCing} & \bA_{\GCing, \GCnot} \\
\bA_{\GCed, \GCnot} & \bA_{\GCnot} & \bA_{\GCnot, \GCnot} \\
\end{bmatrix}
\begin{bmatrix}
\by_{\GCed,t-1} \\ \by_{\GCing,t-1} \\ \by_{-(\GCing \cup \GCed),t-1}
\end{bmatrix}
+ \bu_t,
\end{equation}
where $\bA$ is partitioned conformably with the blocks in $\by_t$. In this case, the best linear predictors in \eqref{eq:GCdef} are given by
\begin{equation*}
\begin{split}
&\mathcal{P} (\by_{\GCed,t} | \by_{t-1}) = \bA_{\GCed, \GCed} \by_{\GCed, t-1} + \bA_{\GCed, \GCing} \by_{\GCing, t-1} + \bA_{\GCed, -(\GCed \cup \GCing)} \by_{-(\GCed \cup \GCing), t-1}, \\
&\mathcal{P} (\by_{\GCed,t} | \by_{\GCing^c,t-1}) = \bA_{\GCed}^* \by_{\GCing^c, t-1}, \text{ where } \bA_{\GCed}^* = \min_{\bA_{\GCed}} \E \left[\left.\norm{\by_{\GCed,t} - \bA_{\GCed} \by_{\GCing^c, t-1}}_2^2 \right. \right].
\end{split}
\end{equation*}
For any arbitrary value of $\by_{t-1}$, these can only coincide if $\bA_{\GCed, \GCing} = \bm 0$. Hence, the null hypothesis of no Granger causality from $\GCing$ to $\GCed$ in the VAR($1$) model can be formulated in terms of $\bA_{\GCed, \GCing} = \bm 0$. This is easily extended to $p>1$ by simply testing if the $(\GCed, \GCing)$-block of all $p$ lag matrices is equal to zero.

In the remainder of the paper, we will be working with a stacked representation of \eqref{eq:dgp} for the variables in $\GCed$. Specifically, let $\bY = \left(\by_{p+1}, \ldots, \by_T\right)^\prime$ and let $\by_{\GCed} = \vecop\left(\bY_{\GCed}\right)$ denote the $\nGCed \times 1$ stacked vector containing all observations corresponding to the variables in $\GCed$. Similarly, let $\bu_{\GCed} = \vecop(\bU_{\GCed})$, where $\bU = \left(\bu_{p+1}, \ldots, \bu_T\right)^\prime$. Let $\bX = \left(\bx_{p}^{(p)}, \ldots, \bx_{T-1}^{(p)}\right)^\prime$ and $\bX^{\otimes} = \bI_{\nGCed} \otimes \bX$, while defining the stacked parameter vector $\bbeta = \vecop((\bA_1, \ldots, \bA_p)^\prime)$. Then we can write
\begin{equation} \label{eq:dgp_i}
\by_{\GCed} = \bX^{\otimes} \bbeta + \bu_{\GCed} = \bX_{GC}^{\otimes} \bbeta_{GC} + \bX_{-GC}^{\otimes} \bbeta_{-GC} + \bu_{\GCed},
\end{equation}
where $\bX_{GC}^{\otimes} = \bI_{\nGCed} \otimes \bX_{GC}$, and $\bX_{GC} = \left(\bx_{\GCing, p}^{(p)}, \ldots, \bx_{\GCing, T-1}^{(p)}\right)^\prime$ contains those columns of $\bX$ corresponding to the potentially Granger causing  variables in $\GCing$; $\bX_{-GC}$ and $\bX_{-GC}^{\otimes}$ are then defined similarly but containing the remaining variables.\footnote{Note that if $\GCed = \{i\}$ for one particular value of interest, then \eqref{eq:dgp_i} simply corresponds to a single equation from the VAR in \eqref{eq:dgp}.} Testing for no Granger causality is then equivalent to testing $H_0: \bbeta_{GC}=\bzero$ against $H_1: \bbeta_{GC}\neq \bzero$.

Define $\nGCing = \abs{\GCing}$ and $\nGCed = \abs{\GCed}$. Note that $\bbeta_{-GC}$ has $\left(K - \nGCing \right) \times \nGCed \times p$ elements, which we assume large through having a large number of variables $K$. On the other hand, throughout the paper we assume that $\nGCing$, $\nGCed$ and $p$ are small, or more precisely, fixed when sample size increases to infinity. As $\bbeta_{GC}$ has $\nGCvars = \nGCing \times \nGCed \times p$ elements, these are also implied to be fixed. While theoretically it is possible to consider an increasing number of elements in $\bbeta_{GC}$ (see Remark \ref{rem:multGC} for details), it would not be required for typical applications. $\GCing$ and $\GCed$ are under the researcher's control and in most applications it is natural to consider a small number of variables of interest; often both $\GCing$ and $\GCed$ will only consist of a single variable, as in our application.

For $p$ it may appear more restrictive to assume it small. However, large $p$ in univariate regressions or small systems often arise from neglected dynamics with omitted variables \citep{hecq2016univariate}. 
As our HD-VAR attempts to include many more variables than typical small systems, we hope to alleviate the omitted variable issue, and thereby also directly making smaller $p$ much more realistic. Of course, $p$ is generally unknown in practice. However, in many applications it is possible to give a reasonable (and small) upper bound on $p$, which is sufficient for our algorithm. If not, $p$ has to be estimated. We discuss two ways in the next section.

\begin{remark}
 Our operational version of Granger causality only considers causality in mean. Additionally, one might argue that considering only linear models is a further restriction on the generality of the concept of Granger causality. However, in our high-dimensional approach linear models are less restrictive as would appear. First, the VAR does not have to be formulated for levels of variables of interest. In fact, in our application we formulate a VAR for (realized) variances, such that we are implicitly testing Granger causality in second moments rather than first moments. Second. the linear VAR model in many cases provides a good approximation to a general nonlinear process via a Wold-type representation argument, see e.g.~\citet{meyer2015on}. Finally, non-linear transformations (such as powers) of the original variables can be added to \eqref{eq:dgp_i}, by which general functional forms can be approximated (even if one then strictly loses the VAR equivalence). While in small systems this is infeasible as it increases the dimensionality dis-proportionally, our high-dimensional approach can handle this without any conceptual issues. In fact, \citet{belloni2014high} explicitly motivate their high-dimensional linear approach as an approximation to a general function; their arguments apply here as well.
\end{remark}

\section{Inference after selection by the lasso} \label{sec:est}
In this section we introduce our inferential procedure to the Granger causality tests in high-dimensional VARs. We first discuss the lasso, which we use in the initial stage to select relevant variables. Next we discuss how naive use of the lasso introduces post-selection problems for inference, and we propose our algorithm to remedy this.

\subsection{The lasso estimator}
As $\bbeta$ is high-dimensional when $Kp$ is large relative to $T$, least squares estimation is not appropriate, and a structure must be imposed on $\bbeta$ to be able to estimate it consistently. We assume \textit{sparsity} of $\bbeta$; that is, we assume that $\bbeta$ can accurately be approximated by a coefficient vector with a (significant) portion of the coefficients equal to zero.

The sparsity assumption validates the use of variable selection methods, thereby reducing the dimensionality of the system without having to sacrifice predictability. For a general $n$-dimensional vector of responses $\by$ and $n \times M$-dimensional matrix of covariates $\bX$, the (weighted) lasso simultaneously performs variable selection and estimation of the parameters by solving
\begin{equation} \label{eq:lasso}
\hat{\bbeta}(\lambda) = \argmin_{\bbeta} \left(\frac{1}{T} \norm{\by- \bX \bbeta}_2^2 + \lambda \sum_{m=1}^{M} \abs{w_m \beta_m} \right),
\end{equation}
where $\lambda$ is a non-negative tuning parameter determining the strength of the penalty, and $\{w_m\}_{m=1}^{M}$ are non-negative weights corresponding to the parameters in $\bbeta$. For the standard lasso the weights are either equal to one, or equal to zero (if this parameter should not be penalized). The notation $\hat{\bbeta}(\lambda)$ highlights that the solution to the minimization problem depends on $\lambda$, which has te be selected as well (see Section \ref{sub:tunpa}). When no confusion can arise, we simply write $\hat{\bbeta}$.

One may also consider the \textit{adaptive lasso} \citep{zou2006adaptive} with parameter-specific weights $w_j$ in \eqref{eq:lasso} based on an initial estimation of $\bbeta$, which is able to delete more irrelevant variables. However, for our purpose such oracle properties are not very relevant; we wish to eliminate the effects of the other ``nuisance'' variables on the relation between the variables tested for Granger causality, but we do not need to identify which of these nuisance variables matter.

Theoretical properties of lasso estimation in stable VAR models have now been studied extensively. We here non-exhaustively mention some of the key results for our setting; see \citet{kock2020penalized} for a thorough review. \citet{kock2015oracle} derive oracle properties of the adaptive lasso for VAR models. \citet{basu2015regularized} establish restricted eigenvalue conditions for VAR models and show their sufficiency for estimation consistency. \citet{medeiros2016l1} relax the Gaussianity assumptions of these papers by considering conditionally heteroskedastic errors, and demonstrate that the adaptive lasso retains oracle properties in time series settings. Finally, \citet{masini2019regularized} derive bounds on estimation errors in approximately sparse VAR models under very general conditions, allowing for heavy tails and dependence in the error terms. In particular, they show that several commonly used volatility processes in financial research satisfy these assmuptions, thereby formally establishing the suitability of the lasso for many financial applications of VAR models.

\subsection{Post-Double-Selection Granger causality test}\label{sec:posi}
\subsubsection{The need for post selection inference}
One might be tempted to simply perform the (adaptive) lasso as in \eqref{eq:lasso} on \eqref{eq:dgp_i}, setting $w_{GC} = 0$, and then testing whether $\bbeta_{GC}=0$, potentially after re-estimating the model by OLS on only the selected variables. However, this ignores the fact that the final, selected, model is random and a function of the data. The randomness contained in the selection step means the post-selection estimators do not converge uniformly to a normal distribution, as the potential omitted variable bias from omitting (weakly) relevant variables in the selection step is too large to maintain uniformly valid inference. 

In a sequence of papers \citep[see e.g.][]{leeb2005model}, Leeb and P\"otscher address these issues, showing that distributions of post-selection estimators only converge point-wise but not uniformly in the parameter space to normal distributions. Therefore, ``standard'' asymptotics fail to deliver a proper approximation of finite-sample behavior due to the presence of small, hard to detect parameters, whose omitted variable bias is too large to ignore asymptotically. As such, post-selection based on oracle properties is only appropriate if one a priori rules out small parameters conditions (via \textit{beta-min} conditions, see e.g.~\citealp{van2011adaptive}) thus obtaining a sharp separation of non-zero from zero coefficients. This is typically far too strong to be reasonable in applications, and methods explicitly accounting for selection are required.

Several approaches to valid post-selection inference, also referred to as \emph{honest inference}, have been developed in recent years based on various philosophies, such as simultaneous inference across models \citep{berk2013valid}, inference conditional on selected models \citep{lee2016exact}, or debiasing (desparsifying) the lasso estimates \citep{van2014asymptotically,zhang2014confidence}. We focus on the double selection approach developed by Belloni, Chernozhukov and co-authors; see e.g.~\citep{belloni2014high} for an overview. This approach is tailored for the lasso, easy to implement, and can be extended to dependent data.

\citet{belloni2014uniform} develop a \textit{post-double-selection} approach to construct uniform inference for treatment effects in partially linear models with high-dimensional controls using the lasso. Two initial lasso estimations of both the outcome and the treatment variable on all the controls are performed, and a final post-selection least squares estimation is conducted of the outcome variable on the treatment variable and all the controls selected in \emph{at least} one of the two steps. The double variable selection step substantially diminishes the omitted variable bias and ensures the errors of the final model are (close enough to) orthogonal with respect to the treatment. The authors proved uniform validity of the procedure under a wide range of DGPs, including heteroskedastic and non-Gaussian errors. 

 \citet{chernozhukov2019lasso} extend the analysis of estimation and inference for highly-dimensional systems in regressions, allowing for (weak) temporal and cross-sectional dependency. Regularization techniques for dimensionality reduction are applied iteratively in the system and the overall penalty is jointly chosen by a block multiplier bootstrap procedure. Oracle properties and bootstrap consistency of the test procedure are derived. Furthermore, simultaneous valid inference is obtained via algorithms employing least square or least absolute deviation after (double) lasso selection step(s). Although our approach is closely related to that of \citet{chernozhukov2019lasso}, it differs in a number of ways. Our method is simpler and faster to implement as it does not rely on bootstrap methods. Also, \citet{chernozhukov2019lasso} focus on general systems of equations and general ways of performing inference, which is different from our specific focus on Granger causality and VAR models. Third, we consider a different set of assumptions to establish the validity of our method, where we specifically focus on the relevance of these assumptions for applications in financial econometrics.

\subsubsection{High-dimensional Granger causality test}
We here describe how to implement the post-double-selection procedure in a VAR context. Let $\bx_{GC,j}$, $j=1,\ldots, \nXGC$, where $\nXGC = p \nGCing$, denote the $j$-th column of $\bX_{GC}$ and consider the partial regressions:
\begin{align}
\by_{\GCed} &= \bX_{-GC}^{\otimes} \bgamma_0 + \be_0, \label{eq:lasso_steps1} \\
\bx_{GC,j} &= \bX_{-GC} \bgamma_j + \be_j, \qquad j = 1, \ldots, \nXGC, \label{eq:lasso_steps2}
\end{align}
where $\bgamma_j, \; j = 0,\ldots, \nXGC$, are the best linear prediction coefficients\footnote{Note that Assumption \ref{as:hlev}(\ref{as:eig}) implies that $\left(\E \bx_{-GC,t-1} \bx_{-GC,t-1} \right)^{-1}$ and hence $\left(\E \bX_{-GC,t-1}^{\otimes} \bX_{-GC,t-1}^{\otimes \prime} \right)^{-1}$ exist.}
\begin{equation*}
\begin{split}
\bgamma_{0} &= \argmin_{\bgamma} \E \norm{\by_{\GCed,t} - \bX_{-GC,t-1}^{\otimes\prime} \bgamma}_2^2 = \left(\E \bX_{-GC,t-1}^{\otimes} \bX_{-GC,t-1}^{\otimes \prime} \right)^{-1} \E \bX_{-GC,t-1}^{\otimes} \by_{i,t}, \\
\bgamma_j &= \argmin_{\bgamma} \E \norm{x_{GC,j,t} - \bx_{-GC,t-1}^\prime \bgamma}_2^2\\
&= \left(\E \bx_{-GC,t-1} \bx_{-GC,t-1}^\prime \right)^{-1} \E \bx_{-GC,t-1} \bx_{GC,j,t}, \qquad j = 1, \ldots, \nXGC,
\end{split}
\end{equation*}
where $\bX_{-GC,t}^{\otimes} = \bI_{\nGCed} \otimes \bx_{-GC,t-1}$.
As the errors $\be_0, \ldots, \be_{\nXGC}$ are orthogonal to $\bX_{-GC}$, partialling out the effects of these variables would allow for a valid test of Granger causality. Of course, \eqref{eq:lasso_steps1} and \eqref{eq:lasso_steps2} are still high-dimensional and cannot be estimated by least squares. However, we can select the relevant variables from lasso estimation of \eqref{eq:lasso_steps1} and \eqref{eq:lasso_steps2} and collect all these for the final estimation of $\by_{\GCed}$ on $\bX_{GC}^{\otimes}$ plus only those relevant variables.

Intuitively, this works because to cause omitted variable bias on the coefficients of $\bX_{GC}$, a particular variable in $\bX_{-GC}$ must have a nonzero coefficient in \emph{both} \eqref{eq:lasso_steps1} and one of the regressions in \eqref{eq:lasso_steps2}. If its coefficient is zero in \eqref{eq:lasso_steps1}, it has no effect on $\by_{\GCed}$ and is therefore not wrongfully omitted. If it has a zero coefficient in all regressions in \eqref{eq:lasso_steps2}, it is not correlated with any variables of interest, and omitting it will not result in a bias. By including all variables that are selected in at least a single of these regressions, we essentially allow for ``one free mistake'' by the lasso in failing to select a relevant variable. That is, omitted variable bias will only occur if the lasso fails to select a relevant variable in both regressions simultaneously. As the probability of this occurring decreases quadratically, this is sufficient to be negligible asymptotically and allow for uniformly valid inference. We provide a formal justification in Section \ref{sec:asy}.

We now state the details of our algorithm which executes the post-double-section along the lines described above, and conclude this section with some remarks.

\begin{algorithm}
\caption{Post-double-selection Granger causality LM (PDS-LM) test} \label{alg:pdslm}
\begin{itemize}
\item[\textbf{[1]}]
Estimate the initial partial regressions in \eqref{eq:lasso_steps1} and \eqref{eq:lasso_steps2} by an appropriate sparsity-inducing estimator such as the (adaptive) lasso, and let $\hat{\bgamma}_0$, \ldots, $\hat{\bgamma}_{\nXGC}$ denote the resulting estimators. Let $\hat{S}_{0}=\{m: \abs{\hat{\gamma}_{m,0}}>0, \; m=1,\ldots, N\}$ and $\hat{S}_{j} = \{m: \abs{\hat{\gamma}_{m,j}}>0, \; m=1,\ldots, \nXGC \}$ for $j=1,\ldots,p$, denote the selected variables in each regression.

\item[\textbf{[2]}] Let $\hat{S}_X = \bigcup_{j=1}^{\nXGC} \hat{S}_j$ denote all variables selected in the regressions for the columns of $\bX_{GC}$, and let  $\hat{S}_X^{\otimes}$ map $\hat{S}_X$ back to $\bX_{-GC}^{\otimes}$ be such that $\bX_{\hat{S}_X^{\otimes}}^{\otimes} = \bI_{\nGCed} \otimes \bX_{\hat{S}_X}$. Collect all variables kept by the lasso in Step [1] in $\hat{S}^{\otimes}= \hat{S}_0 \cup \hat{S}_X^{\otimes}$. Obtain the residuals $\hat{\bxi} = \by_{\GCed} - \bX_{\hat{S}^{\otimes}}^{\otimes} \hat{\bbeta}^{\dagger}$ by OLS estimation. Let $\hat\bXi_{\GCed}$ denote the $T\times \nGCed$-matrix formed from $\hat\bxi$ and construct $\hat{\bSigma}_{\epsilon,\GCed} = \hat{\bXi}_{\GCed}^\prime \hat{\bXi}_{\GCed}/T$ and $\hat{\bSigma}_{\epsilon,\GCed}^{\otimes} = \hat{\bSigma}_{\epsilon,\GCed} \otimes \bI_T$.

\item[\textbf{[3]}] Let $\by_{\nGCed}^* = \left(\hat{\bSigma}_{\epsilon,\GCed}^{\otimes}\right)^{-1/2} \by_{\nGCed}$ and $\bX^{*\otimes} = \left(\hat{\bSigma}_{\epsilon,\GCed}^{\otimes}\right)^{-1/2} \bX^{\otimes}$. Obtain the residuals $\hat{\bxi}^* = \by_{\GCed}^* - \bX_{\hat{S}^{\otimes}}^{*\otimes} \hat{\bbeta}_{FGLS}^{\dagger}$, and regress $\hat{\bxi}^*$ onto the variables retained by the previous regularization steps plus the Granger causality variables, retaining the residuals $
\hat{\bnu}^* = \hat{\bxi}^* - \bX_{\hat{S}\; \cup \;GC}^{*\otimes} \hat{\bbeta}_{FGLS}^*$.
Then obtain the statistic $LM = (\hat{\bxi}^{*\prime} \hat{\bxi}^* - \hat{\bnu}^{*\prime} \hat{\bnu}^*)$.
\item[\textbf{[4a]}] 
Reject $H_0$ if $LM > q_{\chi_{\nGCvars}^2}(1-\alpha)$, where $q_{\chi_{\nGCvars}^2}(1-\alpha)$ is the $1-\alpha$ quantile of the $\chi^2$ distribution with $\nGCvars$ degrees of freedom.
\item[\textbf{[4b]}] 
Reject $H_0$ if $\bigg(\frac{T \nGCed - \hat{s} - \nGCvars}{\nGCvars} \bigg)\bigg(\frac{LM}{T\nGCvars - LM} \bigg) > q_{F_{\nGCvars, \nGCed T - \hat{s} - \nGCvars}} (1-\alpha)$, where $\hat{s} = \abs{\hat{S}^{\otimes}}$ and $q_{F_{\nGCvars, \nGCed T - \hat{s} - \nGCvars}}(1-\alpha)$ is the $1-\alpha$ quantile of the $F$ distribution with $\nGCvars$ and $\nGCed T - \hat{s} - \nGCvars$ degrees of freedom.
\end{itemize}
\end{algorithm}

\begin{remark} \label{rem:otimes}
We perform the initial regressions in terms of $\bX_{GC}$ amd $\bX_{-GC}$ instead of $\bX_{GC}^{\otimes}$ and $\bX_{-GC}^{\otimes}$. The two are equivalent, as the Kronecker product essentially just copies the columns of $\bX$ both in the dependent and explanatory variables. Running the initial regressions in terms of $\bX^{\otimes}$ therefore essentially means running the same regression $\nGCed$ times, which is pointless as the selected variables remain the same in terms of the columns of $\bX_{-GC}$. We therefore perform the regressions just once for each column in $\bX_{GC}$. The construction of $\hat{S}_X^{\otimes}$ ensures that for any selected column $\bx_{-GC,m}$, we select every column of $\bX_{-GC}^{\otimes}$ in which $\bx_{-GC,m}$ appears.
\end{remark}

\begin{remark} \label{rem:FGLS}
The feasible generalized least squares (FGLS) estimation in Step [2] is needed when $\nGCed > 1$ to account for the correlation between equations of the VAR, and the fact we do not have the same selected regressors in each equation, as those coming from \eqref{eq:lasso_steps1} differ. Note that if $\nGCed = 1$, FGLS estimation collapses to the familiar form of the LM statistic. In that case one regresses $\hat{\bxi}$ by OLS onto the variables retained by the previous regularization steps plus the Granger causality variables, and retain the residuals $\hat{\bnu} = \hat{\bxi} - \bX_{\hat{S}\; \cup \;GC}^{\otimes} \hat{\bbeta}^*$, obtaining $R^2 = 1 - \hat{\bnu}^\prime \hat{\bnu} / \hat{\bxi}^\prime \hat{\bxi}$.
\end{remark}

\begin{remark}
Our lasso estimation of an HD-VAR can be interpreted as a general, data-driven, approach to Granger causality testing which encompasses the theory-driven `standard' approach in low-dimensional VARs. In particular, the lasso can be interpreted as imposing (approximate) sparsity over a high-dimensional information set, with the extent and location of the sparsity, or irrelevance, determined in a data-driven way. Conversely, testing Granger causality in a low-dimensional setting can then be interpreted as a priori assuming an extreme degree of sparsity over the same information set; in other words, it amounts to assuming that none of the additional series are relevant. 
\end{remark}

\begin{remark}
Given that we essentially have $\nGCvars = \nGCing \times \nGCed \times p$ steps of selection, it would be more appropriate to refer to our method as ``post-$\nGCvars$-selection'' approach. For expositional simplicity however we stick to the post-double-selection name, as this is the common name for such a procedure, and conveys the essence of our method equally well.
\end{remark}

\begin{remark} \label{rem:multGC}
Although the lasso regressions can handle increasing $\nGCing$, $\nGCed$ or $p$ with any issues, inference becomes more complicated when $\nGCvars$ increases with the sample size as the proposed LM statistic (or similarly a Wald test) will not have a limit distribution anymore. In such a case one could use recently developed Gaussian approximations of maxima of high-dimensional vectors  \citep{chernozhukov2013gaussian,zhang2017gaussian} to base a test statistic on $\max_{m = 1, \ldots, \nGCvars} \abs{\hat{\beta}_{GC,m}^{\textsc{pds}}}$, where $\hat{\bbeta}_{GC}^{\textsc{pds}}$ are the coefficients of $\bX_{GC}^\otimes$ in a regression of $\by_{\GCed}$ on $\bX_{GC}$ and $\bX_{\hat{S}^\otimes}^{\otimes}$ as in \eqref{eq:post-ols}. However, the critical values of this test statistic have to be simulated, which complicates the testing. As we argued in Section \ref{sec:model} that a fixed $\nGCvars$ is a reasonable assumption for typical Granger causality applications, we do not pursue this route.
\end{remark}

\begin{remark} \label{rem:selGC}
In Step [1] we propose not to consider the GC variables in the first regularization and insert them back at Step [2]. Alternatively, the GC variable(s) can be left in the regression, such that, we regress on the full $\bX^\otimes$ matrix. In this case there are then two further possibilities by either penalizing these variables or not. Simulations for these two alternatives have been carried out and in practice we do not find significant differences among the three in terms of size and power. The approach proposed in Step 1 delivers the best results in terms of size.
\end{remark}

\begin{remark} \label{rem:lb}
When the time series length is of same magnitude as the number of covariates, information criteria and time series cross-validation tend to break down and select too many covariates in order to perform a post-selection by OLS. To overcome this issue we propose to place a lower bound on the penalty to ensure that in each selection regression at most $c\, T \nGCed$ variables are selected, for some $0 < c < 1$. In our simulation and empirical studies we set $c=0.5$. 
Note that, as we have $\nGCvars$ selection steps, the possibility remains that different variables are selected in each steps, making the number of variables in the union $\hat{s}$ still too large to perform the post-selection OLS, although this problem is likely to occur far less often. This can be addressed by ensuring that fewer than $\nGCed T/ \nGCvars = T / \nXGC$ variables are selected in each selection step. We do not impose this stricter bound in general, as it will often be much too strict. Instead, we recommend to only address this issue if it arises in practice by an ad-hoc increase of the lower bound on the penalty.\footnote{Although it happens less often, the theoretical plug-in method for the tuning parameter occasionally also selects too many variables to make the post-OLS estimation infeasible. However, for this method no easy solution is available for bounding the penalty. One could increase the constant in the plug-in expression, thus strengthening the penalty, but this would be a rather ad-hoc adjustment. In particular, imposing the lower bound for the other methods only limits the allowed range of the tuning parameter, forcing the minimization to choose another (local) minimum that can still be far away from the boundary and justified graphically. For the plug-in method it is however difficult to justify the right amount of the increase, as the tuning parameter will be fixed to that value, and thus the chosen increase is rather arbitrary.}
\end{remark}

\begin{remark} \label{rem:F}
Although our Granger causality test has a $\chi^2$ distribution under the null hypothesis asymptotically, in smaller samples the test might still suffer from the usual small-sample approximation error. As such we propose a finite-sample correction to the test in Step [3b], which in our simulation studies improved the size of our test.
\end{remark}

\begin{remark} \label{rem:alt1}
Instead of the (adaptive) lasso, other estimators can be used in Step [1] as long as they deliver a sparse coefficient vector. For instance, the elastic net of \citet{zou2005regularization} that adds an $\ell_2$-penalty in addition to the $\ell_1$-penalty of the lasso can be used. The additional penalty ensures that the elastic net is strictly convex, and as a consequence tends to select highly correlated variables as a group together, whereas the lasso would tend to select only one of these variables \citep{zou2005regularization}. Given the typically strong correlations between many economic variables, this appears particularly useful for our context. However, we used the elastic net for both the simulations and the empirical application, and in both cases we found that the results are widely comparable to those of lasso. Therefore we chose to omit them from the paper; they are available upon request.
\end{remark}

\begin{remark}\label{rem:Wald}
One can also perform a standard Wald test of Granger causality instead of the LM test, by regressing the variables of interest on  $\bX_{GC}^{\otimes}$ and $\bX_{\hat{S}^\otimes}^\otimes$, and testing for the significance of the coefficients of $\bX_{GC}^\otimes$. While asymptotically the LM and Wald tests behave equally, differences might arise in small samples. We investigated the Wald version of the test in simulations as well, with results reported in Appendix \ref{app_additionalsim}, Table \ref{tab:3}. In general, differences between the two methods are negligible. However, for the Wald test, occasionally we run into the problem described in Remark \ref{rem:lb}, where even with the imposed lower bound on the penalty, too many variables are selected for performing a post-selection OLS. For this reason we prefer the LM version.
\end{remark}

\subsection{Tuning parameter selection}\label{sub:tunpa}
Appropriate selection of the lasso tuning parameter $\lambda$ in \eqref{eq:lasso} is crucial to achieve good performance. Many different data-driven methods exist giving wildly varying results. We provide a systematic comparison of several popular methods discussed in the literature in a simulation study. To the best of our knowledge, this is the first such comparison in the context of post-selection inference. We now introduce the methods considered in our study.

One option is to minimize an information criterion (IC) to determine an appropriate data-driven $\lambda$. Let $\hat{S}(\lambda) = \left\{m \in \{1,\ldots,Kp\}: \abs{\hat{\beta}_m (\lambda)} > 0 \right\}$ denote the set of active variables in the lasso solution for a given $\lambda$. For a generic response vector $\by$ and predictor matrix $\bX$, the value $\lambda^{IC}$ is found as
\begin{equation*}
\lambda^{IC} = \argmin_{\lambda} \ln  \left(\frac{1}{T} \norm{ \by - \bX \hat{\bbeta}(\lambda)}_2^2 \right) + \frac{C_T}{T} \abs{\hat{S}(\lambda)},
\end{equation*}
where $C_T$ is the penalty specific to each criterion. We consider the \emph{Akaike information criterion} (AIC) by \citet{akaike1974new} with $C_T=2$, the \emph{Bayesian information criterion} (BIC) by \citet{schwarz1978estimating} with $C_T=\ln(T)$, and the \textit{Extended Bayesian information criterion} (EBIC) by \citet{chen2008extended} with $C_T = \ln(T) + 2 \gamma \ln(Kp)$ with $\gamma=0.5$ proposed by \citet{chen2012extended} who argue that BIC fails to select the correct variables when the number of parameters is larger than the sample size.

An alternative approach is to plug in estimates of theoretically optimal values \citep[see e.g.][]{bickel2009simultaneous,belloni2013least,belloni2011square}. The lasso requires that $\lambda\geq c\norm{\bX'\bu}_{\infty}/T$ for some constant $c>0$ with ``high probability''. The central limit theorem motivates a Gaussian approximation where one chooses $\lambda^{th}=\frac{2c\hat{\sigma}}{\sqrt{T}}\Phi^{-1}\bigg(1-\frac{\alpha}{2N}\bigg)$ for a small $\alpha=o(1)$, where $\Phi^{-1}(\cdot)$ is the inverse of standard Gaussian cumulative distribution function and $\hat{\sigma}$ is an estimate the variance of $\bu$. In this paper we set $\alpha =0.05/\ln(T) $ and $c = 0.5$, while we follow \citet{belloni2012sparse} in the estimation of $\sigma$. Specifically, we obtain an initial (conservative) estimate by least squares estimation of $\by$ on the five most correlated regressors. This estimate is then updated iteratively, for details see \citet{belloni2012sparse}.

Perhaps the most popular way to choose the tuning parameter is cross-validation (CV), although CV is not always appropriate in the time series setup without modifications \citep{bergmeir2018note}. To estimate the tuning parameter with CV in a time series setup (TSCV) we use an expanding window out-of-sample forecasting scheme and minimize its squared forecasting error. The rolling window is set up with $80\%$ of the sample for training and $20\%$ for testing.  Cross-validation is appealing since it does not require any plug-in estimates, however, as observed in \citet{chetverikov2020cross} it typically yields small values of $\lambda$ thus still gaining fast convergence rate but at the price of less variable selection.

\begin{remark} \label{rem:p}
Although we assume $p$ fixed, in practice it may still need to be estimated if no reasonable value (or upper bound) can be given. As $p$ determines the number of selection regressions to be conducted, it has to be determined a priori and cannot be integrated in the lasso estimation. It can still be determined though by a (separate) lasso-type algorithm. For example, one may estimate \eqref{eq:dgp_i} with a large initial lag length $p^*$, and let $p$ be determined as the largest lag for which variables are selected, possibly also varying the lag length over variables. For this approach the hierarchical penalties of \citet{nicholson2020high} provide a better option than the regular lasso, as the regular lasso tends to select occasional ``spurious'' high lags, which would have a significant impact on the testing procedure. Alternatively one may marginalize the VAR to a collection of univariate AR($p$) processes, and select the lag length by minimizing an information criterion on the residual covariance matrix. As marginalization increases the lag length, such an approach would yield a simple to compute upper bound on $p$.
\end{remark}

\section{Asymptotic Properties} \label{sec:asy}
In this section we derive the asymptotic properties of our method. We first present and discuss our general high-level assumption under which the properties are derived, and then state our main results.

\begin{assumption}\label{as:hlev}
Let $\delta_T$ and $\Delta_T$ denote sequences such $\delta_T, \Delta_T \rightarrow 0$ as $T \rightarrow \infty$. Then assume that the following conditions are satisfied:
\begin{enumerate}[(a)]
\item\label{as:eig} \textbf{Population Eigenvalues:} Let $\be_t = \left(e_{1,t}, \ldots, e_{\nXGC,t} \right)^\prime$, $\bE = \left(\be_1, \ldots, \be_T \right)^\prime$ and $\bE^{\otimes} = \bI_{\nGCed} \otimes \bE$. Define
\begin{align*}
\bSigma &= \begin{bmatrix}
\bSigma_{GC,GC} & \bSigma_{GC, -GC} \\
\bSigma_{-GC,GC} & \bSigma_{-GC, -GC} \\
\end{bmatrix} = \begin{bmatrix}
\E \left(\bx_{GC,t} \bx_{GC,t}^\prime \right) & \E \left(\bx_{GC,t} \bx_{-GC,t}^\prime \right) \\
\E \left(\bx_{-GC,t} \bx_{GC,t}^\prime \right) & \E \left(\bx_{-GC,t} \bx_{-GC,t}^\prime \right) \\
\end{bmatrix}
\end{align*}
Then there exists a constant $c_L > 0$ not depending on $T$ and $k$ such that $\lambda_{\min} (\bSigma) > c_L$, where $\lambda_{\min}(\bSigma)$ denotes the minimum eigenvalue of $\bSigma$.

\item\label{as:clt} \textbf{Limit Behavior:} Let
\begin{equation*}
\begin{split}
\bE^{\otimes\prime} \bu_{\GCed} / \sqrt{T} &= \vecop(\bE^\prime \bU_{\GCed})/\sqrt{T} = \frac{1}{\sqrt{T}} \sum_{t=p+1}^T \vecop(\be_t \bu_{\GCed,t}^\prime) \xrightarrow{d} N(0, \bOmega),\\
\bE^\prime \bE / T &=  \frac{1}{T} \sum_{t=p+1}^T \be_t \be_t^\prime \xrightarrow{p} \bSigma_{GC|-GC} = \bSigma_{GC,GC} - \bSigma_{GC,-GC} \bSigma_{-GC,-GC}^{-1} \bSigma_{-GC,GC},\\
\bU_{\GCed}^\prime \bU_{\GCed} / T &\xrightarrow{p} \bSigma_{u,\GCed},
\end{split}
\end{equation*}
where $\bOmega = \plim_{T\rightarrow\infty} \left(\bE^{\otimes\prime} \bu_{\GCed} \bu_{\GCed}^\prime \bE^{\otimes} \right)/T$.

\item\label{as:ep} \textbf{Empirical Process:} We have with probability at least $1 - \Delta_T$ that $\norm{\bX_{-GC}^{\prime} \bu_i/\sqrt{T}}_{\infty} \leq \gamma_T$ for all $i \in \GCed$ and $\norm{\bX_{-GC}^\prime \be_j/\sqrt{T}}_{\infty} \leq \gamma_T$ for all $j = 1, \ldots, \nXGC$, with $\be_j$ the $j$-th column of $\bE$, for some deterministic sequence $\gamma_T$ subject to the restrictions in (\ref{as:rates}).

\item\label{as:bound} \textbf{Boundedness:} The (Granger causality) parameters of interest are bounded, that is, there exists a fixed constant $C>0$ such that $\norm{\bbeta_{GC}}_1 \leq C$.

\item\label{as:cons} \textbf{Consistency:} The initial estimators $\hat{\bgamma}_{j}$ are consistent in the prediction sense; specifically, with probability at least $1 - \Delta_T$ we have that
\begin{equation*}
\begin{split}
&\norm{\bX_{-GC}^{\otimes} (\hat{\bgamma}_0 - \bgamma_0)}_2 / \sqrt{T} \leq \delta_T T^{-1/4}, 
&\max_{j = 1, \ldots, \nXGC} \norm{\bX_{-GC} (\hat{\bgamma}_j - \bgamma_j)}_2 / \sqrt{T} \leq \delta_T T^{-1/4}.
\end{split}
\end{equation*}

\item\label{as:spar} \textbf{Sparsity:} Let $S_j = \{m: \gamma_{m,j} \neq 0\}$ denote the sets of active variables in \eqref{eq:lasso_steps1} and \eqref{eq:lasso_steps2} and let $s = \abs{S_0} + \sum_{j=1}^{\nXGC} \abs{S_j}$. Let $\hat{s}$ be as defined in Algorithm 1. Then both the DGP and the initial estimators are sufficiently sparse; in particular, we have that with probability at least $1 - \Delta_T$, $\max(s, \hat{s}) \leq \bar{s}_T$ for a deterministic sequence $\bar{s}_T$ subject to the restrictions in (\ref{as:rates}).

\item\label{as:sev} \textbf{Sparse Eigenvalues:} for any $\bgamma \in \mathbb{R}^{(K - \nGCing)p}$ with $\norm{\bgamma}_0 \leq \bar{s}_T$, we have with probability at least $1 - \Delta_T$ that
$\norm{\bgamma }_2^2 \leq \norm{\bX_{-GC} \bgamma/\sqrt{T}}_2^2 / \phi_{T,\min}^2$,
where $\phi_{T,\min} > 0$ is subject to the restrictions in (\ref{as:rates}).

\item\label{as:rates}\textbf{Rate Conditions:} The deterministic sequences $\bar{s}_T, \gamma_T$ and $\phi_{T,\min}$ introduced above satisfy the restriction $\bar{s}_T \gamma_T / \phi_{T,\min} \leq \delta_T T^{1/4}$.
\end{enumerate}

\end{assumption}

Assumption \ref{as:hlev} is a high-level assumption that allows for much flexibility on the underlying DGP and the used estimators in the first step. We now discuss each part in turn. Part (a) assumes that the minimum eigenvalue of $\bSigma$ is bounded. This is required for application of lasso methods, as well as for the inverse covariance matrix $\bSigma^{-1}$ and the projection coefficients in \eqref{eq:lasso_steps1} and \eqref{eq:lasso_steps2} to exist. Part (\ref{as:clt}) assumes that a central limit theorem and weak law of large numbers hold.  Essentially this require that the process is sufficiently well-behaved in terms of moments and dependence allowed. Although for convenience we assume martingale difference errors in Assumption \ref{as:VAR}, (\ref{as:clt}) holds under much weaker conditions such as mixing errors; see e.g.~\citet[Chapter 14]{davidson1994stochastic}.

Part (\ref{as:ep}) is closely related to (\ref{as:clt}), but additionally controls the tail behavior of the empirical process. Results of this kind are standard in the lasso literature and can be derived using a variety of tail bounds depending on the properties of the random variables of interest, see e.g. \citet{kock2015oracle} and \citet{medeiros2016l1} for results relevant to VAR and time series models. Of particular interest for financial applications, \citet[Proposition 2]{masini2019regularized} show that this condition is satisfied for VAR models with general weakly dependent erorrs that include many popular multivariate volatility models. The boundedness assumption in \eqref{as:bound} is not very restrictive, and with $\nGCvars$ fixed follows directly if the parameter space of $\bbeta$ is a compact set.

Part (\ref{as:cons}) imposes an appropriate consistency rate on the predictions coming from the first-stage estimator. Such prediction consistency is a standard result for lasso estimators; in particular, \citet{wong2020lasso} obtain it for a very general class of VAR models allowing for conditional heteroskedasticity and dependence in the error terms. \citet{adamek2020lasso} derive consistency of the lasso under misspecified time series models, and show that their setting covers (among others) the first-step regressions of the relevant predictors in $\bX_{GC}$ on the other regressions, which are inherently misspecified in a VAR setup due to the missing lags; see their Remark 3 for further details.

Next to consistency, we also require sparsity of the DGP and the estimator, as controlled by part (\ref{as:spar}). The assumption of exact sparsity in the DGP for the initial regressions can be relaxed to approximate sparsity as in \citet{belloni2014inference}. For the sake of expositional clarity we do not work under that assumption here but stick to the simpler exact sparsity. Sparsity of the first-stage estimator is needed in our framework as we perform OLS on the selected variables from the first-stage regressions. If the selected variables are not sparse enough, too many variables will be selected for OLS to be feasible. Sparsity of lasso estimators is analysed in \citet{belloni2013least}, while \citet{kock2015oracle} and \citet{medeiros2016l1} provide results for adaptive lasso for time series. Importantly, we do not require consistent model selection; the selection method used is allowed to make ``persistent'' mistakes, allowing for both variables to be incorrectly included and relevant variables to be missed, as long as the estimator remains sufficiently sparse and consistency is guaranteed. Unlike \citet{belloni2014inference}, we allow for the order of sparsity of the estimator to differ from the true sparsity thereby opening the way for conservative selection procedures.

Given the assumptions above, the eigenvalue assumption in (\ref{as:sev}) becomes almost superfluous, as it is generally needed to establish (\ref{as:cons}) and (\ref{as:spar}) for lasso-type estimators; see e.g. \citet{belloni2013least} and \citet{medeiros2016l1} for details. It requires that for sufficiently sparse vectors, the eigenvalues of the subset of the Gram matrix corresponding to their non-zero support do not decrease to zero too fast. Such assumptions are standard in the lasso literature in various guises as \emph{restricted eigenvalue} conditions, and can typically be derived by making similar conditions on the population covariance matrix $\bSigma_{-GC, -GC}$ coupled with a convergence result of the Gram matrix $\bX_{-GC}^\prime \bX_{GC}$ to $\bSigma_{-GC, -GC}$. \citet{basu2015network}, \citet{masini2019regularized} and \citet{wong2020lasso} establish the plausibility of such restricted eigenvalue conditions for various VAR models. We state the condition here explicitly as it is needed directly in the proofs. 

Finally, note that the restrictions on tail behavior (via $\gamma_T$), sparsity (via $\bar{s}_T$) and minimum eigenvalues (via $\phi_{T,\min}$) are meaningless if no rates on these sequences are imposed. Part (\ref{as:rates}) therefore is the key part which connects all assumptions with explicit rates needed for the validity of the PDS method. The restrictions here represents a trade-off between sparsity, thickness of tails and minimum eigenvalues. For example, if, as often assumed $\phi_{T,\min}$ is fixed and $\bu_t$ is Gaussian, tails are sufficiently thin that $\gamma_T$ can be chosen as roughly the order of $\sqrt{\ln(K^2 p)}$ \citep[cf.][Lemma 4]{kock2015oracle}, leaving room for either almost exponentially large $K$ relative to $T$, or a fairly non-sparse model. On the other hand, if only $m$ moments of $\bu_t$ exist, $\gamma_t$ should be taken roughly of the order $(K^2 p)^{2/m}$ \citep[Lemma 2]{masini2019regularized}, requiring polynomial growth of $K$ compared to $T$ and sparser models. 

The most restrictive and crucial assumption needed on the underlying DGP for satisfying Assumption \ref{as:hlev} is the sparsity of the underlying DGP formulated in part (\ref{as:spar}). The plausibility of this assumption highly depends on the specific application. In many financial applications sparsity (or its approximate version) is natural, for example in portfolio selection when the number of assets is large and the estimation of high-dimensional volatility matrices in financial risk assessment (see \citet{fan2011sparse} for an overview), as well as in our investigation of Granger causality in networks of realized volatilities in Section \ref{sec:appl}. The volatility of one particular stock is likely to have specific channels of contagion rather than affecting the whole stock market at the same time. Shocks to one asset therefore likely propagate through the system via specific channels, which corresponds to sparse lag polynomials. One might worry about systemic shocks affecting many assets; however, the dense covariance matrix $\bSigma_u$ can accommodate simultaneous common shocks. Moreover, the dynamic of such shocks can generally well be captured through a sparse combination of the most important and most affected assets. Similarly, in macroeconomic applications it has been found that a few important variables can capture the effects of unobserved common factors, leading sparse models to perform as well as common factors \citep{demol2008forecasting,smeekes2018macroeconomic}.

We are now ready to state our main asymptotic result of this section in Theorem \ref{th:asdis} which establishes the asymptotic normality of the post-lasso (generalized) least squares estimator. Here we slightly deviate from the LM test in Algorithm \ref{alg:pdslm}; after the double selection procedure carried out in Step [1], we regress the transformed outcome variables $\widetilde \by_t = (\bG_T \otimes \bI_T) \by_{\GCed}$ on both the Granger causing $\widetilde\bX_{GC}^\otimes = (\bG_T \otimes \bI_T) \bX_{GC}^\otimes$ and selected variables $\widetilde \bX_{\hat{S}^\otimes}^\otimes = (\bG_T \otimes \bI_T) \bX_{\hat{S}^{\otimes}}^{\otimes} $
\begin{equation}\label{eq:post-ols}
\widetilde\by_{\GCed} = \widetilde\bX_{GC}^{\otimes} \bbeta_{GC}^{\textsc{pds}} + \widetilde\bX_{\hat{S}^{\otimes}}^{\otimes} \bbeta_{\hat{S}^\otimes}^{\textsc{pds}} + \widetilde\bu_{\GCed}. 
\end{equation}
The transformation by the matrix $\bG_T$ allows for the GLS esitmation needed in the LM procedure by taking $\bG_T = \hat\bSigma_{u,\GCed}^{-1/2}$, while OLS is performed with $\bG_T = \bI_{\nGCed}$. In the latter case the theorem provides the foundation for the Wald test discussed in Remark \ref{rem:Wald} (minus the required variance estimation for that test). We state this result separately as it is interesting in its own right, and can be used to establish validity of other tests such as the Wald test.

\begin{theorem}\label{th:asdis}
Let $\hat{\bbeta}_{GC}^{\textsc{pds}}$ denote the OLS estimator of $\bbeta_{GC}^{\textsc{pds}}$ in \ref{eq:post-ols}. Let $\bG_T$ be any matrix satisfying with probability at least $1-\Delta_T$ that $0 < c_1 \leq \lambda_{\min}(\bG_T^\prime \bG_T) \leq \norm{\bG_T^\prime\bG_T}_{\max} \leq c_2 < \infty$, where $c_1, c_2$ are constants not depending on $T$. Then, uniformly in all DGPs that satisfy Assumption \ref{as:hlev}, we have as $T \rightarrow \infty$,
\begin{equation*}
\sqrt{T} (\hat{\bbeta}_{GC}^{\textsc{pds}} - \bbeta_{GC})\overset{d}{\to}\mathcal{N}\left(\bzero, (\bG^\prime \bG \otimes \bSigma_{GC|-GC})^{-1} \bOmega_{\bG} (\bG^\prime \bG \otimes \bSigma_{GC|-GC})^{-1} \right),
\end{equation*}
where $\bOmega_{\bG} = \plim_{T\rightarrow\infty} \left[(\bG_T^\prime \bG_T \otimes \bE^\prime) \bu_{\GCed} \bu_{\GCed}^\prime (\bG_T^\prime \bG_T \otimes \bE) \right]/T$.
\end{theorem}

Theorem \ref{th:asdis} establishes the asymptotic normality of the post-double-selection OLS estimators. The statement `uniformly in all DGPs that satisfy Assumption \ref{as:hlev}' should be interpreted as the theorem holding uniformly over a parameter space that is defined such that Assumption \ref{as:hlev} holds for all parameters in that parameter space. Importantly, no beta-min conditions on the smallest magnitude of parameters are required, thus alleviating the post-selection inference problem. We refer to Comments 3.4 and 3.5 in \citet{belloni2014inference} for further details regarding the uniformity. The limit distribution of the LM test now follows straightforwardly from Theorem \ref{th:asdis}, and is stated in the corollary below.

\begin{theorem}\label{th:asdis2}
Let $\bbeta_{GC} = \bzero$. Then, uniformly in all DGPs that satisfy Assumption \ref{as:hlev} and for which $\bOmega = \bSigma_{u, \GCed} \otimes \bSigma_{GC|-GC}$, we have that
\begin{align*}
&LM \xrightarrow{d} \chi_{\nGCvars}^2 \qquad \text{as } T \rightarrow \infty.
\end{align*}
\end{theorem}

Theorem \ref{th:asdis2} establishes the limiting distribution of the PDS-LM test under an additional condition on the (co)variances of the partial regression errors, which is satisfied if the errors are iid. To allow for heteroskedaticity the LM test has to be modified, which would only lead to more cumbersome proofs without adding any novelty specific to the high-dimensional case. Therefore we focus on the homoskedastic case here, although we do consider a heteroskedasticity-robust version of the test in the volatility application in Section \ref{sec:appl}.\footnote{Note that this is no different for the Wald test, for which the variance estimation has to be adjusted as well.}

\section{Monte-Carlo Simulations}\label{sec:sim}
We now evaluate the finite-sample performance of our proposed Granger causality test. We consider three Data Generating Processes (DGPs) inspired by \citet{kock2015oracle}: 
\begin{align*}
&\text{DGP1:}& &\by_t=\begin{bmatrix}
0.5
& \ldots &0 \\
\vdots 
& \ddots & \vdots  \\
0
&\ldots & 0.5 
\end{bmatrix}\by_{t-1} +\bepsilon_t,\\
&\text{DGP2:} & &\by_t=\begin{bmatrix}
(-1)^{|i-j|}a^{|i-j|+1}  & \ldots &(-1)^{|i-j|}a^{|i-j|+1} \\
\vdots  & \ddots & \vdots  \\
(-1)^{|i-j|}a^{|i-j|+1}&\ldots & (-1)^{|i-j|}a^{|i-j|+1} 
\end{bmatrix}\by_{t-1} +\bepsilon_t,
\quad\text{with}\; a=0.4,\\
&\text{DGP3:}& &\by_t=\begin{bmatrix}
\bA 
& \ldots &0 \\
\vdots 
& \ddots & \vdots  \\
0
&\ldots & \bA 
\end{bmatrix}\by_{t-1} 
+\bepsilon_t
\quad
\text{with}\; \underbrace{\bA}_{5\times5}=\begin{bmatrix}
0.15&\cdots&0.15\\
\vdots&\ddots&\vdots\\
0.15&\cdots&0.15\\
\end{bmatrix}.
\end{align*}
The diagonal VAR in DGP1 respects the sparsity assumption while in DGP2 the entries are set to decrease exponentially fast in the distance from the main diagonal and hence the sparsity assumption is not met. DGP2 could be empirically motivated by looking e.g. at financial interconnectedness. Financial institution, such as banks, lend to and borrow from one another becoming interconnected through interbank credit exposures. The financial distress experienced by one bank is likely to be most heavily transmitted the closer the connections are as well as less transmitted, the weaker the connections. DGP3 is a block-diagonal system. Such a structure is motivated by e.g. typical quarterly macroeconomic models capturing business cycle dynamic and monetary and fiscal policy effects. One such example is DSGE models, where the dynamic of the economy through time is monitored on quarterly frequency. Note that as written above, DGP1 satisfies the null of no Granger causality from unit 2 to 1, while DGP2 and DGP3 do not. Therefore, we adapt DGP 1 for the power analysis by setting the coefficient in position $(2,1)$ equal to 0.2. Conversely, we set the same coefficient equal to zero for DGP2 and DGP3 for the size analysis.

We choose our series of interest as $\GCed=\{2\}$ and $\GCing = \{1\}$, therby focusing on the case where we have single variables of interest for both elements of the test. Here we consider for simplicity $p=1$ lag, namely the same lag-length as in the DGPs, so $j=1$.
The equation of interest can then be written as
\begin{equation*}
y_{2,t} = \beta_{GC} y_{1,t-1} + \sum_{j=2}^K \beta_{j} y_{j,t-1} + \epsilon_{2,t}.
\end{equation*}
Hence, for each DGP we test
$H_0:\beta_{GC}=0$ against $H_1:\beta_{GC}\neq 0$
using our proposed PDS-LM test.

Table \ref{tab:1} reports the size and power of the test for 1000 replications by using different combinations of time series length $T=(50,100,200,500)$ and number of variables in the system $K=(10,20,50,100)$ and a fixed lag-length $p=1$. All the rejection frequencies are reported using a burn-in period of fifty observations. For each scenario,
AIC, BIC and EBIC are compared with the theoretical choice of the tuning parameter $\lambda^{th}$ and time series cross validation $\lambda^{TSCV}$ as described in Subsection \ref{sub:tunpa}. 

Simulations are also reported for different types of covariance matrices of the error terms. We employ a Toepliz-version for calculating the covariance matrix as $\Sigma_{i,j}=\rho^{|i-j|}$ by using two scenarios of correlation: $\rho=(0,0.7)$. The first case corresponds to no correlation, and is equivalent to set $\bSigma = \bI_{K}$.

In the Appendix we provide some additional simulation results. First, Table \ref{tab:correlated} reports the simulation results for all three DGPs using $\Sigma_{i,j}=0.7^{|i-j|}$. Second, we investigate the Wald version of our test in Table \ref{tab:3}. Third, in Table \ref{tab:4} we investigate the effects of miss-specification of the lag length by estimating the over-specified VAR$(p+1)$ instead of the true-order VAR$(p)$.\footnote{For both the Wald test and the over-specified VAR$(p+1)$ we report the simulations for $\Sigma_{i,j}=0.7^{|i-j|}$ and DGP1 only. Results for the other DGPs are available upon request.} Fourth, in Table \ref{tab:biva} we report the results for the size of a bivariate Granger causality test for a non-sparse DGP when using a standard Wald ($F$) test. This test is obviously sensitive to omitted variable bias, and our goal is to demonstrate its effect. Finally, although all results reported here use the finite sample correction in Step 3b of the algorithm, we also investigated the differences with Step 3a. We comment on these results in the next subsection. All results not reported in this paper are available from the authors upon request.

\begin{sidewaystable}
\begin{threeparttable}
\caption{Simulation results for the PDS-LM Granger causality test}\label{tab:1}
\tiny
\begin{tabular}{ l S[table-format=3.0] cccc c c c c c c cc cccccccccccc}
    \toprule
   
   DGP&{Size/Power}&$\rho$&$T$ &&&50&&&&&100&&&&&200&&&&&500&&&  \\
   \cmidrule(l){1-24} 
&&&K& \multicolumn{5}{c}{AIC \;\;\; BIC \;\;\;EBIC \;\;\;\;$\lambda^{th}$ \; $\lambda^{TSCV}$} & \multicolumn{5}{c}{AIC \;\;\; BIC \;\;\;EBIC \;\;\;\;$\lambda^{th}$ \; $\lambda^{TSCV}$}& \multicolumn{5}{c}{AIC \;\;\; BIC \;\;\;EBIC \;\;\;\;$\lambda^{th}$ \; $\lambda^{TSCV}$}& \multicolumn{5}{c}{AIC \;\;\; BIC \;\;\;EBIC \;\;\;\;$\lambda^{th}$ \; $\lambda^{TSCV}$}\\ 
    \cmidrule(l){4-24}
  &&& 10 &6.7&6.1&5.3&6.8&6.9&6.7&6.1&5.9&6.5&6.4&5.5&4.5&4.5&4.7&5.4&4.3&4.1&4.3&4.2&4.3  \\
    1&{Size}&0&20 &7.3&6.0&6.4&5.9&6.2&6.0&4.7&4.7&5.5&5.8&4.0&4.7&5.4&4.3&4.5&4.3&4.1&4.3&4.1&4.3\\
   &&& 50 &7.0&5.9&5.7&6.4&5.5&7.4&6.4&6.4&6.2&6.6&6.9&5.9&5.9&6.2&6.1&6.8&6.5&6.7&6.8&6.7\\
   &&& 100&7.4&7.2&6.6&NA&6.8&7.2&4.9&4.9&5.9&5.4&6.5&4.7&4.9&5.0&5.2&5.8&3.9&4.0&5.0&4.1 \\
\cmidrule(l){1-24}   

    &&&10 &30.0&31.2&33.3&30.4&30.9&58.3&58.9&60.7&58.5&57.4&89.1&89.3&89.6&89.2&89.5&99.9&99.9&99.9&99.9&99.9  \\
    1&{Power}&0&20 &22.8&26.4&30.2&25.8&24.2&52.8&55.1&57.4&53.9&54.5&85.6&88.0&89.0&86.6&86.1&99.9&100&100&99.9&99.9 \\
    &&&50 &13.6&24.3&33.3&17.9&18.7&38.7&53.8&59.0&46.4&45.5&78.2&85.8&87.0&81.3&80.2&99.9&99.9&99.9&99.9&99.9\\
    &&&100&10.4&20.3&32.1&NA&18.7&20.4&51.9&57.0&31.2&35.8&61.6&85.4&86.7&74.1&73.7&99.5&100&100&99.8&99.7\\
\cmidrule(l){1-24}
&&&10 &6.2&5.5&6.1&6.1&5.9&5.0&4.9&5.4&5.0&5.0&4.7&4.6&4.9&4.6&5.2&3.9&3.9&3.9&3.7&3.7 \\
    2&{Size}&0&20 &7.8&5.5&6.0&5.8&6.5&5.4&4.6&4.7&4.8&5.3&5.9&5.9&5.4&5.7&6.0&4.7&3.8&4.3&4.6&4.7\\
    &&&50 &7.3&5.9&4.7&6.0&7.1&7.9&6.3&6.5&7.3&6.5&7.3&6.1&6.0&6.6&6.6&6.3&6.6&6.4&6.1&6.0\\
    &&&100&6.1&6.9&5.5&NA&6.6&6.8&5.1&6.0&5.3&4.8&5.4&5.5&5.8&4.6&4.9&6.1&4.3&5.0&4.9&4.5\\
\cmidrule(l){1-24}   

    &&&10 &18.0&19.7&21.3&18.2&17.0&37.9&39.3&40.4&38.0&38.4&64.7&64.7&66.9&64.8&65.1&97.4&97.4&97.5&97.4&97.4  \\
    2&{Power}&0&20 &16.0&19.6&24.6&18.8&16.9&35.4&39.8&44.4&37.4&36.7&64.4&67.4&69.7&66.0&65.0&97.2&97.5&97.5&97.5&97.4\\
    &&&50 &8.6&15.2&21.7&12.8&13.1&25.0&36.1&43.2&32.6&31.2&57.0&66.4&71.9&61.8&61.1&95.0&96.2&96.8&96.1&95.4\\
    &&&100&9.2&14.1&25.1&NA&10.1&15.1&34.9&45.9&25.9&25.8&44.8&65.1&74.7&58.0&57.8&94.7&97.3&97.7&96.3&96.5\\
\cmidrule(l){1-24}
&&&10 &5.2&5.0&5.6&5.7&4.6&5.6&4.9&5.7&5.7&6.2&4.0&4.1&6.1&4.1&4.4&4.1&4.0&3.8&3.9&4.1 \\
    3&{Size}&0&20 &4.2&5.1&5.7&5.6&4.9&4.3&4.3&7.2&4.2&3.9&5.2&5.6&9.4&4.7&4.6&4.7&4.4&4.5&4.5&4.8\\
    &&&50 &7.5&6.3&7.4&6.9&6.6&6.4&7.0&9.5&5.9&5.5&6.9&6.9&12.4&6.0&6.6&4.9&5.3&6.6&5.3&5.4\\
    &&&100&7.1&6.7&8.3&NA&7.0&6.2&5.7&8.3&5.7&6.0&4.7&6.2&10.7&4.2&4.7&4.4&5.1&6.5&5.1&4.7\\
\cmidrule(l){1-24}   

    &&&10 &15.4&20.0&23.4&16.2&16.1&31.5&36.4&44.3&32.6&31.4&58.4&61.3&63.7&58.9&59.3&95.2&95.6&95.7&95.5&95.3  \\
    3&{Power}&0&20 &13.4&18.7&26.0&13.8&13.8&29.5&37.0&47.8&30.0&29.8&56.5&62.0&69.8&56.8&55.5&94.0&94.6&94.8&94.4&94.3\\
    &&&50 &11.4&20.7&28.0&12.9&10.6&24.1&39.8&52.3&26.9&27.6&50.3&59.5&73.6&52.1&51.8&91.1&92.7&93.4&92.3&90.9\\
    &&&100&7.9&18.7&26.9&NA&14.1&13.7&42.6&55.0&20.2&22.2&41.4&62.0&75.2&44.9&44.8&90.0&94.0&94.8&91.4&90.2\\

\cmidrule(l){1-24}
\end{tabular}
\begin{tablenotes} 
\scriptsize 
\item Notes: Size and Power for the different DGPs described in Section \ref{sec:sim} are reported for 1000 replications. $T=(50,100,200,500)$ is the time series length, $K=(10,20,50,100)$ the number of variables in the system, the lag-length is fixed to $p=1$. $\rho$ indicates the correlation employed to simulate the time series with the Toeplitz covariance matrix. The different choices of the tuning parameter $\lambda$ are reported as: AIC, BIC, EBIC for information criteria, $\lambda^{th}$ for the theoretical plug-in and TSCV for time series cross-validation as explained in Section \ref{sec:sim}.
\end{tablenotes}
\end{threeparttable}
\end{sidewaystable}

Our proposed approach shows a good performance in terms of size and (unadjusted) power for all DGPs considered. Both for the setting of no correlation and high correlation of errors, sizes are in the vicinity of 5\% and power is increasing with the sample size $T$.

Only moderate size distortion is visible in large systems for small samples (e.g.~$K\ge 50,\; T=50$). As expected, the test procedure works remarkably well for the sparse DGP1 in high dimensions. However, size properties under the non-sparse DGP2 do not deviate much from its sparse counterpart, although for both  DGP2 and DGP3 we do observe a slight deterioration of size when the dimension of the system increases.

Interestingly, the three different information criteria show substantially different behavior. EBIC, due to its very stringent nature, tends to perform well only in very large systems, while it is essentially equal to a bivariate Granger causality test in small systems. We have to add though that the good performance of AIC in particular is somewhat inflated by the imposed lower bound on the penalty; unreported simulations show that without the lower bound AIC performs significantly worse, often selecting too many variables rendering the post-OLS estimation infeasible. The one advantage of using EBIC as information criterion to tune $\lambda$ in the $K>>T$ settings when $T$ is small (e.g. $T= 50,100$) is the possibility to avoid the lower bound on the penalty. However, since this comes at a price of more size distortion, we recommend the use of BIC instead, along with the lower bound on the penalty. When comparing the different choices of the tuning parameter we can narrow down the best performing ones (in terms of size and power) to BIC and $\lambda^{th}$. However, in terms of computational time, estimating the tuning parameter using information criteria is considerably faster.

Comparing our test to the bivariate VAR in Table \ref{tab:biva}, it is clear that our proposed PDS-LM is very robust to omitted variable bias, unlike the bivariate test, whose size distortions increase with both the sample size and the number of variables, with sizes of 45\% observed for the sample sizes we consider in our application in Section \ref{sec:appl}. There we will also further elaborate on this difference between our method and the bivariate test. Table \ref{tab:1} shows that for sample sizes smaller than $T=500$, rarely the power exceeds 90\%. However, one must keep in mind that the powers are not size-adjusted, and thus the high reported power of the low-dimensional test is an artefact of the huge size distortions rather than genuine power. It also seems unreasonable to expect that PDS-LM test has vey high power if $T$ is small; we are still considering large systems with many parameters to estimate, and there seems to be no way around this if one desires to test Granger causality in large systems with many (control) variables. In that sense we may fully expect the bivariate test to also have higher size-corrected power; yet with all its disadvantages and sensitivity to omitted variables this is not a good comparison. All in all, we believe our test still has sufficiently adequate power properties to be useful in practice.

The results of robustness to misspecification of the lag length order with $p=2$ instead of $p=1$ are reported in Table \ref{tab:3} in Appendix \ref{app_additionalsim}. As the size distortions across the range of considered DGPs are only marginally higher for large $K$ and $T$ comparatively small, the test appears to be quite robust to this misspecification. Again, BIC seems to be the best choice for tuning the penalty for all DGPs. Unreported simulations (available upon request) further show that the finite sample adjustment for the test performed in Step 3b of the algorithm is able to substantially reduce size distortions in small samples compared to the asymptotic version of Step 3a.

\section{Networks in Realized Volatilities} \label{sec:appl}
\subsection{Realized Variances}\label{sec:appl1}
We first investigate the volatility transmission in stock return prices using the daily realized variances of 30 US assets.%
\footnote{We would like to thank Marcelo C. Meideiros for providing us with the high frequency data on stock prices that we have used to construct the realized variances. See Table \ref{tab_stocks} for the stocks considered. The R package HDGCvar is available on the GitHub page of the corresponding author (\url{https://github.com/Marga8}).} Both the computational simplicity and the theoretical foundations make realized volatility measures (realized variance, bi-power variation, median realized variance, etc.) very attractive among practitioners and academics for modelling time varying volatilities and monitoring financial risk. We have considered 10-minute realized variances
\begin{equation} \label{eq:RV}
RV10_{t} \equiv \sum_{j=1}^{M} r_{j,t}^{2}, \qquad
r_{j,t} =\ln P_{j,t}-\ln P_{j-1,t},
\end{equation}
using $j=1,\ldots, M$ intraday 10 minutes stock prices $P_{j,t}$. We consider 10 minute returns as this is the frequency that minimizes for our sample the microstructure noise (\citet{mcaleer2008realized}).\footnote{To determine the optimal frequency, we computed realized variances  using different frequencies of 1, 5, 10, 15, 30, 65 and 130 minutes, in addition to the estimation using daily returns. The latter estimation has the advantage of being unbiased but the drawback of being very noisy (\citet{pooter2008predicting}). To find an optimal trade-off between bias and variance \citep[see e.g]{martens2004estimating}), mean, variances and mean squared errors (MSE) were computed for each estimation frequency in a similar way as \citet{pooter2008predicting}, and it was found that the frequency of 10 minutes minimizes the MSE.}
We investigate the period from March 2008 until February 2017 (2236 trading days).

Given the time series of realized volatilities as defined in (\ref{eq:RV}), we employ a multivariate version of the heterogeneous autoregressive model (VHAR) of \citet{corsi2009simple} to model their joint behavior (see also \citet{cubadda201911}). To formally define the VHAR model, we log-transform the series and we stack the logarithmic RV into a vector $y_t$. The VHAR specification is given by the following model:
\begin{equation*}
\by_t = \bc + \bB^{(1)} \by_{t-1} + \bB^{(2)} \by_{t-1}^{(week)} + \bB^{(3)} \by_{t-1}^{(month)} + \bepsilon_t,
\end{equation*}
where
$\by_{t}^{(week)} = \frac{1}{5} \sum_{j=0}^{4} \by_{t-j}$ 
and $\by_{t}^{(month)} = \frac{1}{22} \sum_{j=0}^{21} \by_{t-j}$
are the vectors containing the average volatility over the last 5 (week) and 22 (month) days. Granger causality in this context represents contagion, or spillover, of volatility from one asset to another. To test for the null hypothesis of no Granger causality / no volatility spillovers from $y_{k,t}$ to $y_{i,t}$ against the alternative of spillovers, we test
\begin{equation*}
H_0:\beta^{(1)}_{i,k}=\beta^{(2)}_{i,k}=\beta^{(3)}_{i,k}=0 \qquad \text{vs.} \qquad H_1:\beta^{(1)}_{i,k},\beta^{(2)}_{i,k},\beta^{(3)}_{i,k}\neq0,
\end{equation*}
where $\beta_{i,k}^{(1)}$ is the $(i,k)$-th element of $B^{(1)}$. We perform this test for every $(i,k)$-pair to obtain the full $29\times 29$ network of spillover effects. As heteroskedasticity is likely present in these data, we robustify the PDS-LM procedure by implementing the heteroskedasticity-robust LM test such as for example described in \citet[Ch.~8]{wooldridge2015introductory}. The full algorithm for the heteroskedasticity-robust PDS-LM test is given in Appendix \ref{app_additionappl}.\footnote{In the presence of heteroskedasticity, one might prefer the Wald version of the test, as this can be corrected in the standard way by using heteroskedasticty-robust standard errors. Empirically we found hardly any differences between the LM and Wald versions.}

\begin{figure}
\centering
\subfloat[PDS-LM HVAR\label{fig:im1}]{\includegraphics[width=0.32\textwidth, trim = {3.4cm 3cm 2.2cm 2.6cm}, clip]{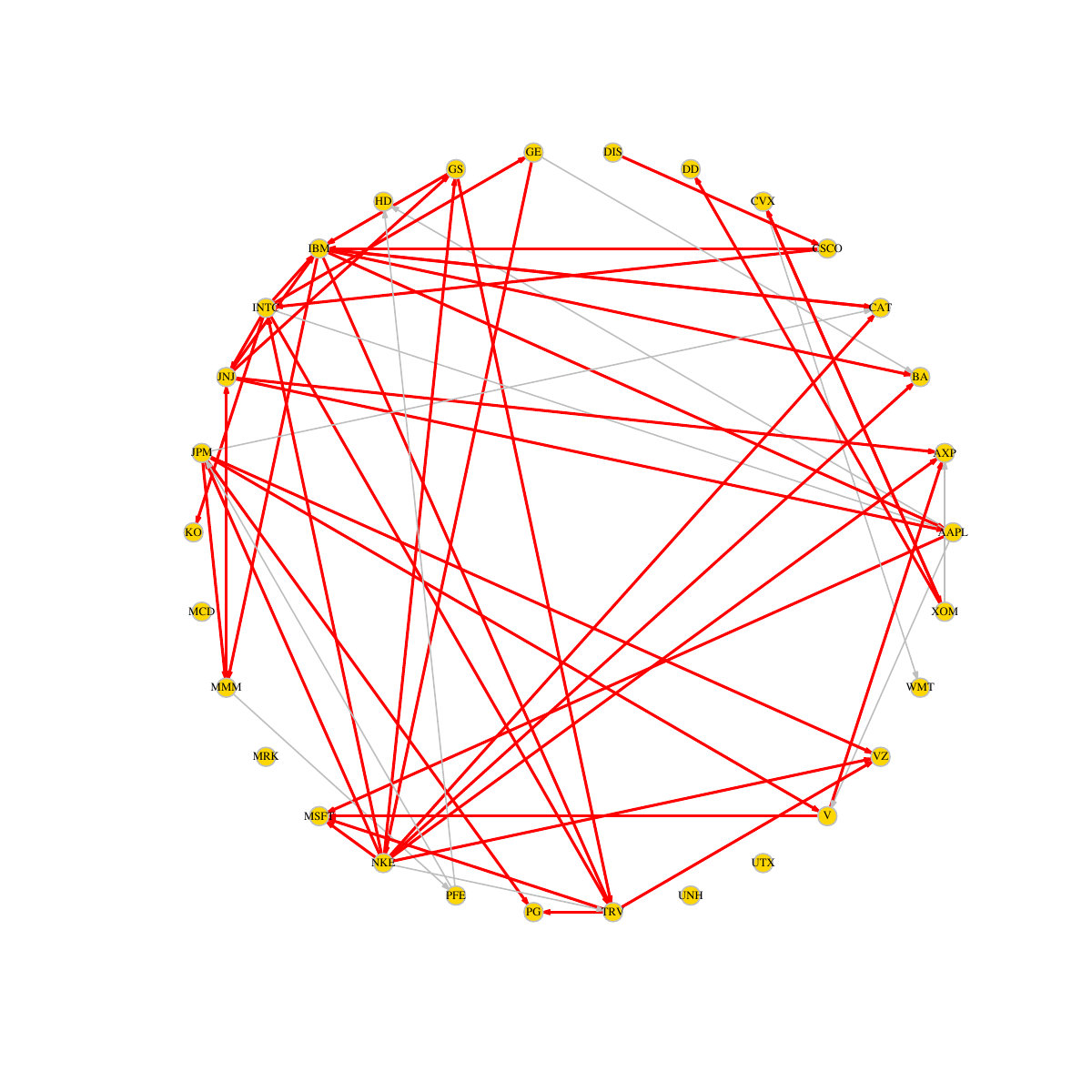}}
\subfloat[BiHVAR\label{fig:im2}]{\includegraphics[width=0.32\textwidth, trim = {3.3cm 3cm 2.3cm 2.6cm}, clip]{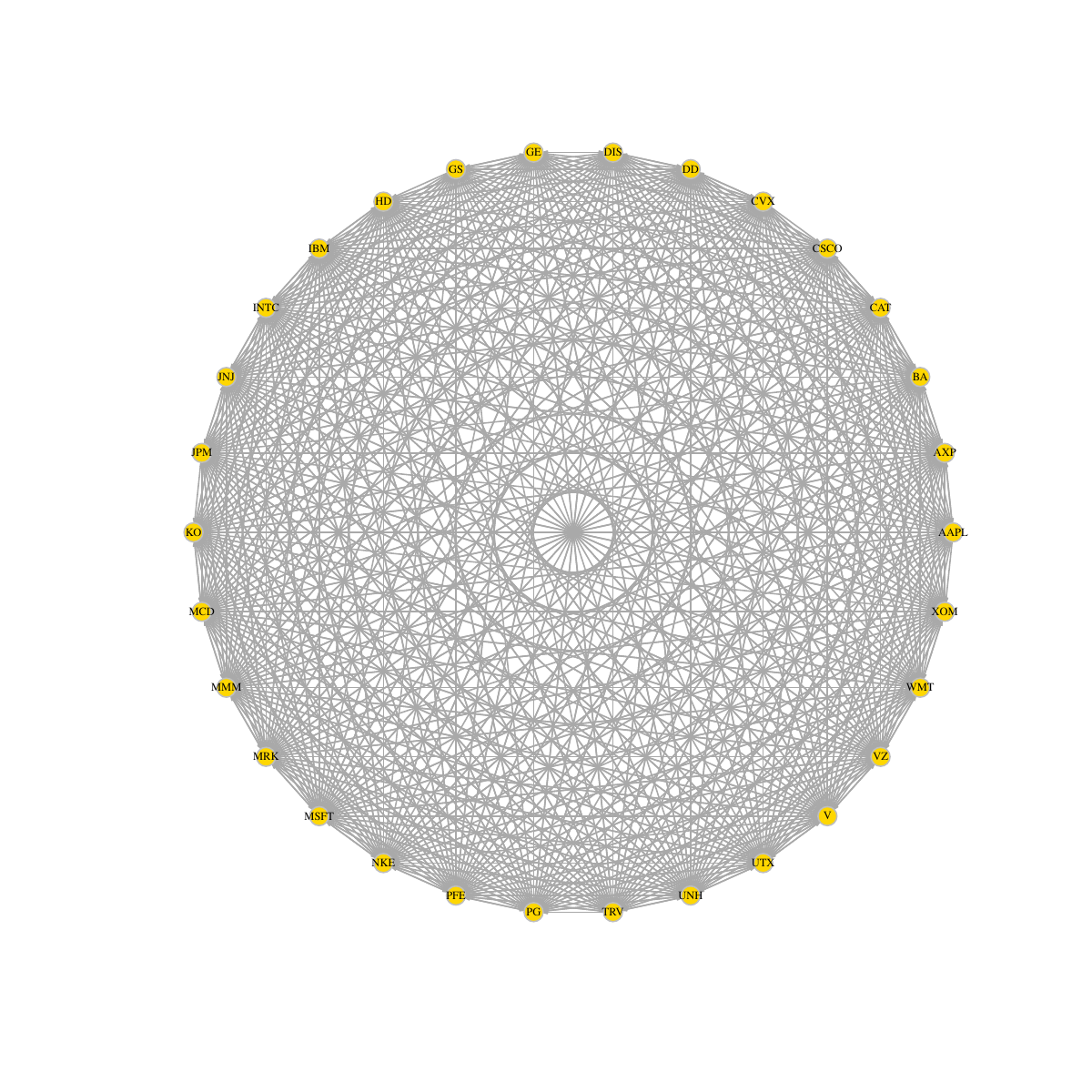}}
\subfloat[FullHVAR\label{fig:im3}]{\includegraphics[width=0.32\textwidth, trim = {3.2cm 3cm 2.4cm 2.6cm}, clip]{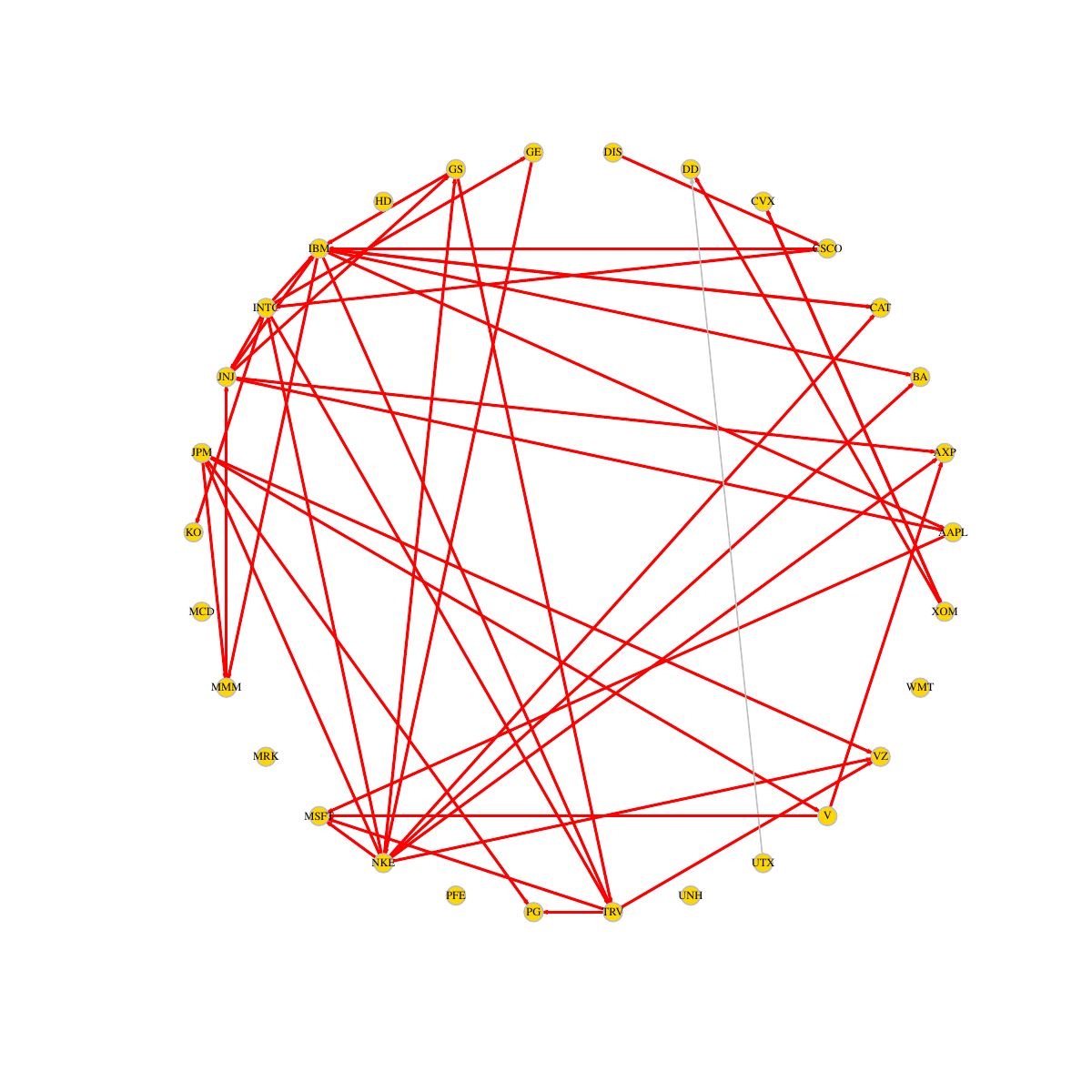}}
\caption{Spillover networks for the full sample period}\label{fig:netw1}
\end{figure}

We now report the results of our spillover tests for the volatility network. We use BIC to select the tuning parameter of the lasso, and perform the Granger causality tests with a 1\% significance level.\footnote{We do not perform a correction for multiple testing, as this would only qualitatively affect our results. Moreover, our goal is not to identify exactly the set of spillovers, but to get a feeling of the relations between two variables at a time. As such, we believe a multiple testing correction is not needed, though it can be easily implemented.} 
Figure \ref{fig:netw1} reports the transmission networks of volatilities estimated with the high-dimensional HVAR (PDS-LM HVAR), bivariate Granger causality tests (BiHVAR) for each pair of stocks, Granger causality tests from a full-system VAR (FullHVAR). The latter is feasible because of our large time series dimension with $T=2236$. For all methods we consider heteroskedasticity-robust variants.


While our PDS-LM method identifies a volatility transmission network which consists of 54 connections and the FullHVAR test picks up 44 connections, the BiHVAR tests detect a network consisting of 803 connections. These differences between our PDS-LM HVAR and the BiHVAR results are in line with our simulation results, confirming that bivariate Granger causality testing in VAR models is seriously affected by omitted variable bias in high-dimensional systems. Given the huge sample size $(T=2236)$ relative to the number of stocks, the FullHVAR is a feasible option, and it is reassuring how similar our PDS-LM HVAR performs compared to the FullHVAR. The similarity is visualized in Figures \ref{fig:netw1}\subref{fig:im1} and \ref{fig:netw1}\subref{fig:im3}, where the connections picked by both methods are highlighted in red. Of the 54 spillovers identified by the PDS-LM HVAR, 43 are also identified by the FullHVAR, while only 1 of the identified spillovers by the FullHVAR is not picked up by the PDS-LM HVAR.

\begin{figure}
\centering
\subfloat[PDS-LM HVAR\label{fig:im4}]{\includegraphics[width=0.32\textwidth, trim = {3.4cm 3cm 2.2cm 2.6cm}, clip]{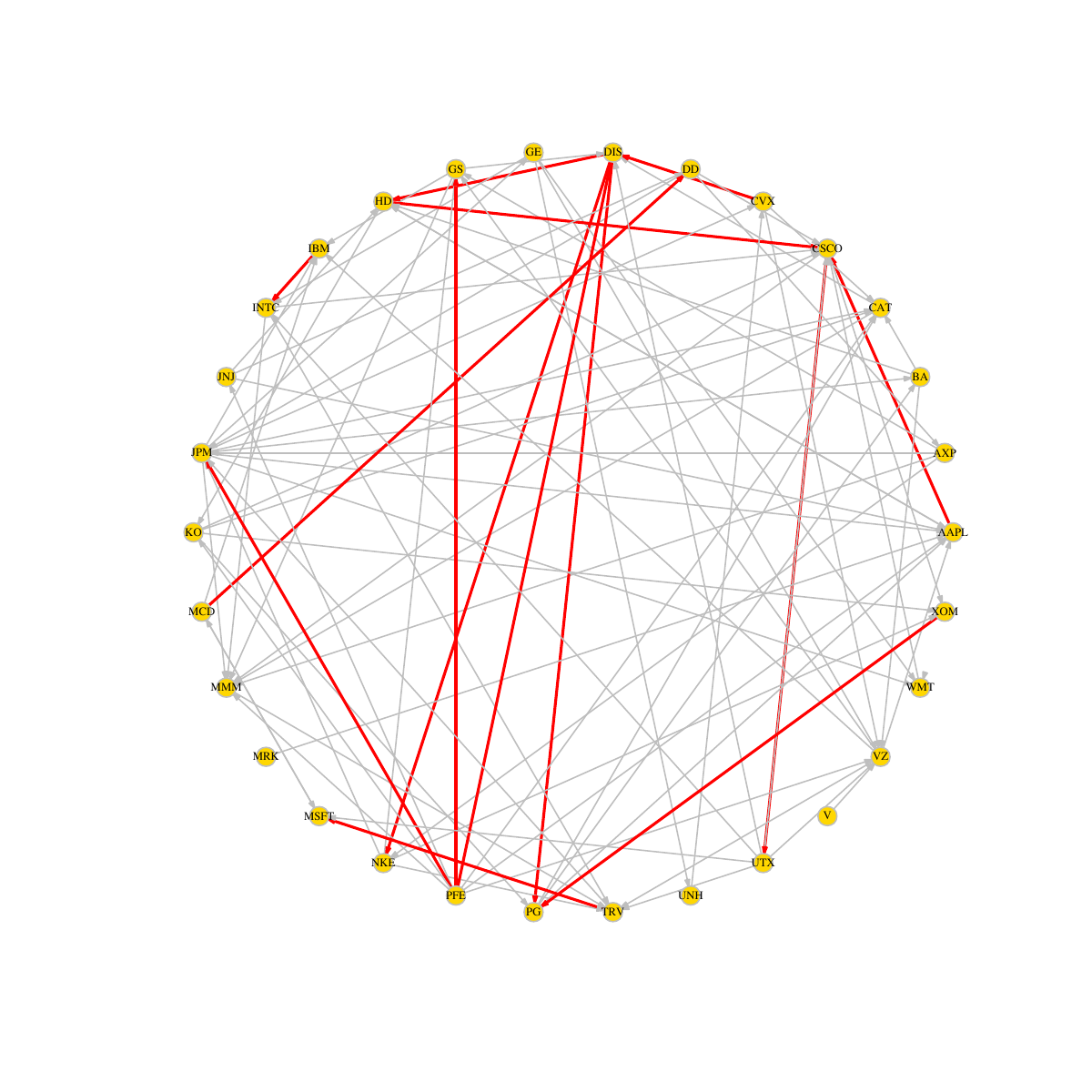}}
\subfloat[BiHVAR\label{fig:im5}]{\includegraphics[width=0.32\textwidth, trim = {3.3cm 3cm 2.3cm 2.6cm}, clip]{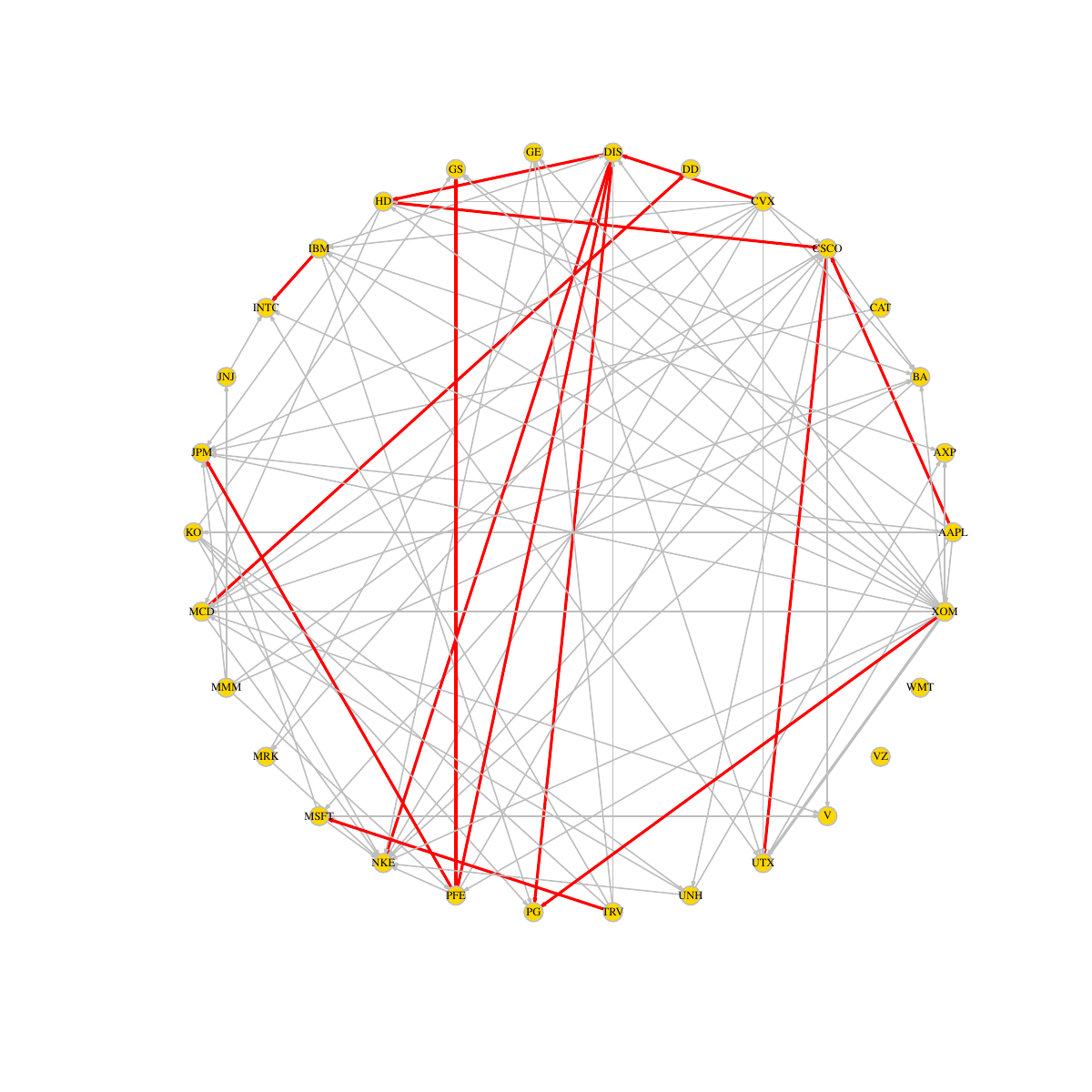}}
\subfloat[FullHVAR\label{fig:im6}]{\includegraphics[width=0.32\textwidth, trim = {3.2cm 3cm 2.4cm 2.6cm}, clip]{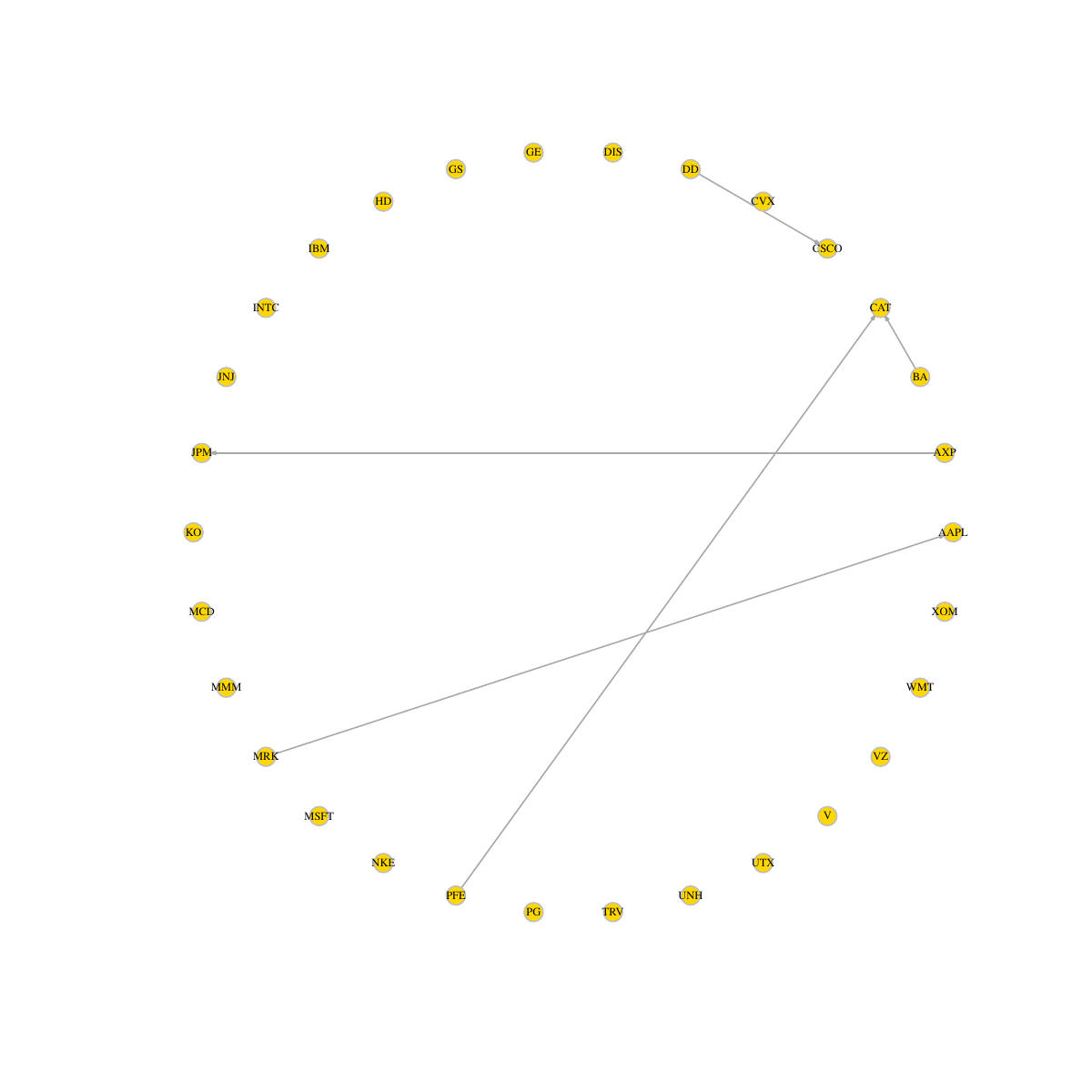}}
\caption{Spillover networks for the 2016-2017 sample period}\label{fig:netw2}
\end{figure}

We also consider a shorter time span, namely the period 2016-2017. Considering a shorter time period makes it more likely that relations remain stable over time. In particular, the chosen period avoids two major events that occurred previously and caused substantial instability on financial markets, namely the global financial crisis of 2008 and the U.S.~debt-ceiling crisis of 2011 \citep{baker2019policy}. It also allows us to study the performance differences among the three methods in a smaller sample of $T=284$ trading days, where the FullHVAR suffers from the curse of dimensionality. We present the results for the PDS-LM HVAR, BiHVAR and FullHVAR in Figure \ref{fig:netw2}. The number of significant transmissions is 91 for the PDS-LM HVAR, 85 for the BiHVAR and only 5 for the FullHVAR. Hence, the FullHVAR breaks down in this setting due to the small sample size and curse of dimensionality. On the other hand, while superficially the PDS-LM HVAR and the BiHVAR appear to perform similarly, they identify mostly different spillovers. The red lines in Figures \ref{fig:netw2}\subref{fig:im4} and \ref{fig:netw2}\subref{fig:im5} show the common connections, which are only 14 out of 91 for the PDS-LM HVAR (85 for the BiHVAR).

As a next step, we use our identified networks to find clusters of closely connected stocks, or communities as they are called in graph theory. Communities are groups of densely connected nodes with fewer connections across groups. In order to represent volatility spillover communities in the graph we use the \citet{newman2004finding} algorithm based on edge-betweenness. The edge betweenness for edge $e$ is defined as
$\sum_{s, t \neq e} \frac { \sigma _ { s t } ( e ) } { \sigma _ { s t } },$
where $\sigma_{st}$ is total number of shortest paths from node
$s$ to node $t$ and $\sigma_{st}(e)$ is the number of shortest paths passing through $e$. The edge with the highest betweenness is sequentially removed and the betweenness is recalculated at each step until the best partitioning of the network is found. 

\begin{figure}
\centering
\subfloat[PDS-LM HVAR\label{fig:im7}]{\includegraphics[width=0.32\textwidth, trim = {3.1cm 2.8cm 1.9cm 2.1cm}, clip]{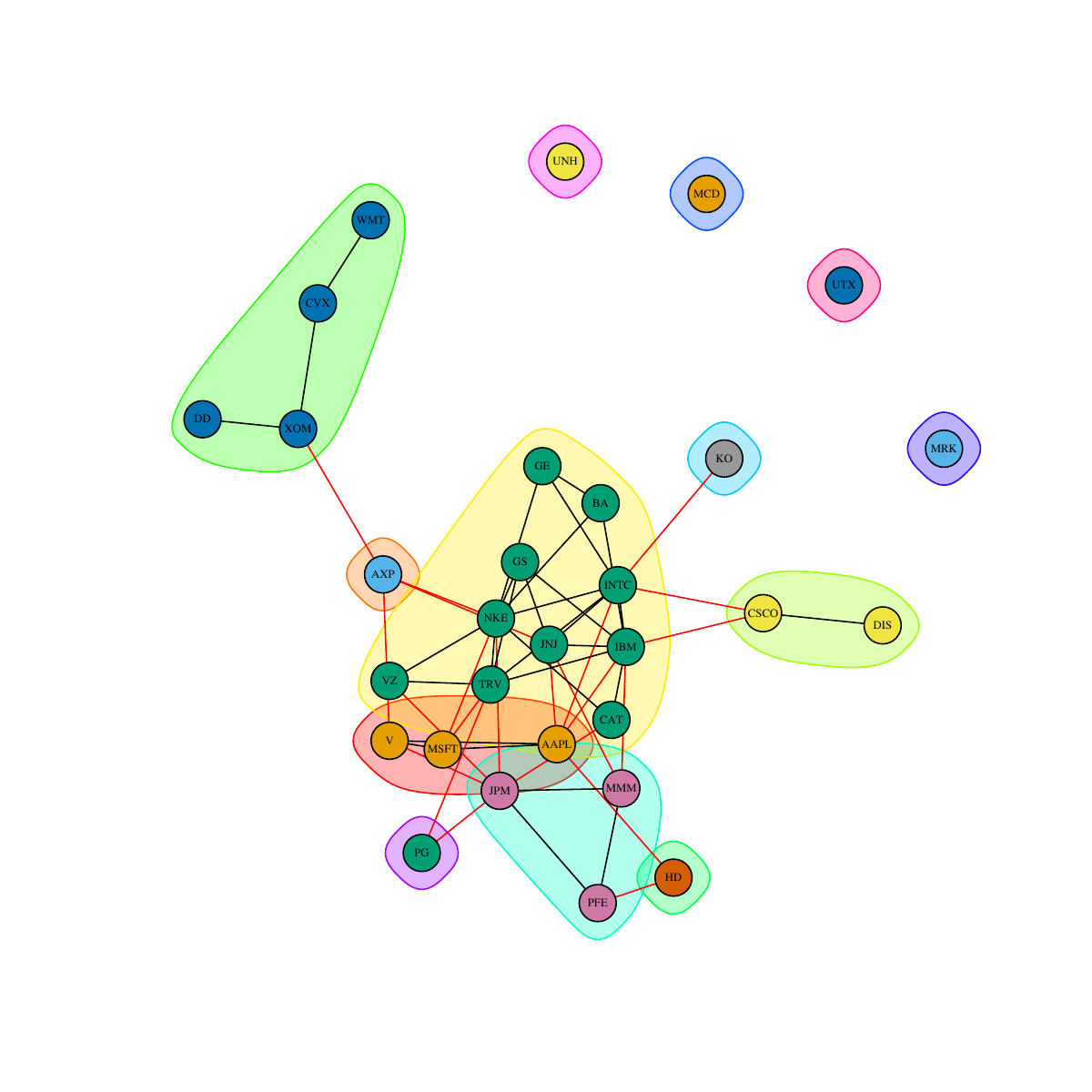}}
\subfloat[BiHVAR\label{fig:im8}]{\includegraphics[width=0.32\textwidth, trim = {3.0cm 2.8cm 2.0cm 2.1cm}, clip]{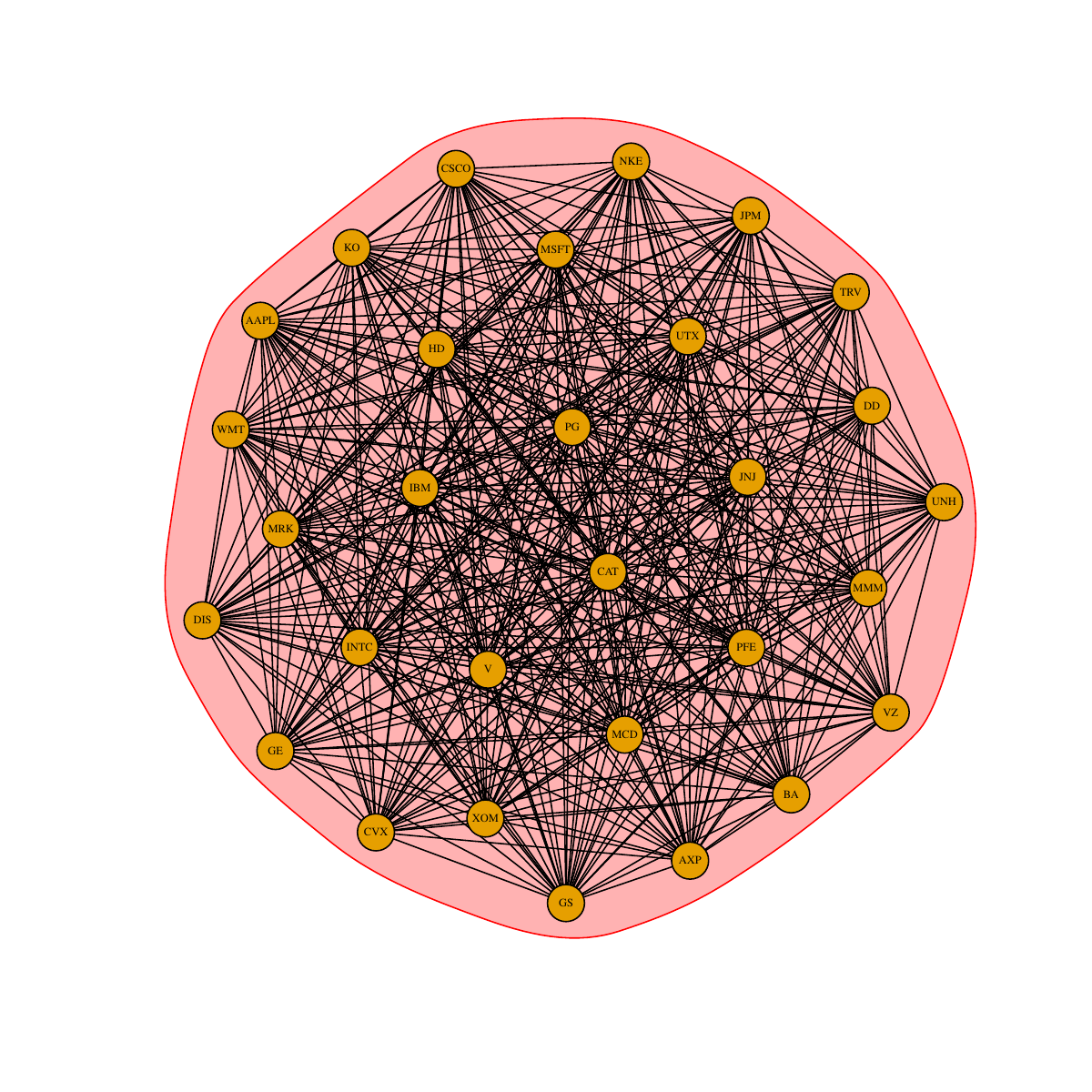}}
\subfloat[FullHVAR\label{fig:im9}]{\includegraphics[width=0.32\textwidth, trim = {2.9cm 2.8cm 2.1cm 2.1cm}, clip]{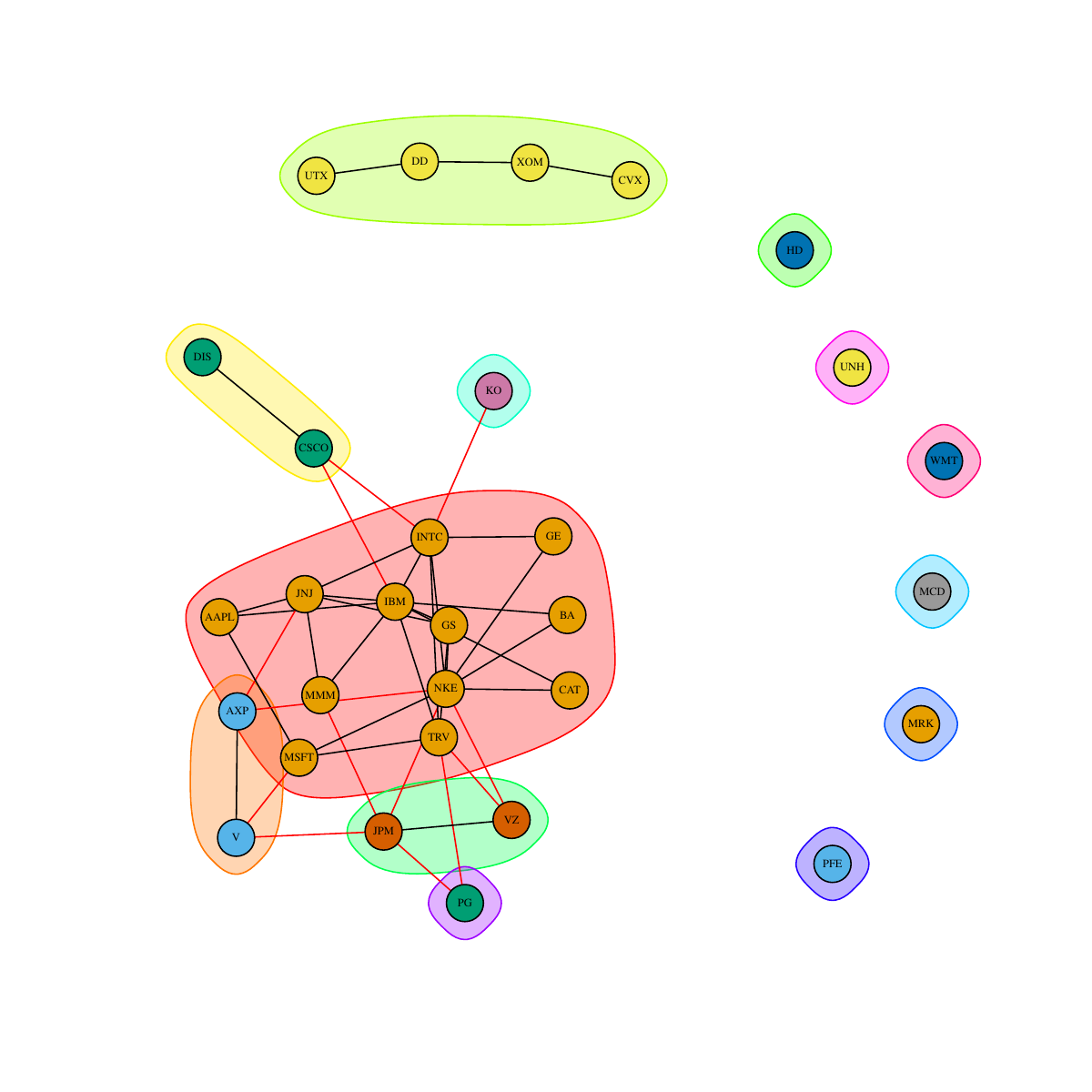}}
\caption{Volatility clusters for the full sample period}\label{fig:netw3}
\end{figure}

Figure \ref{fig:netw3} reports the graphs of the clustered network for the full sample analysis for the PDS-LM VAR, BiHVAR and FullHVAR respectively. The results for the PDS-LM VAR and FullHVAR show similar spillover clustering behavior, as expected. One large big-industry cluster, containing -- among others -- assets such as Johnson \& Johnson (J\&J), IBM, Nike (NKE) and Intel (INTC) dominates the picture being surrounded by small clusters containing 1 to 4 stocks. The PDS-LM VAR and FullHVAR resepctively identify 4 and 6 isolated stocks, which do not have any connections to others. Instead, the BiHVAR finds one single cluster containing all stocks. This reinforces our finding that bivariate Granger causality testing is not informative in high-dimensional systems.

\begin{figure}
\centering
\subfloat[PDS-LM HVAR\label{fig:im10}]{\includegraphics[width=0.32\textwidth, trim = {3.1cm 2.8cm 1.9cm 2.1cm}, clip]{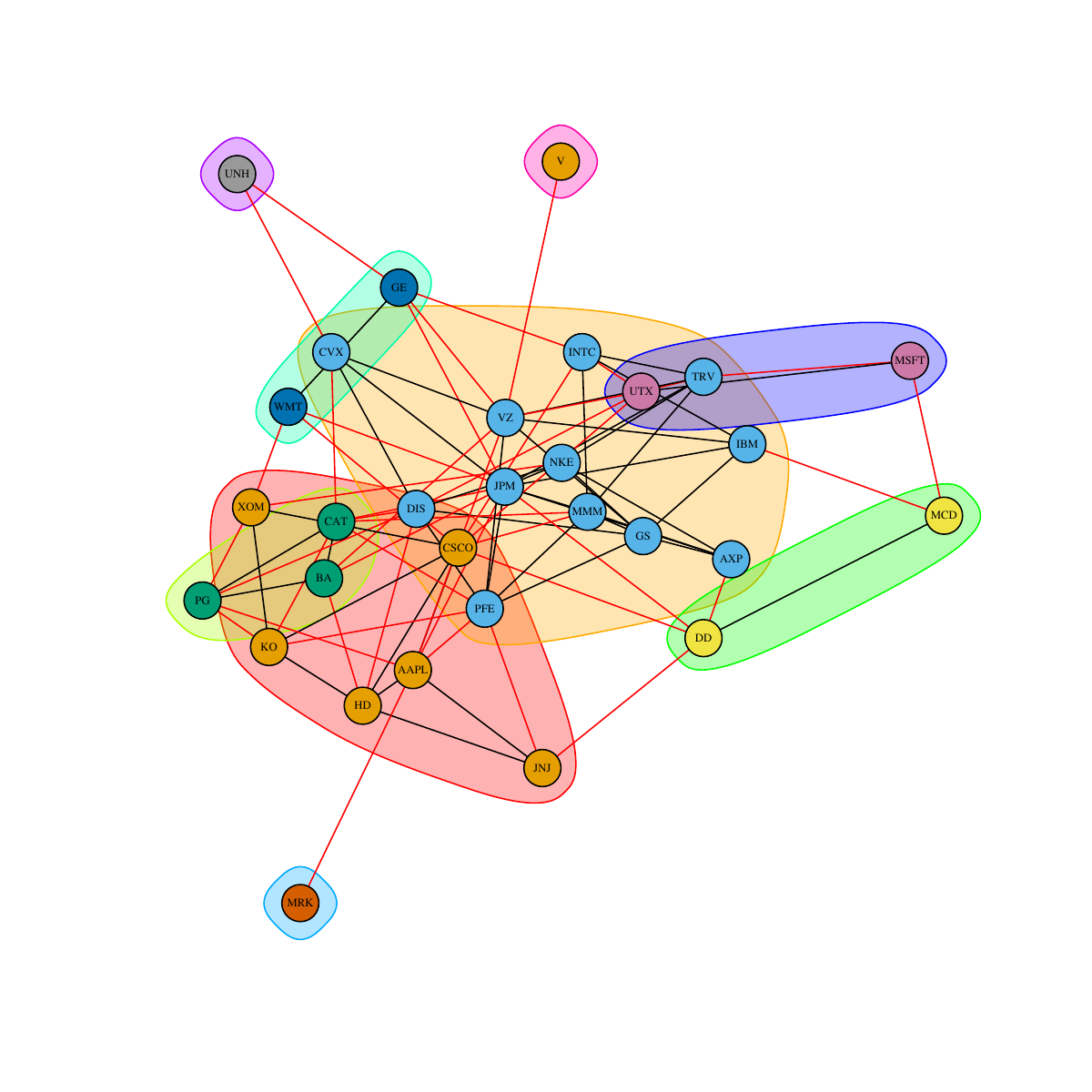}}
\subfloat[BiHVAR\label{fig:im11}]{\includegraphics[width=0.32\textwidth, trim = {3.0cm 2.8cm 2.0cm 2.1cm}, clip]{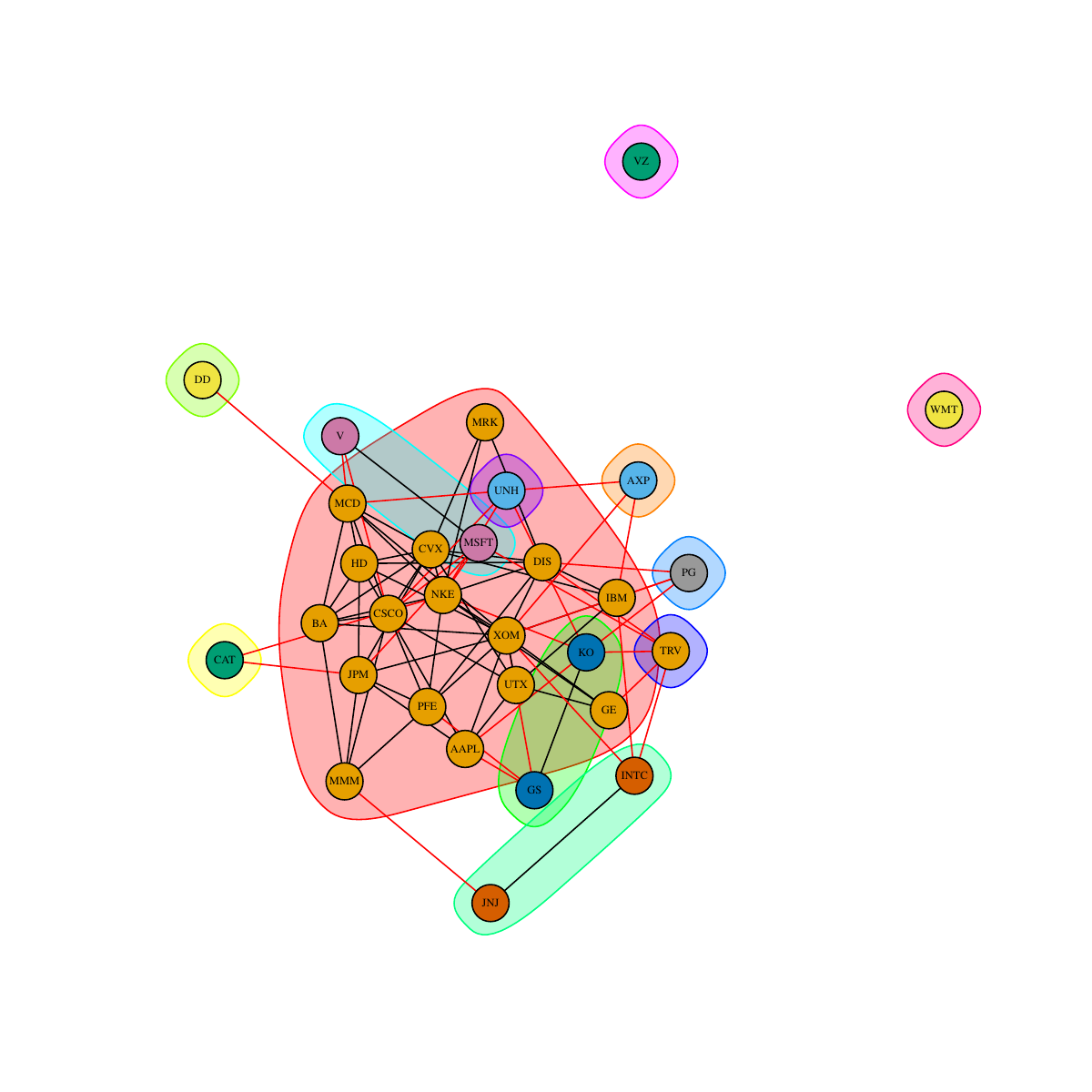}}
\subfloat[FullHVAR\label{fig:im12}]{\includegraphics[width=0.32\textwidth, trim = {2.9cm 2.8cm 2.1cm 2.1cm}, clip]{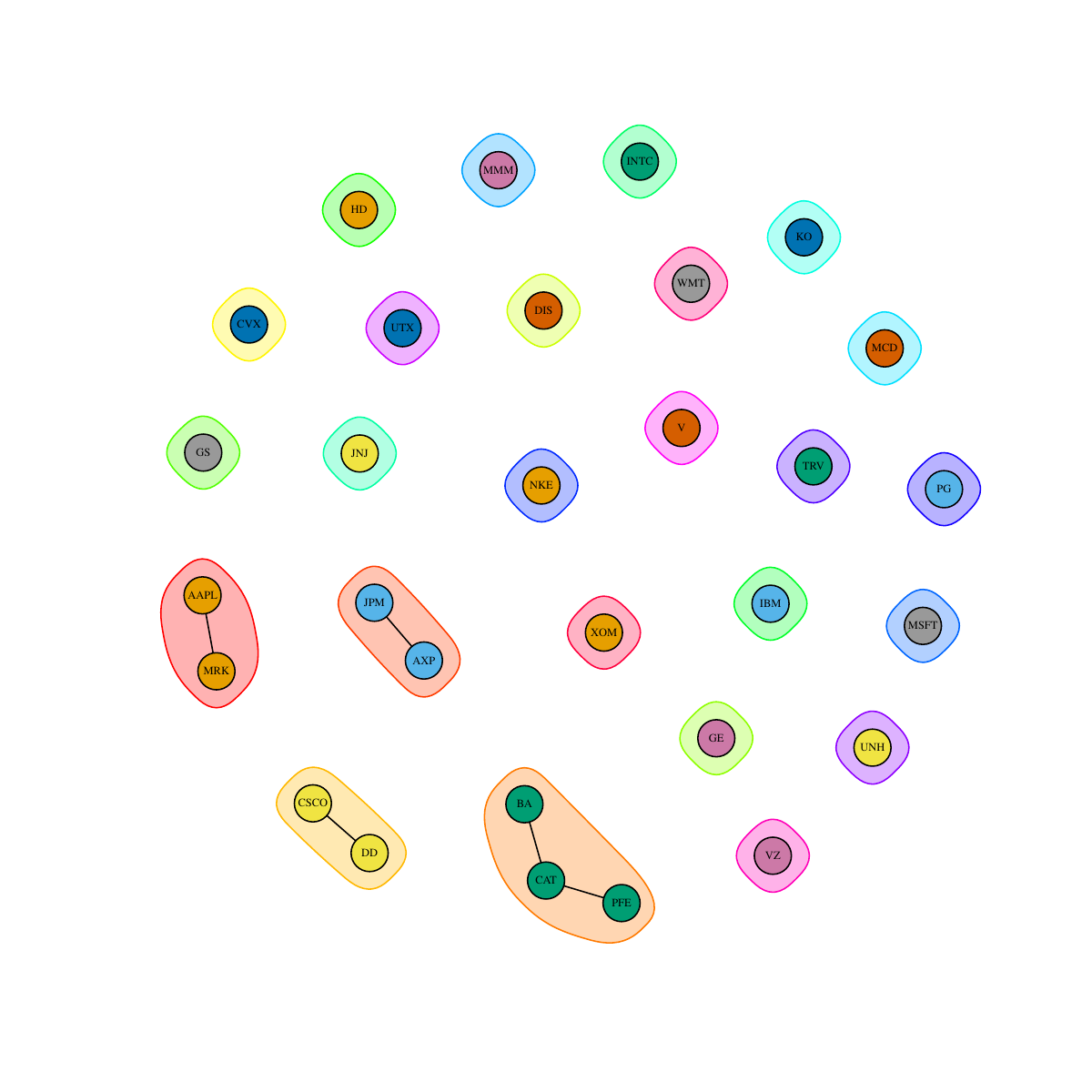}}
\caption{Volatility clusters for the 2016-2017 sample period}\label{fig:netw4}
\end{figure}

The clusters for the analysis done on the smaller 2016-2017 sample are reported in Figures \ref{fig:im10}, \ref{fig:im11} and \ref{fig:im12}. The patterns highlighted in the spillover network graphs re-occur in the clusters. PDS-LM HVAR in Figure \ref{fig:im10} picks up two main clusters of volatility spillovers containing 12 and 6 assets. In addition, four medium size clusters and three single-stock clusters are found. The difference between PDS-LM HVAR and BiHVAR is also reflected in the identified clusters. BiHVAR in Figure \ref{fig:im11} shows only one big conglomerate cluster of stocks linked to three two-stock clusters and 6 single-stock clusters. Finally, the breakdown of FullHVAR shows clearly in the non-informative, mostly unconnected single-stock clusters in Figure \ref{fig:im12}.

\subsection{Realized Variances \& Covariances}
In this subsection we extend our investigation to allow for spillovers from realized correlations to variances. While our application in Section \ref{sec:appl1} was only high-dimensional when we considered the shorter subsample, including correlations, which are of the order $K^2$, put a significantly larger strain on estimation, making the standard full VAR no option.
As elaborated later, it appears quite reasonable to expect changing correlations to also have an affect on the volatilities. By ignoring these in Section \ref{sec:appl1}, we are exposing ourselves again to a potential omitted variable bias. However, our method can directly incorporate these, as we demonstrate here.

While we remain mostly interested in contagion between the 30 realized volatilities, we add the $\frac{30\times 29}{2}=435$ realized correlations between all these assets as control variables. By maintaining our focus on the relations between the variances, our results are directly comparable to Section \ref{sec:appl1} and can be interpreted by assessing how the network changes when the correlations are added as controls in the VAR. Moreover, it also avoids the difficulties of trying to visualize the results from all $(30 \times 435)^2$ possible connections in the large VAR considered here.
In the same way that the realized variances employ high frequency\ data to estimate the integrated variance, the realized covariance (RC hereafter) estimates the integrated covariance of a multivariate diffusion process. Working with the full RC time-varying matrix is important for portfolio allocation and risk management. For a set of $n$ intra-day asset returns at day $t$ observed at $j=1,\ldots,M$ stacked in a column vector $\br_{j,t}$, the realized covariance is obtained such as $\bm{RC}_{t}=\sum_{j=1}^{M}\br_{j,t} \br_{j,t}^\prime$. Note that the realized variances are obviously on the diagonal of RC and that the RC matrix is positive definite when $M>n,$ namely when the number of high frequency observations per day is larger than the number of series. We investigate the same period as before and construct 10-minute realized covariances. Several studies have also proposed a Lasso framework on RC, see for instance \citet{callot2017modeling} and \citet{brito2018forecasting}, although their focus is more on portfolio allocation and forecasting. 

There are two main ways to work with the RC matrix. The first approach stacks realized variances and covariances in a single vector. For instance, \citet{bauer2011forecasting} consider the matrix log transformation of $RC_{t}$ series, a matrix
that they call the log-space volatility.
The drawback of that log transform is that the interpretation of the original series, in our case the volatilities, is lost as the combinations involve nonlinear transforms of both realized variances and covariances. This is not compatible with the aim of this paper.

The second approach uses the log realized volatilities and the correlations separately, as done by for instance \citet{oh2016high}. The underlying idea, following the DCC model of \citet{engle2002dynamic}, is to decompose $ RC_{t}^{(d)}=D_{t}^{(d)}R_{t}^{(d)}D_{t}^{(d)}$ with $D_{t}^{(d)}$ a diagonal matrix with the square root of the individual realized variance and $R_{t}^{(d)}$ the realized correlation matrix. \citet{oh2016high} use the HVAR model structure for each realized volatilities, they consequently assume no Granger causality across volatilities. 

We propose something which is, to some extent, in between these two approaches. 
We look at two separate objects as in the DCC model, but stack the log of the realized variances $\by_{1t}^{(d)^{\prime }}$ 
and $z$-transforms $\by_{2t}^{(d)}=\mathrm{arc}\tanh \left( \mathrm{vech}(\bR_{t}^{(d)})\right) $ of the realized correlations 
in a larger vector $\by_{t}^{(d)}=(\underset{1\times 30}{\underbrace{\by_{1t}^{(d)^{\prime }}}},%
\underset{1\times 435}{\underbrace{\by_{2t}^{(d)^{\prime }}}})^{\prime }$ on which we estimate a VHAR of dimension 465.

\begin{figure}[H]
\centering
\subfloat[PDS-LM HVAR\label{fig:im13}]{\includegraphics[width=0.32\textwidth, trim = {3.1cm 2.8cm 1.9cm 2.1cm}, clip]{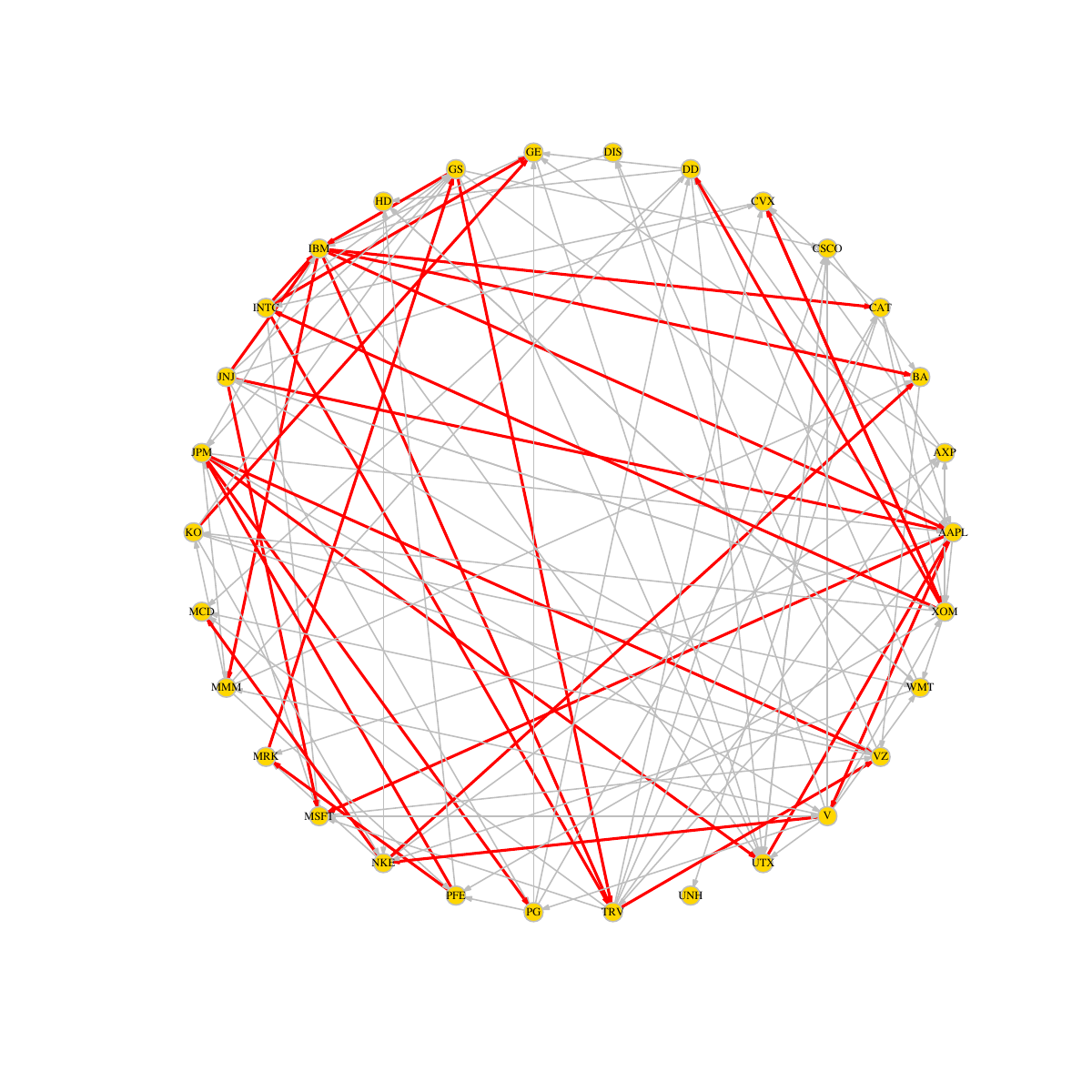}}
\subfloat[PDS-LM HVAR\label{fig:im14}]{\includegraphics[width=0.32\textwidth, trim = {3.0cm 2.8cm 2.0cm 2.1cm}, clip]{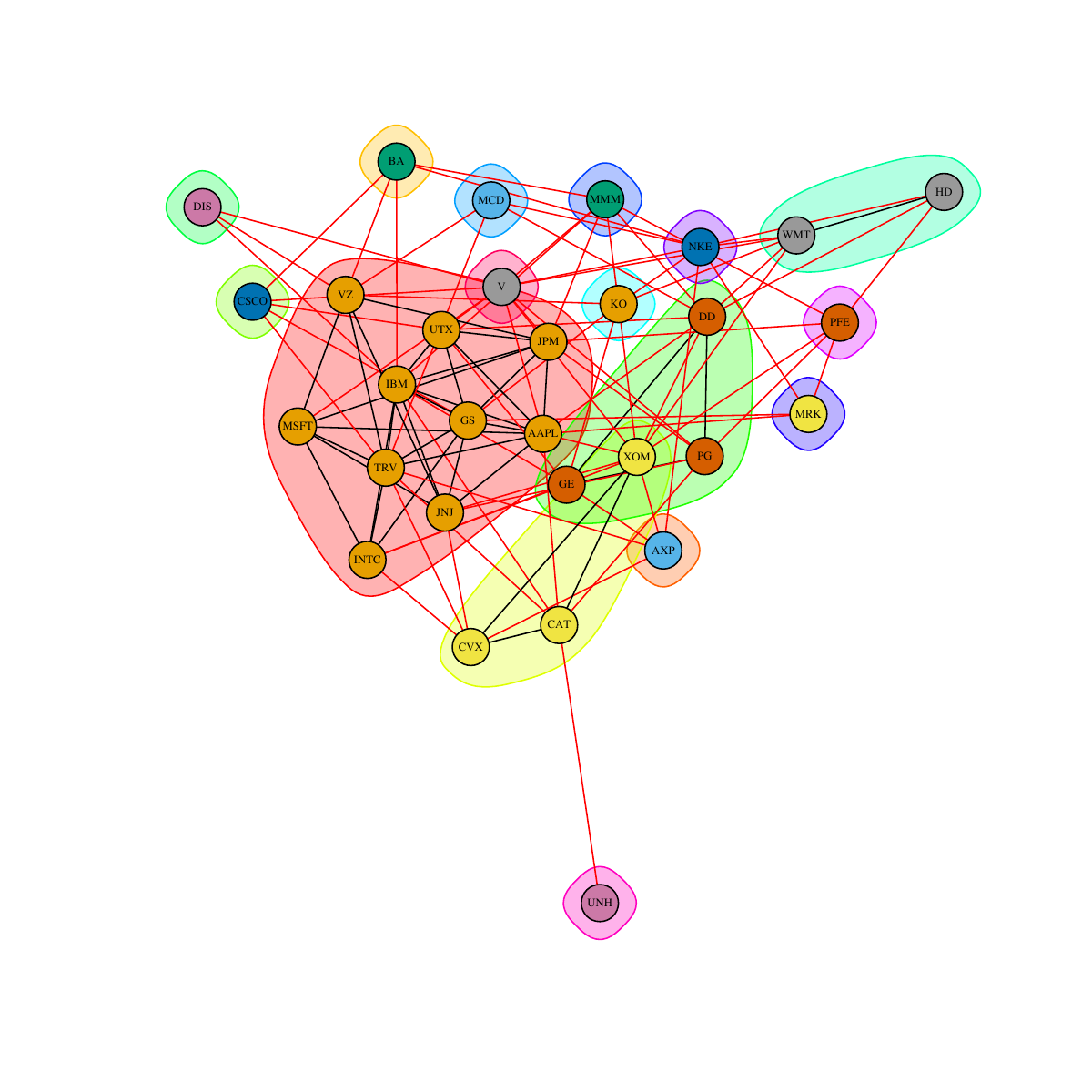}}
\caption{Spillover network and volatility clusters}
\end{figure}

In this HVAR each of the 465 equations depends on 1395 dynamic parameters plus the constant. We focus on the 30 equations corresponding to $\by_{1t}^{(d)}$ volatilities and consequently the bivariate causalities between these realized volatilities as in the previous section. Figure \ref{fig:im13} reports a total of 113 connections, which is about twice the connections in Figure \ref{fig:im1}. In red we highlighted the 31 common connections with Figure \ref{fig:im1}. Interestingly, adding more variables therefore allows us to uncover more relations. It seems that this allows us to uncover partial effects that were previously obscured by counteracting effects of the correlations. Importantly, the number of connections is still far less than compared to the BiHVAR in Figure \ref{fig:im2}, and the PDS-LM HVAR is still able to deliver a clear picture of the causal connections when the system considered is high-dimensional. While the different connections found here obviously also lead to a different clustering, Figure \ref{fig:im14} shows that the clustering is quite similar, certainly regarding qualitative conclusions. 

\section{Conclusion} \label{sec:conc}
We propose an LM test in order to test for Granger causality in high-dimensional VAR models. We employ a post-double selection procedure using the lasso to select the set of relevant covariates in the system. The double selection step allows to substantially reduce the omitted variable bias and thereby allowing for valid post-selection inference on the parameters. 

We provide an extensive simulation study to evaluate the performance of our method in finite samples, paying particular attention to the tuning of the penalty parameter. We compare different information criteria, time series cross-validation and a plug-in method based on theoretical arguments, and find that generally BIC and the theoretically tuned penalty perform best. However, to use information criteria in systems with a significantly larger number of variables than observations, a lower bound on the penalty parameter is needed to prevent too many variables being selected. 
The simulations also show that, when properly tuned, our proposed PDS-LM test attains good results both for size and power under different DGPs. Especially, it is shown to be robust both to non-sparse settings as well as to lag-length overspecification. 

We also empirically investigate the usefulness of our method in 
a study where we apply our PDS-LM method to a high-dimensional VHAR process in order to construct a contagion network of volatility spillovers for 30 large capital stocks, also accounting for effects from changing correlations. We find that by increasing the information set through considering a high-dimensional VAR model instead of bivariate models, we are able to obtain more realistic effects than in low-dimensional models. Furthermore, even when the sample size is not large enough to use standard full-system VAR techniques, our method remains reliable and delivers accurate results.

Note that unlike \citet{belloni2014high}, we do not give a ``truly'' causal interpretation to the established Granger causalities. In how far Granger causality is a useful concept to study true causality is (and has long been) open to debate, see for example \citep{eichler2013causal} and the references therein. Moreover, though it appears desirable and in line with \citeauthor{granger1969investigating}'s (\citeyear{granger1969investigating}) original intentions to make the information set as large as possible, it is well known in the literature on graphical models \citep[see][]{eichler2013causal} for causality that considering only the full model is not sufficient for establishing true causal relations from Granger causal ones. For instance, one-period Granger causality in systems with more than two variables cannot capture indirect causal chains spanning over multiple periods. However, the analysis of the full model is a necessary ingredient for any study of causality in a graphical framework. It would therefore be an interesting avenue for further research to study how the method proposed here could fit into such a graphical framework.

\begin{small}
\onehalfspacing
\bibliographystyle{chicago}
\bibliography{literature}
\end{small}

\begin{appendices}
\numberwithin{table}{section}
\numberwithin{equation}{section}
\numberwithin{lemma}{section}
\section{Proofs}

\begin{lemma} \label{lem:minev}
Let $\bX_{-GC}$ satisfy Assumption \ref{as:hlev}(\ref{as:sev}). Then with probability at least $1 - \Delta_T$, we have that
\begin{equation*}
\norm{\bdelta}_2^2 \leq \bar{s}_T \norm{\widetilde{\bX}_{-GC}^\otimes \bdelta / \sqrt{T}}_2^2 / \phi_{T,\min}^2,
\end{equation*}
for any $\bdelta = (\bdelta_1^\prime, \ldots, \bdelta_{\nGCed}^\prime)^\prime$ such that $\abs{S_{\delta}} \leq \bar{s}_T$, where $S_{\delta} = \bigcup_{i=1}^{\nGCed} \{m: \delta_{i,m} \neq 0 \}$.
\end{lemma}

\begin{proof}[\bf{Proof of Lemma \ref{lem:minev}}]
As before, let $\bX_{S}$ denote the submatrix containing those columns of $\bX_{-GC}$ corresponding to the elements in $S$. It follows from Assumption \ref{as:hlev}(\ref{as:eig}) that for any $\bgamma$ satisfying $\abs{S_{\bgamma}} \leq \bar{s}_T$, we have that $\lambda_{\min} (\bX_{S_{\bgamma}}^\prime \bX_{S_{\bgamma}}/T) \geq \phi_{T,\min}^2$. Then, with probability $1-\Delta_T$ we have that
\begin{equation*}
\begin{split}
&\min_{\bdelta: \abs{S_{\bdelta}} \leq \bar{s}_T}  \norm{\widetilde{\bX}_{-GC}^\otimes \bdelta/ \sqrt{T}}_2^2 / \norm{\bdelta}_2^2 = \min_{\bdelta: \abs{S_{\bdelta}} \leq \bar{s}_T} \bdelta^\prime \left(\bG_T^\prime \bG_T \otimes \bX_{-GC}^\prime \bX_{-GC} / T \right) \bdelta / \norm{\bdelta}_2^2\\
&\quad= \min_{\abs{S} \leq \bar{s}_T} \min_{\bx} \bx^\prime \left(\bG_T^\prime \bG_T \otimes \bX_{S}^\prime \bX_{S} / T \right) \bx / \norm{\bx }_2^2
= \min_{\abs{S} \leq \bar{s}_T} \lambda_{\min} (\bG_T^\prime \bG_T \otimes \bX_S^\prime \bX_S / T) \\
&\quad= \lambda_{\min} (\bG_T^\prime \bG_T) \min_{\abs{S} \leq \bar{s}_T} \lambda_{\min} (\bX_{S}^\prime \bX_{S} / T) \geq C\phi_{T,\min},
\end{split}
\end{equation*}
as $\lambda_{\min} (\bG_T^\prime \bG_T) \geq C > 0$ by assumption. Without loss of generality we may then absorb the constant $C$ into $\phi_{T,\min}$.
\end{proof}

\begin{proof}[\bf{Proof of Theorem \ref{th:asdis}}]
Our proof follows along the lines of the proof of Theorem 1 and 2 of \citet{belloni2014inference}, with the main distinction that we consider multiple variables of interest, and multiple ``treatments'' instead of a single one for each. For this purpose we first define some notation. Let $\bGamma = (\bgamma_1, \ldots, \bgamma_{\nXGC})$ and $\hat{\bGamma} = (\hat{\bgamma}_1, \ldots, \hat{\bgamma}_{\nXGC})$, and let $\bgamma^{\otimes} = \bI_{\nGCed} \otimes \bGamma$. Furthermore, let $\P(\bA) = \bA (\bA^\prime \bA)^{-1} \bA^\prime$ denote the projection on the space spanned by $\bA$ and let $\M (\bA) = \bI - \P(\bA)$ denote the corresponding residual-maker. By standard partitioned regression algebra we get
\begin{equation} \label{eq:FW_betahat}
\begin{split}
\sqrt{T} (\hat{\bbeta}_{GC}^{\textsc{pds}} - \bbeta_{GC}) &= \underbrace{\left( \widetilde\bX_{GC}^{\otimes} \M(\widetilde\bX_{\hat{S}^{\otimes}}^{\otimes}) \widetilde\bX_{GC}^{\otimes} /T \right)^{-1}}_{\bB_T^{-1}}\\
&\quad \times \underbrace{\widetilde\bX_{GC}^{\otimes} \M(\widetilde\bX_{\hat{S}^{\otimes}}) \left[ \widetilde\bX_{-GC}^{\otimes} \bbeta_{-GC} + \widetilde\bu_{\GCed} \right] / \sqrt{T}}_{\ba_T}\\
\end{split}
\end{equation}
where $\widetilde \bX^{\otimes} = \bG_T^\otimes \bX^{\otimes} = \bG_T \otimes \bX$.
We will now show that $\ba_T = \widetilde\bE^{\otimes\prime} \widetilde\bu_{\GCed} / \sqrt{T}$ and $\bB_T = \widetilde\bE^{\otimes\prime} \widetilde\bE^{\otimes} / T + o_p(1)$. Given these results, the limit distribution then follows directly from Assumption \ref{as:hlev}(\ref{as:clt}).

We first consider $\ba_T$. Note that from \eqref{eq:lasso_steps2} we have that $\widetilde\bX_{GC}^{\otimes} = \bG_T \otimes \left[\bX_{-GC} \bGamma + \bE \right] = \widetilde\bX_{-GC}^{\otimes} \bGamma^\otimes + \widetilde\bE^{\otimes}$, and therefore we can write
\begin{equation*}
\begin{split}
\ba_T &= \widetilde\bE^{\otimes\prime} \widetilde\bu_I / \sqrt{T} + \underbrace{\bGamma^{\otimes \prime} \widetilde\bX_{-GC}^{\otimes\prime} \M(\widetilde\bX_{\hat{S}^\otimes}^\otimes) \widetilde\bX_{-GC}^\otimes \bbeta_{-GC} / \sqrt{T}}_{\ba_{T,1}} + \underbrace{\bGamma^{\otimes \prime} \widetilde\bX_{-GC}^{\otimes\prime} \M(\widetilde\bX_{\hat{S}^\otimes}^\otimes) \widetilde\bu_{\GCed} / \sqrt{T}}_{\ba_{T,2}}\\
&\quad + \underbrace{\widetilde\bE^{\otimes\prime} \M(\widetilde\bX_{\hat{S}^\otimes}^\otimes) \widetilde\bX_{-GC}^\otimes \bbeta_{-GC} / \sqrt{T}}_{\ba_{T,3}} - \underbrace{\widetilde\bE^{\otimes\prime} \P(\widetilde\bX_{\hat{S}^\otimes}^\otimes) \widetilde\bu_{\GCed} / \sqrt{T}}_{\ba_{T,4}}\\
\end{split}
\end{equation*}
We will now show that the terms $\ba_{T,1}, \ldots, \ba_{T,4}$ vanish. For $\ba_{T,1}$, note that
\begin{equation*}
\norm{\ba_{T,1}}_2 \leq \sqrt{T} \underbrace{\norm{\M(\widetilde\bX_{\hat{S}^\otimes}^\otimes) \widetilde\bX_{-GC}^\otimes \bGamma^\otimes / \sqrt{T}}_2}_{\norm{\bA_{T,1,1}}_2} \underbrace{\norm{\M(\widetilde\bX_{\hat{S}^\otimes}^\otimes) \widetilde\bX_{-GC}^\otimes \bbeta_{-GC} / \sqrt{T}}_2}_{\norm{\ba_{T,1,2}}_2},
\end{equation*}
where for any matrix $\bM$, the norm $\norm{\cdot}_p$ represents the induced $l_p$-matrix norm $\norm{\bM}_p = \sup_{\bx\neq 0} \norm{\bM \bx}_p / \norm{\bx}_p$. As $\hat{S}_j \subseteq \hat{S}_X$ for all $j=1,\ldots, \nXGC$ and  $\hat{S}_X^{\otimes} \subseteq \hat{S}^\otimes$, the space spanned by $\bG_T^{\otimes} \bX_{\hat{S}_X^\otimes} = \bG_T \otimes \bx_{\hat{S}_X}$ is a subspace of the space spanned by $\bG_T^{\otimes} \bX_{\hat{S}^\otimes}^\otimes$, and therefore $\norm{\M \left(\bG_T^{\otimes} \bX_{\hat{S}^\otimes}^\otimes \right) \by}_2 \leq \norm{\M \left(\bG_T^{\otimes} \bX_{\hat{S}_X^\otimes}^\otimes \right) \by}_2$ for any compatible matrix $\bG_T^{\otimes}$ and vector $\by$. Then, using that
\begin{equation} \label{eq:kron_M}
\begin{split}
&\M \left(\bG_T^{\otimes} \bX_{\hat{S}_X^\otimes}^\otimes \right) = \M\left(\bG_T \otimes \bX_{\hat{S}_X}\right) \widetilde\bX_{-GC}^\otimes \bGamma^\otimes\\
&= \M \left(\bG_T \otimes \bX_{\hat{S}_X} \right) \left(\bG_T \otimes \bX_{-GC} \bGamma \right) = \bG_T \otimes \M \left(\bX_{\hat{S}_X} \right) \bX_{-GC} \bGamma,
\end{split}
\end{equation}
we find that
\begin{equation*}
\begin{split}
\norm{\bA_{T,1,1}}_2 &\leq \norm{\M \left(\widetilde\bX_{ \hat{S}_X^\otimes}^\otimes \right) \widetilde\bX_{-GC}^\otimes \bGamma^\otimes / \sqrt{T}}_2 \leq \norm{\bG_T}_2 \norm{\M(\bX_{\hat{S}_X}) \bX_{-GC} \bGamma / \sqrt{T}}_2 \\
&\leq \norm{\bG_T}_2 \sum_{j=1}^{\nXGC} \norm{\M(\bX_{\hat{S}_X}) \bX_{-GC} \bgamma_j/ \sqrt{T}}_2 \leq \norm{\bG_T}_2 \sum_{j=1}^{\nXGC} \norm{\M(\bX_{\hat{S_j}}) \bX_{-GC} \bgamma_j/ \sqrt{T}}_2.
\end{split}
\end{equation*}
Then,
\begin{equation}\label{eq:MXb}
\begin{split}
\norm{\M(\bX_{\hat{S}_j}) \bX_{-GC} \bgamma_j/ \sqrt{T}}_2 &= \min_{\gamma: \gamma_m = 0, m \notin \hat{S}_j} \norm{\bX_{-GC} \bgamma_j - \bX_{\hat{S_j}}\bgamma }_2/ \sqrt{T} \\
&\leq \norm{\bX_{-GC} (\bgamma_j - \hat{\bgamma}_j) }_2 / \sqrt{T}, \quad\qquad j = 0,\ldots, N_X,
\end{split}
\end{equation}
as $\hat{S}_j = \{m: \hat{\gamma}_{m,j} \neq 0\}$ and therefore the constraint in the minimization is satisfied. It then follows from Assumption \ref{as:hlev}(\ref{as:cons}) that $\norm{\bA_{T,1,1}}_2 \leq \nGCvars \delta_T T^{-1/4}$ with probability $1 - \Delta_T$.

For $\ba_{T,1,2}$, from the definition of the best linear predictor it directly follows that
\begin{equation*}
\begin{split}
\bgamma_0 &= \left(\E \bX_{-GC}^{\otimes\prime} \bX_{-GC}^\otimes \right)^{-1} \E \bX_{-GC}^\prime (\bX_{GC}^\otimes \bbeta_{GC} + \bX_{-GC}^\otimes \bbeta_{-GC} + \bu_{\GCed}) = \bGamma^\otimes \bbeta_{GC} + \bbeta_{-GC},
\end{split}
\end{equation*}
such that we can substitute $\bbeta_{-GC} = \bgamma_0 - \bGamma^\otimes \bbeta_{GC}$ in $\ba_{T,1,2}$ to find
\begin{equation*}
\begin{split}
\norm{\ba_{T,1,2}}_2 &\leq \norm{\M\left(\widetilde\bX_{\hat{S}^\otimes}^\otimes \right) \widetilde\bX_{-GC}^\otimes \bgamma_0/\sqrt{T}}_2 + \norm{\bA_{T,1,1}}_2 \norm{\bbeta_{GC}}_2,
\end{split}
\end{equation*}
where the negligibility of the second term follows directly from the result above plus Assumption \ref{as:hlev}(\ref{as:bound}). As $\hat{S}_0 \subseteq \hat{S}^\otimes$, the first term can be bounded by
\begin{equation*}
\begin{split}
&\norm{\M \left(\widetilde\bX_{\hat{S}^\otimes}^\otimes \right) \widetilde\bX_{-GC}^\otimes \bgamma_0/\sqrt{T}}_2 \leq \norm{\M \left(\widetilde\bX_{\hat{S}_0}^\otimes \right) \widetilde\bX_{-GC}^\otimes \bgamma_0/\sqrt{T}}_2 \\
&\quad\leq \norm{\bG_T}_2 \norm{\bX_{-GC}^\otimes (\bgamma_0 - \hat\bgamma_0)}_2 / \sqrt{T} \leq \sqrt{\nGCed} \delta_T T^{-1/4},
\end{split}
\end{equation*}
where we use that, for $\hat{\bbeta}_{\bG_T} = \M(\bG_T \bX) \bG_T \by$,
\begin{equation*}
\norm{\M(\bG_T \bX) \bG_T \by}_2 = \norm{\bG_T \left(\by - \bX \hat{\bbeta}_{\bG_T} \right)}_2 \leq \norm{\bG_T \left(\by - \bX \hat{\bbeta}_{\bI} \right)}_2 \leq \norm{\bG_T}_2 \norm{\by - \bX \hat{\bbeta}_{\bI}}_2.
\end{equation*}
It then follows directly that $\norm{\ba_{T,1}}_2 = O_p (\delta_T^2) = o_p(1)$.

For $\ba_{T,2}$, let $\bgamma_j$ denote the $j$-th column of $\bGamma^\otimes$ and define the noiseless generalized least squares estimator
\begin{equation} \label{eq:tildebeta}
\check{\bgamma}_{j, S}^\otimes = \argmin_{\bgamma:\gamma_m = 0, m \notin S} \norm{\widetilde \bX_{-GC}^\otimes \bgamma_j^\otimes - \widetilde\bX_{-GC}^\otimes \bgamma}_2^2, \qquad j = 1,\ldots,\nGCvars,
\end{equation}
for any compatible index set $S$, and let $\check{\bGamma}_{S}^\otimes = \left(\check{\bgamma}_{1, S}^\otimes, \ldots, \check{\bgamma}_{\nXGC, S}^\otimes \right)$, such that $\M\left(\widetilde\bX_{S}^\otimes\right) \widetilde\bX_{-GC}^\otimes \bGamma^\otimes = \widetilde\bX_{-GC}^\otimes \left(\bGamma^\otimes - \check{\bGamma}_{S}^\otimes \right)$. Then, with probability $1- \Delta_T$,
\begin{align*}
\norm{\ba_{T,2}}_1 &= \norm{\left(\check{\bGamma}_{\hat{S}^\otimes}^\otimes - \bGamma^\otimes \right)^\prime \widetilde\bX_{-GC}^{\otimes\prime} \widetilde\bu_I /\sqrt{T}}_1 \overset{(1)}{\leq} \sum_{j=1}^{\nGCvars} \norm{\check{\bgamma}_{j,\hat{S}^\otimes}^\otimes - \bgamma_j^\otimes}_1 \norm{\widetilde\bX_{-GC}^{\otimes\prime} \widetilde\bu_I /\sqrt{T}}_\infty\\
&\overset{(ii)}{\leq} \gamma_T \sum_{j=1}^{\nGCvars} \norm{\check{\bgamma}_{j,\hat{S}^\otimes}^\otimes - \bgamma_j^\otimes}_1 \overset{(iii)}{\leq} \sqrt{\bar{s}_T} \gamma_T \sum_{j=1}^{\nGCvars} \norm{\widetilde\bX_{-GC}^\otimes \left(\check{\bgamma}_{j, \hat{S}^\otimes}^\otimes - \bgamma_j^\otimes \right)/T}_2 / \phi_{T, \min} \\
&\overset{(iv)}{\leq} \sqrt{\bar{s}_T} \gamma_T \sum_{j=1}^{\nGCvars} \norm{\widetilde\bX_{-GC}^\otimes \left(\hat{\bgamma}_{j}^\otimes - \bgamma_j^\otimes \right)/T}_2 / \phi_{T, \min}\\
&\overset{(v)}{\leq} \frac{\sqrt{\bar{s}_T} \gamma_T}{\phi_{T, \min}} \norm{\bG_T}_2 N_I \sum_{j=1}^{\nXGC} \norm{\bX_{-GC} \left(\hat{\bgamma}_j - \bgamma_j \right)/T}_2
\overset{(vi)}{\leq} \nGCed^{3/2} \nXGC \frac{\sqrt{\bar{s}_T} \gamma_T }{\phi_{T, \min}} \delta_T T^{-1/4} \leq \delta_T^2.
\end{align*}
Here inequality $(i)$ uses that 
\begin{equation} \label{eq:dualnorm}
\norm{\bA \bx }_1 = \sum_{j=1}^m \abs{\ba_{j\cdot} x_i} \leq \norm{\bx}_{\infty} \sum_{j=1}^m \norm{\ba_{j\cdot}}_1
\end{equation}
from the dual norm inequality, where $\bA$ is a generic $m\times n$ matrix $\bA$ with $j$-th row denoted as $\ba_{j\cdot}$, and a $n \times 1$ vector $\bx$. Letting $\norm{\bA}_{\max} = \max_{i,j} \abs{a_{ij}}$, Step $(ii)$ follows from the fact that $\norm{\widetilde\bX_{-GC}^{\otimes\prime} \widetilde\bu_I / \sqrt{T}}_\infty = \norm{(\bG_T^\prime\bG_T \otimes \bX_{-GC}^\prime) \bu_I / \sqrt{T}}_\infty \leq \norm{\bG_T^\prime \bG_T}_{\max} \norm{\bX_{-GC}^{\otimes\prime}  \bu_I / \sqrt{T}}_\infty \leq \gamma_T$ by Assumption \ref{as:hlev}(\ref{as:ep}), while $(iii)$ follows from bounding the $l_1$-norm by the $l_2$-norm and applying Lemma \ref{lem:minev}. $(iv)$ follows from the definition of $\check{\bgamma}_{\hat{S}}$ as minimizer of the sum of squares and $(v)$ from the properties of the Kronecker product. Finally $(vi)$ follows from Assumption \ref{as:hlev}(\ref{as:cons}).

For $\ba_{T,3}$, define $\check{\bgamma}_{0, S} = \argmin_{\bgamma:\gamma_m = 0, m \notin S} \norm{\widetilde\bX_{-GC}^\otimes \bgamma_0 - \widetilde\bX_{-GC}^\otimes \bgamma}_2^2$ analogously to \eqref{eq:tildebeta}. Then we have with probability $1 - \Delta_T$
\begin{equation*}
\begin{split}
\norm{\ba_{T,3}}_1 &\overset{(i)}{\leq} \norm{\widetilde\bE^{\otimes \prime} \M(\widetilde\bX_{\hat{S}^\otimes}^\otimes) \widetilde\bX_{-GC}^\otimes \bgamma_0 /\sqrt{T}}_1 + \norm{\widetilde\bE^{\otimes\prime} \M(\widetilde\bX_{\hat{S}^\otimes}^\otimes) \widetilde\bX_{-GC}^\otimes \bGamma^\otimes \bbeta_{GC} /\sqrt{T}}_1\\
&\overset{(ii)}{\leq} \norm{\widetilde\bE^{\otimes\prime} \widetilde\bX_{-GC}^\otimes \left( \check{\bgamma}_{0,\hat{S}} - \bgamma_0 \right) /\sqrt{T}}_1 + \norm{\widetilde\bE^{\otimes\prime} \widetilde\bX_{-GC}^\otimes \left(\check{\bGamma}_{\hat{S}^\otimes}^\otimes - \bGamma^\otimes \right) \bbeta_{GC} /\sqrt{T}}_1\\
&\overset{(iii)}{\leq} \norm{\check{\bgamma}_{0,\hat{S}} - \bgamma_0}_1 \sum_{j=1}^{\nGCvars} \norm{\widetilde\bX_{-GC}^{\otimes\prime} \widetilde\be_j^\otimes / \sqrt{T}}_\infty + \norm{\bbeta_{GC}}_\infty \sum_{j=1}^{\nGCvars} \norm{\widetilde\bX_{-GC}^{\otimes\prime} \widetilde\be_j^\otimes / \sqrt{T}}_{\infty} \sum_{j=1}^{\nGCvars} \norm{\check{\bgamma}_{j, \hat{S}^\otimes}^\otimes - \bgamma_j}_1\\
&\overset{(iv)}{\leq} \sum_{j=1}^{\nGCvars} \norm{\widetilde\bX_{-GC}^{\otimes\prime} \widetilde\be_j^\otimes / \sqrt{T}}_{\infty} \left[\norm{\check{\bgamma}_{0,\hat{S}} - \bgamma_0}_1 + C \nGCed^{3/2} \nXGC \frac{\sqrt{\bar{s}_T}}{\phi_{T, \min}} \delta_T T^{-1/4} \right]\\
&\overset{(v)}{\leq} \sum_{j=1}^{\nGCvars} \norm{\widetilde\bX_{-GC}^{\otimes\prime} \widetilde\be_j^\otimes / \sqrt{T}}_{\infty} \left[\frac{\sqrt{\bar{s}_T} \norm{\bG_T}_2}{\phi_{T,\min}} \norm{\bX_{-GC}^\otimes \left( \check{\bgamma}_0 - \bgamma_0 \right)/\sqrt{T}}_2 + C \nGCed^{3/2} \nXGC\frac{\sqrt{\bar{s}_T}}{\phi_{T, \min}} \delta_T T^{-1/4} \right]\\
&\overset{(vi)}{\leq} N_{GC} \gamma_T \left[\frac{\sqrt{\bar{s}_T} \sqrt{N_i}}{\phi_{T,\min}} \delta_T T^{-1/4} + C \nGCed^{3/2} \nXGC\frac{\sqrt{\bar{s}_T}}{\phi_{T, \min}} \delta_T T^{-1/4} \right] \leq C \nXGC \nGCed^{3/2} \frac{\sqrt{\bar{s}_T} \gamma_T}{\phi_{T, \min}} \delta_T T^{-1/4} \leq \delta_T^{2}.
\end{split}
\end{equation*}
Inequality $(i)$ follows from the fact that $\bbeta_{-GC} = \bgamma_0 - \bGamma^\otimes \bbeta_{GC}$, while $(ii)$ follows from the definition of $\tilde{\bgamma}_{0, S}$ and \eqref{eq:tildebeta}. For the first term in (iii) we use \eqref{eq:dualnorm} whereas for the second term we apply it twice to get 
\begin{equation} \label{eq:dualnormtwice}
\norm{\bB \bA \bx}_1 \leq \norm{\bx}_\infty \sum_{i=1}^p \norm{\bb_{i\cdot} \bA}_1 \leq \norm{\bx}_\infty \sum_{i=1}^p \norm{\bb_{i\cdot}}_\infty \sum_{j=1}^m \norm{\ba_{\cdot j}}_1
\end{equation}
for any $p \times n$ matrix $\bB$. Step $(iv)$ follows from Assumption \ref{as:hlev}(\ref{as:bound}) and the results for $\ba_{T,2}$, while Step $(v)$ applies the same arguments as used therein to $\norm{\check{\bgamma}_{0,\hat{S}} - \bgamma_0}_1$. Finally, Step $(vi)$ follows analogoulsy to Step $(ii)$ for $\ba_{T,2}$ by noting that $\norm{\widetilde\bX_{-GC}^{\otimes\prime} \widetilde\be_j^\otimes / \sqrt{T}}_\infty \leq \norm{\bG_T^\prime \bG_T}_{\max} \norm{\bX_{-GC}^{\prime} \be_j / \sqrt{T}}_\infty \leq \gamma_T$, plus using the bound from Assumption \ref{as:hlev}\ref{as:ep}).

Finally, we consider $\ba_{T,4}$. We get
\begin{align*}
\norm{\ba_{T,4}}_1 &\overset{(i)}{\leq} \norm{\widetilde\bX_{\hat{S}^\otimes}^{\otimes\prime} \widetilde\bu_I /\sqrt{T}}_\infty \sum_{j=1}^{\nGCvars} \norm{\widetilde\be_j^{\otimes\prime} \widetilde\bX_{\hat{S}^\otimes}^{\otimes}  \left(\widetilde \bX_{\hat{S}^\otimes}^{\otimes\prime} \widetilde\bX_{\hat{S}^\otimes}^\otimes \right)^{-1}}_1\\
&\overset{(ii)}{\leq} \gamma_T \sqrt{\bar{s}_T} \sum_{j=1}^{\nGCvars} \norm{\widetilde\be_j^{\otimes\prime} \widetilde\bX_{\hat{S}^\otimes}^{\otimes}  \left(\widetilde \bX_{\hat{S}^\otimes}^{\otimes\prime} \widetilde\bX_{\hat{S}^\otimes}^\otimes \right)^{-1}}_2\\
&\overset{(iii)}{\leq} \gamma_T \bar{s}_T \norm{\left(\widetilde\bX_{\hat{S}^\otimes}^{\otimes\prime} \widetilde\bX_{\hat{S}^\otimes}^\otimes / T \right)^{-1}}_2 \sum_{j=1}^{\nGCvars} \norm{\widetilde\bX_{\hat{S}^\otimes}^{\otimes\prime} \widetilde\be_j^{\otimes} / \sqrt{T} }_\infty / \sqrt{T} \\
&\overset{(iv)}{\leq} \nGCvars \gamma_T^2 \bar{s}_T T^{-1/2} / \phi_{T,\min} \leq \delta_T^2,
\end{align*}
where step $(i)$ follows from \eqref{eq:dualnorm}. For $(ii)$ we bound the $l_1$-norm with the $l_2$-norm, using that $\widetilde\bX_{\hat{S}^\otimes}^\otimes$ contains a subset of the variables in $\widetilde\bX_{-GC}^\otimes$ and therefore $\norm{\widetilde\bX_{\hat{S}^\otimes}^{\otimes\prime} \widetilde\bu_{\GCed}^\otimes}_\infty \leq \norm{\widetilde\bX_{-GC}^{\otimes\prime} \widetilde\bu_{\GCed}^\otimes}_\infty$ and apply Assumption \ref{as:hlev}(\ref{as:ep}). Step $(iii)$ follows from the Cauchy-Schwarz inequality, bounding the $l_2$-norm by the $l_\infty$-norm and reasoning as for Step $(ii)$ that $\norm{\widetilde\bX_{\hat{S}^\otimes}^{\otimes\prime} \widetilde\be_j^\otimes}_\infty \leq \norm{\widetilde\bX_{-GC}^{\otimes\prime} \widetilde\be_j^\otimes}_\infty$. Finally $(iv)$ follows from Assumption \ref{as:hlev}(\ref{as:ep}) and Lemma \ref{lem:minev}.

We next consider $\bB_T$.  Using that $\widetilde\bX_{GC}^\otimes = \widetilde\bX_{-GC}^\otimes \bGamma^\otimes + \widetilde\bE^\otimes$, we  write
\begin{equation*}
\begin{split}
\bB_T &= \widetilde\bE^{\otimes\prime} \widetilde\bE^\otimes / T + \underbrace{\bGamma^{\otimes\prime} \widetilde\bX_{-GC}^{\otimes\prime} \M(\bX_{\hat{S}^\otimes}^\otimes) \widetilde\bX_{-GC}^\otimes \bGamma^\otimes / T}_{\bB_{T,1}} + \underbrace{\widetilde\bGamma^{\otimes\prime} \widetilde\bX_{-GC}^{\otimes\prime} \M(\widetilde\bX_{\hat{S}^\otimes}^\otimes) \widetilde\bE^\otimes / T}_{\bB_{T,2}}\\
&\quad + \underbrace{\widetilde\bE^{\otimes\prime} \M(\bX_{\hat{S}^\otimes}^\otimes) \widetilde\bX_{-GC}^\otimes \bGamma^\otimes / T}_{\bB_{T,2}^\prime} - \underbrace{\widetilde\bE^{\otimes\prime} \P(\widetilde\bX_{\hat{S}^\otimes}^\otimes) \widetilde\bE^\otimes / T}_{\bB_{T,3}}.\\
\end{split}
\end{equation*}
These terms can be handled as the terms for $\ba_T$. In particular, with probability $1-\Delta_T$, $\norm{\bB_{T,1}}_2 \leq \norm{\bA_{T,1,1}}_2^2 \leq \delta_T^2 T^{-1/2}$, $\norm{\bB_{T,2}}_2 \leq \delta_T^2 T^{-1/2}$ using the same steps as for $\ba_{T,2}$, and $\norm{\bB_{T,3}}_2 \leq \delta_T T^{-1/2}$ analogously to $\ba_{T,4}$.

This shows that $\ba_T = \widetilde\bE^{\otimes\prime} \widetilde\bu_{\GCed} / \sqrt{T}$ and $\bB_T = \widetilde\bE^{\otimes\prime} \widetilde\bE^{\otimes} / T + o_p(1)$. It then follows directly from Assumption \ref{as:hlev}(\ref{as:clt}) that
\begin{align*}
\sqrt{T} \left(\hat{\bbeta}_{GC}^{\textsc{pds}} - \bbeta_{GC} \right) &= \left(\bG_T^\prime \bG_T \otimes \bE^\prime \bE \right)^{-1} \bE^\otimes (\bG_T^\prime \bG_T \otimes \bI_T) \bu_I + o_p(1) \\
&\xrightarrow{d} \mathcal{N} \left(\bm 0, (\bG^\prime \bG \otimes \bSigma_{GC|-GC})^{-1} \bOmega_{\bG} (\bG^\prime \bG \otimes \bSigma_{GC|-GC})^{-1} \right). \qedhere
\end{align*}

\end{proof}

\begin{proof}[\bf{Proof of Theorem \ref{th:asdis2}}]
By partitioned regression algebra, we find that
\begin{equation*}
\begin{split}
LM = \hat{\bxi}^{*\prime} \hat{\bxi}^* - \hat{\bnu}^{*\prime} \hat{\bnu}^* &= \underbrace{\by_I^{*\prime} \M(\bX_{\hat{S}^\otimes}^{*\otimes}) \bX_{GC}^{*\otimes}}_{\ba_T^{*\prime}} \underbrace{\left[\bX_{GC}^{*\otimes\prime} \M(\bX_{\hat{S}^\otimes}^{*\otimes}) \bX_{GC}^{*\otimes} \right]^{-1}}_{\bB_T^{*-1}} \underbrace{\bX_{GC}^{*\otimes\prime} \M(\bX_{\hat{S}^\otimes}^{*\otimes}) \by_I^*}_{\ba_T^*}.
\end{split}
\end{equation*}
Note that $\ba_{T}^*$ and $\bB_T^*$ are special cases of their counterparts in the proof of Theorem \ref{th:asdis} with $\bG_T = \hat{\bSigma}_{T,I}^{-1/2}$. We now show that this choice of $\bG_T$ satisfies the conditions of Theorem \ref{th:asdis}. We do this by proving that $\bG_T$ converges to $\bG = \bSigma_{u,I}^{-1/2}$, and this satisfies the conditions in the theorem.

Consider one particular element $(i,j)$ of $\hat{\bSigma}_{u,I}$, say $\hat{\sigma}_{u,ij}$. Let $\hat{S}_{0,\GCed_i}$ denote the variables selected in $\hat{S}_0$ corresponding to the equation for variable $\by_{\GCed_i}$, where $\GCed = \{\GCed_1, \ldots, \GCed_{\nGCed} \}$, and let $\hat{S}_i = \left(\bigcup_{j=1}^{\nXGC} \hat{S}_j \right) \cup \hat{S}_{0,i}$ denote all variables selected that are relevant for $\by_{\GCed_i}$. We can then write
\begin{equation*}
\begin{split}
\hat{\bsigma}_{u,ij} &= \hat{\bxi}_{\GCed_i}^\prime \hat{\bxi}_{\GCed_j}/ T = \by_{\GCed_i}^\prime \M(\bX_{\hat{S}_i}) \M(\bX_{\hat{S}_j}) \by_{\GCed_j} \\
&= \bu_{\GCed_i}^\prime \bu_{\GCed_j} / T + \underbrace{\bbeta_{-GC,i}^\prime \bX_{-GC}^{\prime} \M(\bX_{\hat{S}_i}) \M(\bX_{\hat{S}_j}) \bX_{-GC} \bbeta_{-GC,j}/T}_{d_{T,ij,1}}\\
&\quad - \underbrace{\bu_{\GCed_i}^\prime \M(\bX_{\hat{S}_i}) \M(\bX_{\hat{S}_j}) \bX_{-GC} \bbeta_{-GC,j}/T}_{d_{T,ij,2}} - \underbrace{\bu_{\GCed_j}^\prime \M(\bX_{\hat{S}_j}) \M(\bX_{\hat{S}_i}) \bX_{-GC} \bbeta_{-GC,i}/T}_{d_{T,ji,2}}\\
&\quad + \underbrace{\bu_{\GCed_i}^\prime \left[\bI_T - \M(\bX_{\hat{S}_i}) \M(\bX_{\hat{S}_j}) \right] \bu_{\GCed_j} / T}_{d_{T,ij,3}},
\end{split}
\end{equation*}
where (under $H_0$) we write $\by_{\GCed_i} = \bX_{-GC} \bbeta_{-GC,i} + \bu_{\GCed_i}$.

We can use the same reasoning as used in the proof of Theorem \ref{th:asdis} to prove that the terms $d_{T,ij,k}$, $k=1,2,3$ are negligible. Let $\gamma_{0,i}$ denote the sub-vector of $\bgamma_0$ corresponding to unit $i$, and note that under the null hypothesis $\bgamma_{0,i} = \bbeta_{-GC,i}$. Define $\bar{\bgamma}_{i,S} = \argmin_{\bgamma: \gamma_m = 0, m \notin S} \norm{\bX_{-GC} \gamma_{0,i} - \bX_{-GC} \bgamma}_2^2$. Then, as $\hat{S}_{0,i} \subseteq \hat{S}_i$, we get that
\begin{equation*}
\begin{split}
\abs{d_{T,ij,1}} &\leq \norm{\M(\bX_{\hat{S}_i}) \bX_{-GC} \bbeta_{-GC,i}}_2 \norm{\M(\bX_{\hat{S}_j}) \bX_{-GC} \bbeta_{-GC,j}}_2 / T\\
&= \norm{\bX_{-GC} (\hat{\bgamma}_{0,\hat{S}_i} - \bgamma_{0,i})/\sqrt{T}}_2 \norm{\bX_{-GC} (\hat{\bgamma}_{0,\hat{S}_i} - \bgamma_{0,j}) /\sqrt{T}}_2 \\
&\leq \norm{\bX_{-GC} (\hat{\bgamma}_{0,i} - \bgamma_{0,i})/\sqrt{T}}_2 \norm{\bX_{-GC} (\hat{\bgamma}_{0,j} - \bgamma_{0,j}) /\sqrt{T}}_2 \leq \delta_T^2 T^{-1/2}.
\end{split}
\end{equation*}
Similarly,
\begin{equation*}
\begin{split}
\abs{d_{T,ij,2}} &= \abs{\bu_{\GCed_i}^\prime \M(\bX_{\hat{S}_i \cup \hat{S}_j}) \bX_{-GC} \bbeta_{-GC,j} / T} = \abs{\bu_{\GCed_i}^\prime \bX_{-GC} (\hat{\bgamma}_{0,\hat{S}_i\cup \hat{S}_j} - \bgamma_{0,j}) / T} \\
&\leq \norm{\bu_{\GCed_i} \bX_{-GC} /\sqrt{T}}_\infty \norm{\hat{\bgamma}_{0,\hat{S}_i\cup \hat{S}_j} - \bgamma_{0,j}}_1 /\sqrt{T}\\
&\leq \frac{\sqrt{\bar{s}_T} \gamma_T}{ \phi_{T,\min}} \norm{\bX_{-GC} (\hat{\bgamma}_{0,j} - \bgamma_{0,j}) /\sqrt{T}}_2 / \sqrt{T} \leq \frac{\sqrt{\bar{s}_T} \gamma_T}{ \phi_{T,\min}} \delta_T T^{-3/4}.
\end{split}
\end{equation*}
Finally,
\begin{equation*}
\begin{split}
\abs{d_{T,ij,3}} &= \abs{\bu_{\GCed_i}^\prime \P(\bX_{\hat{S}_i\cup\hat{S}_j})  \bu_{\GCed_j} / T} \leq \sqrt{\bar{s}_T} \norm{\bX_{\hat{S}_i\cup\hat{S}_j}^\prime \bu_{\GCed_j}}_\infty \norm{ \bu_{\GCed_i}^\prime \bX_{-GC} \left(\bX_{\hat{S}_i\cup\hat{S}_j}^\prime \bX_{\hat{S}_i\cup\hat{S}_j}\right)^{-1}}_2/T\\
&\leq \bar{s}_T \norm{\bX_{-GC}^\prime \bu_{\GCed_i} /\sqrt{T}}_\infty \norm{\bX_{-GC}^\prime \bu_{\GCed_j} /\sqrt{T}}_\infty \norm{ \left(\bX_{\hat{S}_i\cup\hat{S}_j}^\prime \bX_{\hat{S}_i\cup\hat{S}_j}/T\right)^{-1}}_2 / T\\
&\leq \bar{s}_T \gamma_T^2 T^{-1} / \phi_{T,\min},
\end{split}
\end{equation*}
where all bounds hold with probability at least $1 - \Delta_T$. Similarly, by Assumption \ref{as:hlev}(\ref{as:clt}) we know that there exist a sequence $\delta_T \rightarrow 0$, such that with probability $1 - \Delta_T$, we have that $\abs{\bu_{\GCed_i}^\prime \bu_{\GCed_j} - \sigma_{u,ij}} \leq \delta_T$. As $\bSigma{u,\GCed}$ only contains a finite number ($\nGCed^2$) elements, we may then conclude that with probability at least $1-\Delta_T$, it holds that $\norm{\hat{\bSigma}_{u,\GCed} - \bSigma}_{2} \leq \delta_T$.

We have that $0 < c_L \leq \lambda_{\min} (\bSigma_{u,\GCed}) \leq \lambda_{\max} (\bSigma_{u,\GCed}) \leq c_u < \infty$, where the lower bound follows from Assumption \ref{as:hlev}(\ref{as:eig}) and the upper bound from the fact that $\bSigma_{u,I}$ has a finite number of elements. As
\begin{equation*}
\begin{split}
&\bx^\prime \hat{\bSigma}_{u,\GCed} \bx \leq \bx^\prime \bSigma_{u,\GCed} \bx + \abs{\bx^\prime \left(\hat{\bSigma}_{u,\GCed} \bx - \bx^\prime \bSigma_{u,\GCed} \right) \bx} \leq \norm{\bx}_2^2 \lambda_{\max} (\bSigma_{u,\GCed}) + \norm{\bx}_2 \norm{\hat{\bSigma}_{u,\GCed} - \bSigma_{u,\GCed}}_2^2,\\
&\bx^\prime \hat{\bSigma}_{u,\GCed} \bx \geq \bx^\prime \bSigma_{u,\GCed} \bx - \abs{\bx^\prime \left(\hat{\bSigma}_{u,\GCed} \bx - \bx^\prime \bSigma_{u,\GCed} \right) \bx} \leq \norm{\bx}_2^2 \lambda_{\min} (\bSigma_{u,\GCed}) + \norm{\bx}_2 \norm{\hat{\bSigma}_{u,\GCed} - \bSigma_{u,\GCed}}_2^2,
\end{split}
\end{equation*}
the established the consistency of $\hat{\bSigma}_{u,\GCed}$ then directly yields that $C_1 - \delta_T \leq \lambda_{\min} (\hat{\bSigma}_{u,\GCed}) \leq \lambda_{\min} (\hat{\bSigma}_{u,\GCed}) \leq C_2 + \delta_T$ with probability $1 - \Delta_T$. It then also follows that with probability $1 - \Delta_T$ we can find a $0 < C_1 \leq C_2 < \infty$ such that $c_1 \leq 1/\lambda_{\max} (\hat{\bSigma}_{u,I}) = \lambda_{\min} (\hat{\bSigma}_{u,I}^{-1}) \leq  \lambda_{\max} (\hat{\bSigma}_{u,I}^{-1}) = 1/\lambda_{\min} (\hat{\bSigma}_{u,I}) \leq c_2$, such that the conditions of Theorem \ref{th:asdis} are satisfied for $\bG_T = \hat{\bSigma}_{u,\GCed}^{-1/2}$.

With this choice of $\bG_T$ and $\bOmega = \bSigma_{u,I} \otimes \bSigma_{GC|-GC}$, we have that $\bOmega_{\bG} = \bI_{\nXGC} \otimes \bSigma_{GC|-GC}$. Letting $\bZ_{\nGCvars} \sim \mathcal{N}(\bzero, \bI_{\nGCvars})$, it then follows that 
\begin{align*}
LM &= \ba_T^{*\prime} \bB_T^{*-1} \ba_T^* \xrightarrow{d} \bZ_{\nGCvars}^\prime \left(\bI_{\nXGC} \otimes \bSigma_{GC|-GC} \right)^{1/2\prime} \\
&\quad \times \left( \bI_{\nXGC} \otimes \bSigma_{u,I} \otimes \bSigma_{GC|-GC} \right)^{-1} \left(\bI_{\nXGC} \otimes \bSigma_{GC|-GC} \right)^{1/2} \bZ_{\nGCvars}\\
&= \bZ_{\nGCvars}^\prime \bZ_{\nGCvars} = \chi_{\nGCvars}^2.\qedhere
\end{align*}

\end{proof}

\clearpage

\section{Additional Simulation Results}\label{app_additionalsim}

\begin{sidewaystable}
\begin{threeparttable}
\caption{Simulation results for the PDS-LM Granger causality test ($\rho=0.7$)}\label{tab:correlated}
\tiny
\begin{tabular}{ l S[table-format=3.0] cccc c c c c c c cc cccccccccccc}
    \toprule
   
   DGP&{Size/Power}&$\rho$&$T$ &&&50&&&&&100&&&&&200&&&&&500&&&  \\
   \cmidrule(l){1-24} 
&&&K& \multicolumn{5}{c}{AIC \;\;\; BIC \;\;\;EBIC \;\;\;\;$\lambda^{th}$ \; $\lambda^{TSCV}$} & \multicolumn{5}{c}{AIC \;\;\; BIC \;\;\;EBIC \;\;\;\;$\lambda^{th}$ \; $\lambda^{TSCV}$}& \multicolumn{5}{c}{AIC \;\;\; BIC \;\;\;EBIC \;\;\;\;$\lambda^{th}$ \; $\lambda^{TSCV}$}& \multicolumn{5}{c}{AIC \;\;\; BIC \;\;\;EBIC \;\;\;\;$\lambda^{th}$ \; $\lambda^{TSCV}$}\\ 
    \cmidrule(l){1-24}
&&& 10 &5.8&5.1&5.4&6.1&5.7&5.8&5.7&5.2&5.2&5.8&5.1&4.6&4.6&5.0&4.5&4.3&4.0&4.2&4.0&4.2  \\
    1&{Size}&$0.7$&20&7.3&6.4&6.1&6.0&7.2&6.0&6.1&6.2&6.3&6.5&5.1&5.5&5.5&5.8&5.9&5.0&4.9&4.9&4.4&5.0\\
    &&&50 &7.6&5.5&6.0&7.5&6.8&6.4&7.3&5.8&7.3&6.7&6.5&4.9&4.2&6.2&5.5&5.7&4.8&4.8&5.2&5.3\\
    &&&100&6.4&8.1&6.6&NA&8.0&7.7&6.4&6.2&7.1&6.8&6.9&5.8&4.4&5.3&4.6&5.6&3.6&3.6&4.2&3.8\\
\cmidrule(l){1-24}   

    &&&10 &19.0&18.2&18.8&20.3&19.5&33.7&34.2&34.3&35.2&33.4&56.6&57.1&56.8&57.8&55.6&93.7&94.2&94.2&94.3&93.3  \\
    1&{Power}&$0.7$&20 &15.3&18.3&19.1&18.5&17.9&29.2&30.7&31.6&30.7&29.8&54.4&56.8&56.6&56.3&54.8&92.8&94.1&94.1&93.8&93.4 \\
    &&&50 &10.3&14.1&19.3&15.2&14.6&24.0&30.4&32.7&31.1&30.3&50.1&57.1&58.8&56.7&51.7&90.7&92.0&92.5&92.1&91.4\\
    &&&100&9.0&14.0&19.2&NA&15.4&13.6&29.2&33.6&25.4&24.1&34.1&53.6&55.7&49.4&48.9&88.1&94.0&94.5&92.4&91.1\\
\cmidrule(l){1-24}
&&&10 &6.4&5.8&5.5&5.8&6.8&5.0&5.1&5.3&5.0&5.5&5.2&4.9&5.0&4.5&5.1&4.7&5.0&4.9&4.5&5.1 \\
    2&{Size}&0.7&20 &7.5&6.5&5.6&5.9&5.5&4.8&5.5&5.3&5.8&5.2&5.1&5.0&4.8&5.0&5.8&4.8&4.5&4.9&4.4&6.3\\
    &&&50 &6.8&5.9&6.8&6.0&6.2&7.4&5.3&6.1&5.2&6.3&5.8&5.9&5.7&5.2&6.2&5.6&5.5&5.6&5.9&6.5\\
    &&&100&8.1&6.5&6.8&6.9&6.5&8.1&5.8&6.2&7.0&5.9&6.7&6.1&6.3&5.5&5.9&5.3&4.4&4.0&4.5&5.0\\
\cmidrule(l){1-24}   

    &&&10 &15.4&15.1&16.0&14.5&14.3&26.0&28.1&28.8&26.6&26.5&47.4&49.7&51.6&49.2&51.0&83.2&84.6&85.3&84.6&86.2\\
    2&{Power}&0.7&20 &12.6&15.8&19.3&15.7&15.5&26.5&27.9&30.0&26.7&26.9&48.4&51.6&53.2&51.1&51.8&83.1&85.3&86.7&85.4&85.9\\
    &&&50 &9.6&15.4&18.4&13.0&12.2&19.1&28.5&31.5&26.1&26.3&40.3&50.5&52.2&47.0&48.9&83.9&87.4&88.8&87.2&87.2\\
    &&&100&10.3&16.9&21.3&10.7&11.5&12.6&29.3&32.7&21.0&21.4&32.5&53.0&56.0&45.4&49.8&78.7&88.6&89.1&85.4&85.4\\
\cmidrule(l){1-24}
&&&10 &4.8&4.8&4.7&4.3&5.5&5.3&4.6&4.8&4.6&5.1&5.5&5.5&5.6&5.6&5.4&4.8&4.8&5.0&4.8&5.1\\
    3&{Size}&0.7&20 &5.6&5.4&5.2&5.1&5.6&5.4&5.8&5.6&5.2&5.2&4.4&4.0&4.0&4.1&4.5&4.5&4.0&4.2&4.2&4.4\\
    &&&50 &8.3&6.0&5.3&5.5&7.0&6.1&5.9&5.2&5.9&6.4&5.6&4.8&4.6&5.2&5.1&6.2&6.8&6.7&6.9&6.7\\
    &&&100&7.6&6.3&6.3&5.6&6.8&7.2&5.2&5.2&6.1&6.4&5.6&5.8&5.8&6.2&4.7&3.6&4.1&3.7&4.4&4.6\\
\cmidrule(l){1-24}   

    &&&10 &9.9&9.8&10.3&10.3&10.7&16.2&16.6&16.4&16.8&16.9&31.5&31.3&31.4&31.6&31.5&68.4&68.9&68.9&68.9&68.8 \\
    3&{Power}&0.7&20 &8.1&8.8&8.2&9.1&9.5&15.7&16.4&16.2&16.4&16.8&29.2&29.0&29.0&28.7&28.3&66.5&68.5&69.4&68.8&67.9\\
    &&&50 &9.7&9.4&10.1&9.7&10.2&15.7&16.1&17.4&16.2&16.6&31.3&31.2&31.7&31.8&29.9&65.1&66.6&67.4&66.9&65.0\\
    &&&100&8.4&9.4&9.3&9.7&9.9&12.0&15.3&17.7&14.6&15.6&26.4&30.9&31.9&30.2&29.3&66.2&67.8&68.3&68.3&66.5\\
\cmidrule(l){1-24}
\end{tabular}
\begin{tablenotes} 
\scriptsize 
\item Notes: Size and Power for the different DGPs described in Section 4.1 are reported for 1000 replications. $T=(50,100,200,500)$ is the time series length, $K=(10,20,50,100)$ the number of variables in the system, the lag-length is fixed to $p=1$. $\rho$ indicates the correlation employed to simulate the time series with the Toeplitz covariance matrix. The different choices of the tuning parameter $\lambda$ are reported as: AIC, BIC, EBIC for information criteria, $\lambda^{th}$ for the theoretical plug-in and TSCV for time series cross-validation as explained in Section \ref{sec:sim}.   
\end{tablenotes}
\end{threeparttable}
\end{sidewaystable}

\begin{sidewaystable}
\begin{threeparttable}
\caption{Simulation results for the PDS-WALD Granger causality test}\label{tab:3}
\tiny
\begin{tabular}{ l S[table-format=3.0] cccc c c c c c c cc cccccccccccc}
    \toprule
   DGP&{Size/Power}&$\rho$&$T$ &&&50&&&&&100&&&&&200&&&&&500&&&  \\
   \cmidrule(l){1-24} 
&&&K& \multicolumn{5}{c}{AIC \;\;\; BIC \;\;\;EBIC \;\;\;\;$\lambda^{th}$ \; $\lambda^{TSCV}$} & \multicolumn{5}{c}{AIC \;\;\; BIC \;\;\;EBIC \;\;\;\;$\lambda^{th}$ \; $\lambda^{TSCV}$}& \multicolumn{5}{c}{AIC \;\;\; BIC \;\;\;EBIC \;\;\;\;$\lambda^{th}$ \; $\lambda^{TSCV}$}& \multicolumn{5}{c}{AIC \;\;\; BIC \;\;\;EBIC \;\;\;\;$\lambda^{th}$ \; $\lambda^{TSCV}$}\\ 
    \cmidrule(l){4-24}
  &&& 10 &7.1&6.6&6.0&7.0&7.6&7.0&6.4&6.1&6.8&7.1&5.5&4.3&4.6&4.7&6.9&4.3&4.1&4.3&4.3&4.8  \\
    1&{Size}&0&20&8.0&6.4&6.8&6.5&6.4&6.2&4.9&4.9&5.7&6.9&4.2&4.8&5.5&4.4&5.0&4.3&4.1&4.3&4.1&4.7 \\
   &&& 50 &7.2&6.2&6.2&7.2&6.2&7.6&6.4&6.6&6.4&7.1&7.1&5.9&5.9&6.3&7.0&6.8&6.5&6.7&6.8&7.1\\
   &&& 100&7.1&7.4&7.3&NA&7.3&8.0&5.2&5.1&6.1&5.9&6.6&4.7&4.9&5.0&5.3&5.8&3.9&4.0&5.2&4.5\\
\cmidrule(l){1-24}   
    &&&10 &31.3&31.8&34.0&31.3&33.8&58.6&59.5&61.2&58.9&58.3&89.1&89.3&89.6&89.3&89.6&99.9&99.9&99.9&99.9&99.9 \\
    1&{Power}&0&20&23.9&27.6&30.9&27.3&25.3&53.4&55.9&58.4&54.6&55.0&85.6&88.2&89.1&86.8&86.1&99.9&100&100&99.9&99.8 \\
    &&&50 &15.3&26.0&34.9&19.0&20.4&39.9&54.5&60.0&46.7&46.3&78.8&86.1&87.3&81.4&80.3&99.9&99.9&99.9&99.9&99.9\\
    &&&100&13.0&21.9&33.1&NA&20.0&21.6&52.6&57.3&31.3&36.7&61.7&85.9&87.3&74.2&74.2&99.5&100&100&99.8&99.7\\
\cmidrule(l){1-24}
&&&10 &6.5&6.1&6.5&6.4&6.6&5.1&5.1&5.8&5.1&6.2&5.1&4.8&5.2&4.8&6.3&3.9&3.9&3.9&3.7&4.2  \\
    2&{Size}&0&20&8.4&6.0&6.7&6.5&7.4&5.3&4.8&5.1&5.2&6.0&5.9&6.0&5.8&5.8&6.9&4.7&3.8&4.3&4.6&5.1 \\
    &&&50 &8.3&6.2&5.1&6.7&7.6&8.4&6.4&6.7&7.4&7.4&7.6&6.4&6.2&6.8&7.7&6.4&6.6&6.4&6.2&6.4\\
    &&&100&6.5&7.6&5.9&NA&7.2&7.4&5.3&6.2&5.6&5.2&5.4&5.6&5.9&4.7&5.1&6.1&4.3&5.1&5.0&4.8\\
\cmidrule(l){1-24}   
    &&&10 &18.7&20.2&21.6&18.9&19.2&38.3&39.6&40.9&38.7&40.3&65.2&65.2&67.4&65.0&65.5&97.4&97.4&97.6&97.3&97.6   \\
    2&{Power}&0&20&16.7&20.9&25.4&19.7&19.0&35.8&40.2&44.9&38.1&38.1&64.8&67.5&69.8&66.2&65.5&97.2&97.5&97.4&97.5&97.4  \\
    &&&50 &10.1&15.8&22.7&13.8&14.4&25.4&36.8&43.8&33.6&32.1&57.2&66.8&72.1&62.1&61.6&95.0&96.2&96.8&96.1&95.4 \\
    &&&100&10.0&14.6&25.9&NA&11.4&16.6&35.1&46.5&27.3&26.2&45.1&65.3&74.9&57.3&57.9&94.7&97.3&97.7&96.3&96.4 \\
\cmidrule(l){1-24}
&&&10 &5.5&5.6&6.1&5.8&5.2&6.0&5.1&5.9&5.9&6.3&3.9&4.1&6.1&4.2&4.6&4.1&4.1&3.9&4.0&4.2 \\
    3&{Size}&0&20 &4.8&5.5&6.3&5.3&5.5&4.7&4.4&7.4&4.3&4.3&5.4&5.6&9.6&4.7&4.8&4.7&4.4&4.3&4.6&4.8\\
    &&&50 &7.9&6.7&7.8&7.4&7.0&6.7&7.2&9.6&6.1&5.6&7.0&6.9&12.7&6.1&6.8&5.0&5.3&6.6&5.3&5.4\\
    &&&100&7.4&7.0&8.4&NA&7.7&7.0&6.0&8.6&5.9&6.0&4.7&6.3&10.8&4.2&4.8&4.4&5.1&6.7&5.1&4.8\\
\cmidrule(l){1-24}   
    &&&10 &16.0&20.7&24.5&16.6&17.4&32.1&36.8&44.6&32.9&31.9&58.6&61.4&63.8&59.1&59.6&95.2&95.6&95.7&95.5&95.3 \\
    3&{Power}&0&20 &14.3&19.9&27.2&14.2&14.8&29.9&37.8&48.8&30.6&30.3&56.7&62.2&70.0&57.2&55.7&94.1&94.6&94.8&94.5&94.3 \\
    &&&50 &12.6&21.7&28.8&13.7&11.8&24.5&40.5&52.8&27.3&27.9&50.6&59.7&73.8&52.5&51.8&91.2&92.8&93.4&92.4&90.9\\
    &&&100&9.8&20.0&27.7&NA&14.8&15.4&43.0&55.5&20.3&23.0&41.7&62.2&75.2&45.3&45.0&90.0&94.2&95.0&91.5&90.2\\
\cmidrule(l){1-24}
&&& 10 &6.2&5.5&5.9&6.4&5.6&5.9&5.9&5.4&5.4&5.2&5.1&4.7&4.7&5.1&4.4&4.5&4.1&4.2&4.1&4.2 \\
    1&{Size}&$0.7$&20 &7.8&6.4&6.2&6.4&7.1&6.5&6.2&6.2&6.4&6.3&5.1&5.5&5.5&5.8&5.7&5.1&4.9&4.9&4.6&5.1 \\
    &&&50 &9.0&6.4&6.9&8.3&6.4&6.9&7.5&6.1&7.5&6.6&6.6&5.1&4.3&6.5&5.5&5.7&4.8&4.9&5.2&5.3\\
    &&&100&8.2&8.8&6.9&7.5&8.1&8.3&6.5&6.4&7.4&6.5&7.0&5.8&4.6&5.3&4.6&5.6&3.6&3.7&4.2&4.7\\
\cmidrule(l){1-24}   
    &&&10 &19.7&18.9&19.8&21.1&19.3&34.7&35.2&35.2&35.9&33.9&56.7&57.3&57.1&57.8&55.7&93.7&94.2&94.2&94.3&93.9 \\
    1&{Power}&$0.7$&20 &16.2&19.3&20.0&19.0&17.9&29.7&31.0&32.2&31.5&29.8&54.7&57.1&56.8&56.5&55.3&92.8&94.2&94.2&93.5&93.4\\
    &&&50 &11.3&14.9&20.1&16.4&15.1&24.9&30.8&33.7&31.9&30.8&50.6&57.3&59.0&56.9&52.2&90.7&92.0&92.5&92.1&91.4\\
    &&&100&9.6&14.9&20.3&12.4&15.4&15.4&29.6&34.0&25.8&24.6&34.5&53.8&56.0&49.7&49.3&88.2&94.1&94.6&92.4&91.2\\
\cmidrule(l){1-24}
\end{tabular}
\begin{tablenotes} 
\scriptsize 
\item Notes: Size and Power for the different DGPs described in Section \ref{sec:sim} are reported for 1000 replications. $T=(50,100,200,500)$ is the time series length, $K=(10,20,50,100)$ the number of variables in the system, the lag-length is fixed to $p=1$. $\rho$ indicates the correlation employed to simulate the time series with the Toeplitz covariance matrix. NAs are placed whenever the post-OLS estimation was not feasible due to $\hat{s}>T$. The different choices of the tuning parameter $\lambda$ are reported as: AIC, BIC, EBIC for information criteria, $\lambda^{th}$ for the theoretical plug-in and TSCV for time series cross-validation as explained in Section \ref{sec:sim}. 
    \end{tablenotes}
\end{threeparttable}
\end{sidewaystable}
\begin{sidewaystable}
\begin{threeparttable}
\caption{Simulation results for the PDS-LM Granger causality test (Overspecified lag-length)}\label{tab:4}
\tiny
\begin{tabular}{ l S[table-format=3.0] cccc c c c c c c cc cccccccccccc}
    \toprule
   DGP&{Size/Power}&$\rho$&$T$ &&&50&&&&&100&&&&&200&&&&&500&&&  \\
   \cmidrule(l){1-24} 
&&&K& \multicolumn{5}{c}{AIC \;\;\; BIC \;\;\;EBIC \;\;\;\;$\lambda^{th}$ \; $\lambda^{TSCV}$} & \multicolumn{5}{c}{AIC \;\;\; BIC \;\;\;EBIC \;\;\;\;$\lambda^{th}$ \; $\lambda^{TSCV}$}& \multicolumn{5}{c}{AIC \;\;\; BIC \;\;\;EBIC \;\;\;\;$\lambda^{th}$ \; $\lambda^{TSCV}$}& \multicolumn{5}{c}{AIC \;\;\; BIC \;\;\;EBIC \;\;\;\;$\lambda^{th}$ \; $\lambda^{TSCV}$}\\ 
    \cmidrule(l){4-24}
  1&{Size}&0&10&7.0&6.4&6.3&6.7&7.2&5.9&6.3&5.8&6.7&5.7&5.6&5.0&5.1&5.2&5.4&5.3&5.4&5.4&5.2&5.6 \\
   &&& 20 &8.4&7.5&6.7&8.3&7.3&6.3&6.0&5.6&4.9&6.0&5.8&5.6&5.6&5.7&5.7&5.0&3.7&3.7&4.7&4.3 \\
   &&& 50 &NA&5.4&4.6&NA&NA&8.7&5.0&5.4&6.7&5.9&7.4&5.2&4.9&6.6&6.1&5.7&4.7&5.0&4.7&4.9\\
    &&&100&NA&NA&5.2&NA&NA&NA&5.0&5.2&NA&NA&7.0&4.3&4.0&5.9&5.5&5.9&5.1&5.3&5.6&4.3\\
\cmidrule(l){1-24}   
    1&{Power}&0&10 &20.2&23.3&24.7&22.5&21.0&45.4&48.7&50.0&47.2&44.8&82.3&83.2&83.8&82.6&81.5&99.7&99.8&99.8&99.8&99.7 \\
    &&&20 &13.8&17.4&22.6&16.5&15.1&33.0&44.3&47.5&38.2&38.1&72.9&78.9&79.4&75.2&74.4&99.6&99.6&99.6&99.6&99.6 \\
    &&&50 &NA&NA&23.8&NA&NA&16.6&43.0&47.8&29.9&29.6&55.9&77.3&79.8&68.2&66.9&98.0&99.7&99.7&99.0&98.9\\
    &&&100&NA&NA&23.9&NA&NA&NA&37.2&45.5&NA&NA&20.0&76.3&79.1&53.7&53.1&96.0&99.8&99.8&99.2&94.2\\
\cmidrule(l){1-24}
2&{Size}&0&10&5.4&5.0&4.3&5.2&6.0&6.8&6.9&6.4&6.8&5.9&5.8&5.3&6.0&5.4&5.8&5.1&5.1&5.0&5.1&4.8 \\
    &&&20 &7.3&5.6&5.7&7.2&6.6&5.1&4.5&5.0&5.0&5.5&4.9&5.0&5.2&5.6&5.2&4.4&4.4&4.7&4.2&4.1\\
    &&&50 &NA&5.4&5.3&NA&NA&6.4&5.3&6.3&4.7&4.9&8.3&5.2&5.8&5.2&6.1&6.3&5.2&4.8&4.2&4.5\\
    &&&100&NA&NA&5.7&NA&NA&NA&4.6&5.4&NA&4.6&7.6&5.0&5.9&5.0&6.5&6.1&5.1&4.9&4.5&5.4 \\
\cmidrule(l){1-24}   
    2&{Power}&0&10 &13.1&14.7&14.8&14.5&13.6&29.0&29.9&31.3&29.1&28.7&52.7&52.5&55.0&53.5&53.9&94.1&94.3&94.4&94.5&93.8\\
    &&&20 &10.8&15.7&18.7&15.4&13.7&24.0&28.0&34.2&26.8&26.2&52.5&54.9&58.2&54.3&51.7&92.6&92.7&93.0&92.8&91.9 \\ 
    &&&50 &NA&NA&14.9&NA&NA&9.2&26.5&33.7&23.5&20.5&38.4&53.0&61.0&50.0&46.5&88.7&92.1&93.5&91.6&90.3 \\
    &&&100 &NA&NA&20.3&NA&NA&NA&26.9&37.9&NA&5.7&13.4&54.8&62.7&42.3&13.6&83.9&93.6&95.2&91.4&89.2\\
\cmidrule(l){1-24}
3&{Size}&0&10 &5.6&5.9&7.9&5.8&4.8&6.0&6.4&9.1&6.0&6.1&4.7&5.1&8.2&4.9&4.9&5.1&5.3&4.9&5.1&5.0 \\
    &&&20 &6.4&5.5&7.7&6.3&5.8&5.3&5.7&8.2&3.8&4.6&5.8&5.3&9.0&5.1&5.6&4.2&3.5&3.6&3.5&3.7 \\
    &&&50 &NA&6.8&7.7&NA&NA&7.9&6.8&11.8&5.9&5.0&6.2&6.2&13.1&5.2&5.6&4.8&4.5&5.5&4.6&5.1 \\
    &&&100&NA&NA&7.7&NA&NA&NA&6.9&11.4&NA&6.3&7.2&6.3&12.2&5.4&5.4&3.7&5.0&5.5&4.3&3.2 \\
\cmidrule(l){1-24}   
    3&{Power}&0&10 &10.0&12.2&16.5&9.8&10.1&22.1&24.8&31.7&22.0&21.7&43.2&44.0&50.0&43.8&43.4&87.4&87.4&87.8&87.3&86.8  \\
    &&&20 &8.5&13.4&19.9&9.7&9.4&15.5&23.2&37.0&18.4&18.4&39.8&43.4&53.6&41.2&39.9&85.3&86.8&87.0&86.3&84.9 \\
    &&&50 &NA&NA&20.5&NA&NA&10.4&26.3&40.1&18.4&17.9&32.2&45.8&59.0&38.3&35.5&78.3&81.6&82.6&79.9&79.4\\
    &&&100 &NA&NA&20.8&NA&NA&NA&26.4&40.2&NA&NA&14.1&46.5&63.8&27.3&30.0&73.0&83.9&85.6&80.9&77.2\\
\cmidrule(l){1-24}
    1&{Size}&0.7&10 &7.6&6.9&5.9&6.6&6.5&5.5&5.5&5.7&4.7&6.2&4.7&4.6&5.1&4.9&5.4&4.8&4.8&4.8&4.5&4.5 \\
    &&&20&6.9&7.3&7.4&7.7&6.9&6.3&7.1&5.8&7.2&6.0&5.1&4.7&4.9&5.6&5.6&5.4&5.6&5.3&6.0&5.1 \\
    &&&50 &NA&NA&7.2&NA&NA&8.6&6.2&5.5&7.2&6.3&7.8&4.6&4.4&5.9&5.0&6.5&5.3&4.9&5.6&5.8\\
    &&&100 &NA&NA&6.5&NA&NA&NA&6.1&5.8&NA&5.2&6.3&5.1&4.5&4.5&5.3&5.9&4.3&4.6&3.9&4.3 \\
    \cmidrule(l){1-24}   
    1&{Power}&0.7&10 &13.0&13.6&14.4&14.4&14.2&22.7&24.0&25.3&24.7&24.0&42.1&43.4&43.9&43.4&42.7&86.0&87.5&87.5&87.6&86.6 \\
    &&&20 &10.2&11.5&12.5&12.8&13.2&19.6&21.1&20.9&21.1&20.4&39.4&42.3&43.2&42.0&40.9&84.8&86.8&86.7&86.4&85.0 \\
    &&&50 &NA&NA&15.0&NA&NA&12.6&22.3&23.6&21.6&20.2&30.6&43.6&46.4&41.8&38.8&80.7&85.8&86.0&84.4&82.8\\
    &&&100 &NA&NA&13.2&NA&NA&NA&19.9&23.1&NA&NA&14.3&41.3&44.2&35.0&34.7&72.3&85.2&86.1&82.8&81.5\\
    \cmidrule(l){1-24} 
\end{tabular}
\begin{tablenotes} 
      \scriptsize 
      \item Notes: Size and Power for the different DGPs described in Section \ref{sec:sim} are reported for 1000 replications. $T=(50,100,200,500)$ is the time series length, $K=(10,20,50,100)$ the number of variables in the system, the lag-length is fixed to $p=1$. $\rho$ indicates the correlation employed to simulate the time series with the Toeplitz covariance matrix. NAs are placed whenever the post-OLS estimation was not feasible due to $\hat{s}>T$. The different choices of the tuning parameter $\lambda$ are reported as: AIC, BIC, EBIC for information criteria, $\lambda^{th}$ for the theoretical plug-in and TSCV for time series cross-validation as explained in Section \ref{sec:sim}. 
    \end{tablenotes}
\end{threeparttable} 
\end{sidewaystable}

\begin{center}
\centering
\begin{threeparttable}
\caption{Simulation results for the bivariate Granger causality test}\label{tab:biva}
\tiny
\begin{tabular}{ l S[table-format=3.0] cccc c c c c }
    \toprule
   
   DGP&{Size/Power}&$\rho$&${K\char`\\ T}$&50&100&200&500\\
   \cmidrule(l){1-8} 
&&& 10 &5.9&6.6&7.8&11.8 \\
    2&{Size}&0&20 &5.6&5.9&7.8&11.8\\
   &&& 50 &4.3&7.0&9.7&14.5\\
   &&& 100&5.5&6.7&8.9&13.9 \\
\cmidrule(l){1-8}
\end{tabular}
\begin{tablenotes} 
\scriptsize 
\item Notes: Size is reported for DGP 2, as described in Section \ref{sec:sim}, for 1000 replications. $T=(50,100,200,500)$ is the time series length, $K=(10,20,50,100)$ the number of variables in the system, the lag length is fixed to $p=1$. $\rho$ indicates the correlation employed to simulate the time series with the Toeplitz covariance matrix.
    \end{tablenotes}
\end{threeparttable}
 \end{center}
\clearpage

\section{Additional Material for the Empirical Application}\label{app_additionappl}



\begin{algorithm}
\caption{Heteroskedasticity-robust PDS-LM Granger causality test} \label{alg:HR-pdslm}
\begin{itemize}
\item[\textbf{[1]}] Obtain $\bX^{*\otimes}$ and $\hat{\bxi}^*$ as in Algorithm \ref{alg:pdslm}, and obtain $\hat{\bE}^{*\otimes} = \bX_{GC}^{*\otimes} - \bX_{\hat{S}^\otimes}^{*\otimes} \hat{\bGamma}^{*\otimes}$ as the residuals from the multivariate OLS regression of $\bX_{GC}^{*\otimes}$ on $\bX_{\hat{S}^\otimes}^{*\otimes}$.
\item[\textbf{[2]}] Compute element-wise products $\hat{\bpi}_j = \hat{\be}_j^{*\otimes} \odot \hat\bxi^*$ for $j=1,\ldots, \nGCvars$. Regress a vector of ones on $\hat{\bPi} = (\hat{\bpi}_1, \ldots, \hat{\bpi}_{\nGCvars})$ and compute $T \nGCed R^2$ from this regression.
\item[\textbf{[3]}] Reject $H_0$ if $T \nGCed R^2 > q_{\chi_{\nGCvars}^2}(1-\alpha)$, where $q_{\chi_{\nGCvars}^2}(1-\alpha)$ is the $1-\alpha$ quantile of the $\chi^2$ distribution with $\nGCvars$ degrees of freedom.
\end{itemize}
\end{algorithm}






\begin{table}[H]  %
\begin{center}
\centering
\begin{threeparttable}
\caption{Stocks used in Section \ref{sec:appl}}
\label{tab_stocks}
\scriptsize
\begin{tabular}{ l S[table-format=3.0] cccccc}
    \cmidrule(l){1-6} 
   {N.}&{Symbol}& {Issue name}&{N.}&{Symbol}& {Issue name} \\
   \cmidrule(l){1-6} 
1&{AAPL}& {APPLE INC} &16& KO&COCA-COLA CO \\
2&{AXP}& {AMERICAN EXPRESS CO}&17&  MCD&MCDONALD'S CORP\\
3&{BA}& {BOEING CO}&18 &MMM&3M\\ 
4&{CAT}& {CATERPILLAR}&19&MRK&MERCK \& CO\\
5&{CSCO}& {CISCO SYSTEMS}&20&  MSFT&MICROSOFT CORPORATION\\
6&{CVX}& {CHEVRON CORP}&21 &NKE&NIKE INC \\
7&{DD}& {DOW CHEMICAL COMPANY}&22&PFE&PFIZER INC\\
8&{DIS}& {WALT DISNEY CO}&23& PG&PROCTER \& GAMBLE CO \\

9&{GE}& {GENERAL ELEC} &24&TRV&TRAVELERS COMPANIES INC\\
10&{GS}& {GOLDMAN SACHS GROUP INC}&25& UNH&UNITEDHEALTH GROUP INC\\
11&{HD}& {HOME DEPOT INC}  &26&UTX &UNITED TECHNOLOGIES CORPORATION\\
12&{IBM}& {INTL BUS MACHINE} &27& V&VISA INC\\ 
13&{INTC}& {INTEL CORP} &28&VZ& VERIZON COMMUNICATIONS INC\\
14&JNJ& JOHNSON \&JOHNSON&29&WMT&WALMART INC \\
15& JPM&JPMORGAN CHASE \& CO&30&XOM&EXXON MOBIL CORPORATION\\

 \bottomrule 
 \end{tabular}
\begin{tablenotes} 
\scriptsize 
\item 
    \end{tablenotes}
\end{threeparttable}
 \end{center}
 \end{table}

\clearpage

\end{appendices}
\end{document}